\documentstyle[12pt,fleqn,epsf]{report}
\newcommand{\ria}{\rightarrow}
\newcommand{\Ria}{\Rightarrow}
\newcommand{\Lonlr}{\Longleftrightarrow}
\newcommand{\ro}{\varrho}
\renewcommand{\baselinestretch}{1.2}
\newcommand{\p}{\prec}
\renewcommand{\baselinestretch}{1.2}

\def\Re{\mathop{\mbox{Re}}}

\newcommand{\bea}{\begin{eqnarray}}
\newcommand{\beq}{\begin{equation}}
\newcommand{\eea}{\end{eqnarray}}
\newcommand{\eeq}{\end{equation}}

\begin{document}

\renewcommand{\thepage}{}

\begin{center}
\vspace{2cm}


\LARGE
{\bf Causal Sets, a Possible Interpretation for the Black Hole Entropy,
and Related Topics }

\vspace{3cm}

\large
Thesis Submitted in Partial Fulfillment  \\
of the Requirements for the degree\\
``Doctor Philosophi\ae''\\

\vspace{3cm}
\begin{center}
\begin{tabbing}
\hspace{8cm} \= hspace{6cm} \kill
\Large{\tt CANDIDATE}  \>  \Large{\tt SUPERVISORS}\\ \\
\Large{\tt Djamel DOU} \> \Large{\tt Supervisor:Prof.R.Sorkin} \\
                      \>  \Large{\tt Co-Supervisor:Dr.R.Percacci} 
\end{tabbing}
\end{center}

\begin{picture}(160,5)
\put(0,0){\line(1,0){160}}
\end{picture}
\vspace{1cm}

\vfill{\bf Trieste,  October 1999} \\

\end{center}

\newpage

\setcounter{page}{1}

\renewcommand{\thepage}{\roman{page}}
\newpage

\baselineskip=7mm
\begin{center}
\begin{flushright}
{\it
{\LARGE
{\bf \emph{}
  \\
}
}
}

\end{flushright}
\end{center}
\begin{center}
\newpage
{\Large{\bf Acknowledgments}}
\end{center}
I would like to thank my supervisors Dr.R.Percacci for his continued support
throughout the period of my study, and Prof.R.Sorkin with whom I have
worked closely in the final year of my Ph.D, and who provided the main
focus of my thesis.I should not forget Prof.J.Starthdee who was not just my 
diploma course dissertation supervisor but has always been a sourse of 
inspiration and encouragement.  

I would like also to thank the General Relativity
Group of Syracuse University for their kind hospitality at the initial
stage of the work. It is also pleasure to thank Abdus Salam Center for
Theoretical Physics, especially the High Energy Group and all the
professors of H.E.P Diploma course 94-95. I don't forget all my friends
and brothers from ICTP,  SISSA and Syracuse University, however the space left
in this  page  is too small to mention all the names....
$
$
\emph{To my parents, my sister  and my son, in the memory of Marouan.}

\tableofcontents

\newpage

\baselineskip=7mm

\begin{center}
{\Large {\bf Abstract}}
\end{center} 
The Causal Set hypothesis asserts that  spacetime, ultimately, is
discrete and
its underlying structure is that of a locally finite partial ordered set,
and macroscopic causality reflects a deeper notion of order in terms of
which   all the geometrical structure of spacetime must find their
ultimate expression. After reviewing the main aspects of
Causal Sets  Kinematics, and the recently developed Stochastic Dynamics.
We concentrate on possible  implications  in  the fields of cosmology and 
black holes. In the context of black hole, we propose a possible
interpretation of the entropy as the number of links crossing the horizon. 


\setcounter{page}{1}
\renewcommand{\thepage}{\arabic{page}}
\newpage

\bibliographystyle{plain}

\baselineskip=8mm

\chapter*{Introduction}
\addcontentsline{toc}{chapter}{Introduction}
General relativity (G.R) and quantum mechanics (Q.M) are the two major
pieces of our understanding  of the physical world. These two
theories are consistent  with the facts they were created to explain.
Quantum Mechanics or Standard Model of particle physics has found a
dramatic empirical success, showing that quantum field theory (QFT) is
capable of describing all accessible fundamental physics, or at least all
the non-gravitational physics. General relativity is capable of describing
all the large scale phenomena, and  it is, perhaps, the best tested
theory ever constructed in the history of physics. These two theories
offer us the best confirmed set of fundamental rules. More
importantly, there
aren't
today
experimental facts that openly challenge or escape this set of fundamental
laws. On the down side, when we  try to combine these two elements 
  we run into apparently insurmountable technical and conceptual
problems \cite{s,i, r} .

With the exception of the cosmological constant problem,  the need for
combining  the two theories cannot
be addressed directly to any observed property of the world that can
interplay
general relativity and quantum theory. This  stems from the fact
that
Planck length -defined using dimensional analysis
as:$=(G\hbar/c^3)^{1/2}$-
has an
extremely small value, which is well beyond the range of any foreseeable
laboratory-based experiments. And it seems that the only physical regime
where the effects of quantum gravity might be studied directly is in
the immediate post big-bang era of the universe which is  not the easiest
thing to probe experimentally. 

The motivations for  studying quantum
gravity or looking for  a more  fundamental  structure for
space-time 
are more of internal nature:for
example, the search for mathematical consistency, the desire for a
unified
theory of all forces, or the implementation of various quasi-philosophical
views on the nature of space and time e.g why do we live in 4-d
spacetime?
why is the universe  so big....etc \cite{i}. The theoretical
consequences of these two theories, and most  importantly the discovery of
the
quantum induced radiation by black hole, were  major reasons for studying
quantum gravity  to understand the end state of gravitational
collapse and the so-called information loss
puzzle that accompanies it,  and the origin 
of the black hole thermodynamics. On the other hand and interestingly, the
conceptual and
interpretive problems that confront
the quantization of gravity in any standard way, are the same as those
that
have plagued the foundations of quantum mechanics in general, ranging
from the measurement problem to the meaning of probability.

All this has led to the belief
 that a consistent theory that would combine quantum
mechanics and G.R may require a radical revision of our most fundamental
concept of spacetime and substance \cite{s,i}. 

Over the  last three decades the subject of quantum gravity  witnessed
much progress along different lines and the emergence of many  ideas that  
range from  very
conservative to  very radical ones . It is not, of course,  the aim of this
introduction to mention all these developments, but it is fair to say
that  String Theory
stands as the most developed  the idea and the one that  attracted
most \cite{r, pol}.The
recent understanding of its non-perturbative aspects, makes it the most
promising candidate for unified theory of all forces, hence  providing
a
quantum theory of gravity \cite{pol}. This of course by no means leaves no
room for
other approaches. At the end there is no reason to expect that different
approaches are necessarily  exclusive. On the other hand, it is also fair
to say that none of the present approaches has provided a satisfactory
answer
to
any of the long standing questions, for which those theories were first
created \cite{r}.\footnote{Well, string theory was not created to
provide an
answer to  those questions!} 

One of the approaches to quantum
gravity that witnessed a considerable progress and  has started getting
more
attention, and gaining popularity is the causal set approach. 

The causal set hypothesis asserts that  space time , ultimately, is discrete
and that its underlying structure is that of a locally finite, partial
ordered set which continue to make sense even when the standard
geometrical
picture ceases to do so.  

The causal set   idea  was first considered in the quantum gravity
context first by t'Hooft \cite{th}, however without being developed to
any
extent, Meyrheim considered independently the same idea  (may be
 from a pure geometrical point view) and developed
what on may call Statistical Geometry \cite{m} , although the line of
development
of
Meyrheim overlaps with many aspects of the Kinematics of the causal set
the underlying idea is  dynamically different  from the causal set one,
and
the issues  with which the causal set set is mainly concerned  were not
dealt with. 

It was not until the late 80's that the idea of causal set was taken up
seriously and studied systematically as an approach to quantum gravity. 

The causal set approach, and to many respect, can be said to be
less
developed compared with other approaches , as String theory,or Loop
Gravity, it
experienced a considerable progress on its kinematic aspects and
a significant advance has recently been made along the dynamical front.

The  basic motivational observation behind the causal set idea range from
physical-conceptual, and technical, to pure mathematical ones \cite{s,
spicmen, mey, luca}.

\section*{ Physical  Motivations}
The main reasons that make people find an underlying
discreteness of spacetime more natural than the persistence of a continuum
down to arbitrarily
small sizes and short time, can be summarized in what Sorkin has called
 the three (or four) infinities \cite{spro1}.

The first one is the infinity(ies) that plague all renormalizable quantum
field theories, 
when one tries to make prediction. These infinities can be dealt via
Renormalization , however this procedure has been for many workers not
satisfactory, and only an indication that something is going  wrong
at some energy
scale\footnote{The Renormalization can be better understood by treating
the theory as an effective theory valid up to some a small distance
cutoff and prediction can be made along the same line as the standard
renormalization procedure, if the energy probed is much smaller than the
cutoff [ Dirac]},  the non-perturbative study of $\phi^4$ theory which is an
essential ingredient of the standard model has shown (although not in
an absolutely conclusive way \cite{Wein}) that this theory can't be
consistent above some energy
 scale
unless it is trivial, the same results are believed to hold for any theory
which is not asymptotically free, Abelian gauge theory is an example, in
this sense Standard Model can't have any predictive power above some
energy scale . On the other hand the Renormalization procedure which
enable us
  to make predictions, fails to give an
unambiguous
results when gravity is included ( not necessary quantized), unless the
metric of the spacetime background is static or stationary, but there is
no
reason why a semi-classical metric should have this property \cite{i}. 

The second type of infinity  is the one that ruin any procedure to quantize
gravity,
and in this case one cannot apply the renormalization procedure that works
for what are usually called renormalizable theories, and here one is
not
able to make any prediction at all. However it should be noted that
when 
gravity is understood as an effective theory with   cutoff around the
Planck scale one is able to do a quantum realistic calculation and make
prediction \cite{do, reu, dong}

The third infinity is ,maybe, less appreciated than the others, and
often overlooked,
this
infinity occurs whenever one tries to account for the for contribution to
the black hole entropy, which can not be excluded on any physical ground,
unless a
short distance cutoff is introduced those contribution are inevitably
infinite. And it seems that this infinity is closely related to the
infinity met when one tries to quantize gravity \cite{suss}.

 The fourth infinity is  singularities in classical general
relativity  that are inevitable 
in many physically reasonable contexts, inside the black
hole or in
the big-bang, in the singularities our laws break down and one fails
to make any prediction.
 
Above these infinities there are more reasons that point towards
discreteness .
The fact that the we
have a dimensional scale, the Planck scale, can be understood as an
evidence
for discreteness
underlying the fundamental structure of "spacetime", this scale can't emerge
from a continuous picture "smooth manifold" of spacetime, (at least with
relatively simple topology), and would turn out be zero or infinity.
The recent development of string theory and the recent calculation of
loop gravity, are all pointing towards some kind of discreteness, 
or fuzziness of spacetime. 
The uncertainty principle combine with G.R connection
between mass and spacetime -curvature  in such a way that it is
impossible to measure the metric on a sub-Planckian scale, without the
apparatus collapsing into black, leading to sort of fuzziness in the
space, and  the continuum geometric picture of
spacetime ceases to make sense.

\section*{Mathematical Motivations }
Here I give a brief account for some known, although
insufficiently
appreciated, pure mathematical results showing how the classical
spacetime's
causal structure comes  very close to determining its  entire geometry , 
i.e
Topology, differential structure, and the conformal metric.Those results
once combined with the above physical motivations seem to lead naturally
to the Causal Set picture.     

In  general relativity a space time is usually assumed   to be a
4-d
connected  $ C^\infty$ Hausdorff manifold $M$ endowed with a Lorentzian
metric $g$ and time orientation.
Once the time orientation is given one can speak about future and
past.                                                             

A point $x$ is said to be in the causal future (past) of a point $y$
and we write $x \in J^+(y)$ $( x\in J^-(y))$  if there is a future (past)
directed causal curve (Time like or null) from $x$ to $y$.

Other basic definition is of the  chronological future (past) of  a point
$x$, denoted by $I^+(x) $ ($ I^-(x) $), defined as the set of points which
can
be connected to $x$ by future (past) directed timelike curve.

The causal relations defined above can be used to put a topology on $M$
called the $Alexandroff$ $topology$, this is the topology in which a set 
is
defined to be open if and only if it is the union of one or more sets of
the form $ I^+(x)\cap I^-(y), x,y \in M$. As $I^+(x) \cap I^-(y)$ is open
in
the manifold topology , any set which is open in the Alexandroff
topology
will be open in the manifold topology, the converse is not necessarily
true.
 However if the strong causality condition holds for $M$, \footnote{A
spacetime $M$ is said to be strongly causal if for all $p \in M$ and every
neighborhood $O$ of $p$, there exists  a neighborhood $V$ of $p$
contained in $O$ such that no causal curve intersects $V$ more than once}
 the Alexandroff topology  coincides with
the manifold topology of $M$, since $M$ can be covered by local causal 
neighborhoods as one can find about any point $x\in M$ a local causal
neighborhood.This means that if the strong causality holds, one can
determine
the topological structure of the spacetime by observation of causal
relationships, stated otherwise, the topology of spacetime is already
coded in its causal structure.

It is a standard result that by knowing  which points can communicate with
a
given  point $p$ one can determine the null cone in the tangent space of
$p$ .Once the null cone is known,  the metric can be determined
up to conformal factor.                                     

The above statements can be strengthed (relaxing the strong causality
condition)  and made more precise by the two following theorems .

$Theorem$\cite{hkm} : Let $(M_1,g_1)$ and $(M_2,g_2)$ be spacetimes and
$f:M_1\ria M_2$
a homeomorphism where both $f$ and $f^{-1}$ preserve future (past)
directed
continuous null geodesics.Then $f$ is a smooth conformal isometry.

$Theorem$\cite{mala} : Suppose $(M_1,g)$ and $(M_2,g)$ are past \emph{and}
future
distinguishing\footnote{ a spacetime is said to be future (resp.past)
distinguishing iff for all $x$ and $y$:$I^+(x) = I^+(y) \Ria x=y$
(resp.$I^- (x)= I^-(y) \Ria x=y$).}                 
spacetimes and $f:M_1 \ria M_2$  is a causal
isomorphism.Then $f$ is a homeomorphism .

Now the second theorem asserts that tow spacetimes having the same
causal structure (There is a causal isomorphism between them), and both
future and past distinguishing , then they must be homeomorphic, hence
topologically equivalent, and this isomorphism with its inverse can be
shown
to preserve future (past) directed continuous curves which in turn would
imply that the isomorphism, being also a homeomorphism, preserve null
geodesics, and by the first theorem must be a conformal smooth isometry. 
This result is of great interest in its own right since \cite{mala,bs},
  the bottom line of the above assertions is that the  two
spacetimes have the same conformal structure. So given a
spacetime obeying suitable smoothness and causality conditions one can
retain from all its structure only the information embodied in the
causal order .Then one can recover from the causal order not only the
topology of the spacetime but also its differential structure, and the
conformal metric.

Now what is a causal order? It is simply a partial ordering between
points in the spacetime, and the above construction can be turned "on
its head" with the partial ordering relation construed as primitive, and
in fact it is natural to guess that in reality one should derive from
the causal order the metric rather than the other way around.
The problem with this new construction is the lack of information to
determine the conformal factor of the metric.In other words, we get from
the causal relations the metric, but without its associated volume.
There seems to be no way  to over come this problem within the
context of the continuous space time (in fact, this is all what can hope
for since all conformally equivalent Lorentz metric on a manifold induce
the same causal structure), but as we discussed in the physical
motivations, there are many reasons to doubt that spacetime is truly
continuous. If instead we postulate that a finite volume of spacetime
contains (a large ) but finite number of points (events or space time
atoms) then
we can -as Riemann suggested- measure the volume of  a region in
spacetime by merely counting the number of points it contains, provided
some density set by some fundamental scale \footnote{We will turn to
this point in the coming sections since as we shall see will turn out to
be a little subtle.}. 

So by putting the conceptual and the mathematical motivations together we
arrive naturally to a new "substance" (Structure) underlying spacetime
is what
Riemann might have called an "ordered discrete manifold" but we will call
"causal set " or what is known in discrete mathematics as Partially 
Ordered
Sets ( Posets). In this view the volume is a number, and macroscopic
causality reflects a deeper notion of order in terms of which all the
geometrical structure of spacetime must find their ultimate expression.

At the end of this section it is intriguing to note, assuming the that
this proposal is step in the
right road, how the
ultimate (fundamental) rules of space-time, may
find their ultimate expression in such a simple  mathematical object as
the Partially Ordered Sets.

$$
$$

This thesis is organized as follow, in the first chapter we review the
Kinematic aspects of the causal set that was developed by 
Sorkin
and Bombelli, Daughton, Meyer including some relevant recent results.
In the second chapter, after reviewing the
dynamical aspects including some old proposals and the recently developed
stochastic  dynamics, we outline possible  cosmological
implications
of this
dynamics. The third chapter after brief review of the main aspects
of
Black Hole thermodynamics and the notion of Entanglement Entropy , we
propose a possible interpretation of the black hole entropy and we
interpret it on
the light of entanglement entropy and information theory and we conclude by 
some remarks on other possible interpretions of the result obtained.

 In appendix A  a technique is developed for calculating
volumes needed to ensure some causality conditions in 4-d flat spacetime.

Appendix B contains a detail account for the derivation of the results 
 that appeared in Chapter 3.

\chapter{Kinematics Of Causal Sets}
In this chapter, I  review the basic definitions and terminology of posets
relevant for causal set theory, focusing mainly on
the physical aspects  which have been developed by Bombelli, Meyer,
Sorkin and Daughton \cite{bs, luca, mey,daug}.

As in any new developed physical theory, the first step towards a complete
 understanding of its physical insight is to understand 
 the Kinematics. Here by the Kinematics of causal sets we mean
 the study of the general structure of  causal sets, more precisely  to
see how to make contact between causets and spacetime. As it turned out,
concepts such as length, topology, and dimension make  little sense for generic causal set; 
so it is necessary to understand in what circumstance they do emerge
 in a purely causal manner. 
 
In order to be able to address  such question and  many other
physically relevant questions we will encounter later on,
new mathematical concepts had to be introduced, above the already existing
ones in the theory of partial ordered sets which in itself forms a
category .

\section{ Basic definitions and New concepts}

A \emph{partially ordered set} (or a poset for short) ${\cal P}$ is a
set endowed with a relation  $\p$ satisfying the following axioms:

(1) reflexive : $ \forall$ $ p$ $\in$ ${\cal P}$ , $p$ $\p$ $p$

(2) transitive: $ \forall$ $ p,q,r$ $\in {\cal P}$ , $p$ $\p$ $q$
$\p$ $r$ 
$\Ria$ $p$ $\p$ $r$;

(3) antisymmetric : $\forall$  $p,q$ $\in$ $P$ , $p$ $\p$ $q$ and
$q$ 
$\p$ $p$ $\Ria$ $p$ $=$ $q$.

Notice here that the refelxivity axiom is a matter of convention and we
could have used instead the irreflexive convention\footnote{The
irreflexive
convention is  the one used most in  simulations.} 
                             
If $p\p q$  , $p$ ($q$) is said to be in  the past (future) of $q$ ($p$).

The future (past) of an element $p$ is the set of  all element which are in the
 future
(past) of $p$.

An  \emph{interval} (\emph{Alexandrov set}) $J(p,q)$ defined by two 
elements
$p,q$, 
with $p\p q$, is the intersection of the future of $p$ with the past of $q$:
\beq
J(p,q):= \{r\mid p\p r\p q\}.
\eeq

An interval is a special case of \emph{induced subposet}, a subset
${\cal P}'
 \subset {\cal P}$ in which two elements are related, $ p\p q$, iff
they are 
related in ${\cal P}$.

An element is called \emph{maximal (minimal)} if there is no elements to
its  future (past). 

A $chain$ is a subset of $P$ in which any two elements are related ( a 
totally or linearly ordered subset).

A $locally $ $finite$ poset is a poset ${\cal P} $ such that all the
Alexondrov 
sets are finite (have finite cardinality) .

A useful concept for the description of locally finite posets is the 
$link$,
if the interval $J(p,q)$  contains no element except $p$ and $q$
themselves, and we
 say there is a link between 
$p$ and 
$q$, or $q$ cover $p$ and we write $p\p\cdot q$, this notion is similar to 
that of the nearest neighbor for  lattices embedded in manifold with 
positive-definite metrics. It should be remarked that, for a general 
poset, there is no metric meaning associated with this notion of closeness 
in the partial order, although in some cases it is related to a notion of 
closeness in a Lorentzian metric.

The knowledge of all links is equivalent to knowledge of all relations 
among elements: $p\p q$ iff there are elements $q_1, q_2,,......q_n$ such 
that  $ p\p q_1\p\cdot q_2\p\cdot,,.....\p\cdot q_n\p \cdot q$.

A $Path$ between $p$ and $q$  is a $maximal$  $chain$ between these 
elements, i.e, a chain made of links, like $ p\p q_1\p. 
q_2\p\cdot,,,.....\p\cdot q_n\p\cdot q$.Two paths between $p$ and $q$ need
not 
have the 
same $length$  (number of links), and we will call "maximal path" one 
with the maximum length.
 
A \emph{connected poset} is one for which there is at least one path which
can be 
followed to go "continuously" from any given  element to  any
other element allowing backward.

A $null $ $path$ is path which is also the interval formed by its
endpoints (in Minkowski space, the only points causally related to two 
null-related points are those on the null geodesic joining them).

An $antichain$ is any subset of a poset ${\cal P}$ in which no two
elements are 
related; whereas a \emph{maximal antichain} is an antichain such that no 
further element can be added to it.

The $width$ of  a poset is the size of 
the largest antichain, and the $height$ is the length of the longest 
path.

A $join-independent$ subset of ${\cal P}$ is a subposet ${\cal P'}$
of ${\cal P}$ such 
that, for every $p'\in {\cal P'}$, there is a $p\in P$ , with
$p'\not\prec p$,
but 
$p$ is to the future of every other element of ${\cal P'}$ (such a
${\cal P'}$ is 
always an antichain).  The $breadth$ of ${\cal P}$ is the size of its
largest 
join-independent subset.

A $loop$ is a pair of paths between two elements, which do not meet 
except at their endpoints.

$EXAMPLES$

(a) the totally , or linearly , ordered set with $n$ elements,

$$
{\cal P}^{l}_N =: \{ a_i ; i= 1,...N | a_i\p a_{i+1} ; i = 1,...,N-1\}
;
$$
(b) the "fence" poset of size $2N$,
$$
{\cal P}^{f}_{2N} =: \{ a_i, b_i ; i= 1,...N | a_i\p b_{i} ,\forall i;
a_i\p 
b_{i-1}\} ; 
$$
(c) the set $2^{[n]}$ of all subsets of $[n]$ ( a set of n-elements) 
ordered by inclusion i.e., $A \p B$ $\Lonlr $ $A \subset B$ as sets; 
this is called the binomial poset $B_n$ .The binomial posets are of a 
particular importance, because of the results known about them, and the
possibility of realizing any poset as
 a subposet of some binomial poset .
 
A $ Hasse $ $diagram$ for a poset is a one-to-one mapping from the 
poset to the two dimensional Euclidean space such that each element is
mapped 
to a point and a link to a line, with a point $p$ is placed above $q$ 
if $q\p p$; Fig (1.1) illustrates this definition for the following poset.
$$
S= \{ a, b, c, d, e, f\}
$$
With the partial order :
$$
a\p \{ b,c,d, e, f\}; 
b\p \{ d, e\};
c\p \{ d, e, f\};
d\p \{ e\}.
$$
\begin{figure}[t]
\epsfxsize=3cm
\centerline{\epsfbox{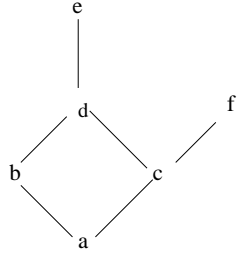}}
\caption{Hasse diagram for the poset $S$ defined below.}
\end{figure}

Given the above examples of partial ordered sets it is  natural to ask how 
many different posets one can construct with $n$ elements. This question 
turns out to be  
an unsolved problem in combinatorics  for a generic $n$: in fact the 
answer is 
known only  for $n< 14$ and for large $n$ and asymptotic formula is 
available.   
\beq
 D_n \approx C 2^{\frac{n^2 +6n}{4} } e^n n^{-n-1} 
\eeq

 where  $D_n$ being the number of different posets and $C$ is a number of
order one.

Having introduced the above terminology we are now ready to give a precise
definition of our central object. 

A \emph{Causal set} (or \emph{causet} for short) is a locally finite
connected poset.

From here on we will only be interested in causets.

\section{Realizations of posets}

It is often useful to think of a poset in terms of some realization of
it,
by realization of a poset we mean a mapping of poset to another 
mathematical object preserving  all the order  
existing in the poset. 
Here one should bear in mind that  our aim is not to look for applications of 
posets or to find their realizations,  the realization serves only as a 
mathematical tool to gain more insight to the posets structures through the 
known 
mathematical structure of the realization, and at the end we should be 
able to formulate the physics of causal set without 
any reference to their realization. However as a first step towards the 
understanding of the laws (Kinematics and Dynamics) of causal sets, 
 we will often think of a poset ${\cal P}$ as realized in terms of
points in a 
Lorentzian manifold $({\cal M},g)$ 
this realization we will call a conformal\footnote{We use the term
conformal to emphasize that only the conformal metric is used.}
realization 
of ${\cal P}$. In a physical term, since 
the low energy physics of the causal set dynamics, (or whatever theory
which
is  taken to be a candidate  for the  "quantum" theory of gravity) must
give us the  the Lorentzian geometry , and as well known the low energy
physics may 
constraint very much the high energy physics, so it is natural to start by 
understanding the causal realization.

 A $Conformal$ $realization$ is defined as follow,

let ${\cal P}$ be a poset and $({\cal M},g)$ a Lorentzian spacetime with
metric $g$, a 
causal realization is a map $f : {\cal P}\ria M $  with $f(p)\in
J^-(f(q))$ iff 
$p\p q$.

Another useful and common realization is the \emph{linear realization} in
which elements of ${\cal P}$ are
mapped to points in $\Re^n$ for some $n$ and the ordering is
reproduced
by
the partial ordering induced by the coordinates: $f :{\cal P}\ria
\Re^n$,
with
$f^i(p)\le f^i(q)$, $\forall i$ iff $p \p q$.

\section{Causal sets and differentiable manifolds}

 Recall  that our a proposal is to take this causal set as
the matter underlying spacetime and as mentioned in the introduction our
large scale perception of space time is that of a continuous manifold .
The explanation of the emergence of this structure is by now
 one of the (if not the most ) most fundamental questions in theoretical
physics. 
Answering this question in the casual set
approach
would
need a full understanding of the dynamics of causal set, but what has
been done in the context of causal set  is to try to address more
moderate   questions, as for instance, to what
 extent the elementary structure of causets could give rise to Lorentzian
Manifold in some suitable approximation? 
and others questions we will encounter later on.

I first start by giving some definitions which are used in formulating
such questions in a causal set terms.

\textbf{Definition}

A causal realization of a poset ${\cal P}$ in a spacetime
$({\cal M},g)$ is said to be a
\emph{faithful embedding} if:

(1) The embedded points are distributed  uniformly with respect to the
volume form on $({\cal M},g)$ with density $\ro$ and

(2) The characteristic scale over which the continuous geometry varies
appreciably is everywhere much greater than spacing between the embedded
points.
 
And $({\cal M},g)$                                               is
said to be
associated to ${\cal P}$.

Let me now try to explain and motivate the above definition.

By uniformly distributed points we mean if one take an arbitrary
Alexondrov neighborhood of size $V$ in ${\cal M}$, the  number
of
embedded points in it is $\ro V$, within the Poisson-type fluctuation
which could be expected from a random "sprinkling" of points, so  the
probability distribution for
having $n$ points in the neighborhood is a given by Poisson distribution ,
\beq
{\cal P}(n)= \frac{(\ro V)^n e^{-\ro V}}{n!}
\eeq

that is  to say that the image of the causal set looks as an outcome of
stochastic process :  points sprinkled uniformly and independently,
i.e, there is no preferred region in the manifold as far as the density is
concerned, the probability of
finding a point in region of finite volume depends only on the volume of
the region.

The second condition can be taken simply to require that each embedded 
point have a suitable neighborhood, which is approximately flat, see
[Bombelli \cite{luca} for more discussion of this point and possible
implications].

The above definition for a faithful embedding seems to be the only one
consistent with our intuition and general covariance for a manifold to be
a continuum approximation to a causal set, this can be seen as follow.  

Two elements related in the causal set their image will be causally
related; unrelated elements will have a space like related images. 
The
order relation in the causal  corresponds to the time orientation in the
manifold.

If two
elements are linked in the casual set then their images under an embedding
will be nearest neighbors :the Alexandrov neighborhood they determine will
contain the image of no other element in the causal set.

A chain is embedded as a sequence of causally related events -there is
a causal curve passing through them.

If one tries to define reasonable  non-uniform 
Lorentzian density in invariant sense will be forced in the end to have a
uniform density everywhere, since even a small Alexandrov neighborhood
can
extend between "far apart' regions in the manifold (think for instance of
Alexondrov neighborhood  between two points which look approximately
null in some
frame, for a Minkowski spacetime) and to  produce a 
varying density in invariant
 way one would have to make it uniform in all the direction, 
arbitrarily close to the null ones,and since the light cones of all 
points meet\footnote{This may not actually be true, but the argument can 
be pushed further}, thus we end up with
the density
cannot vary at all\footnote{In fact, it was known that a Lorentz
invariant
lattice would have to be random \cite{tdl}.}. 

Note here that, this condition justifies our use of the locally finite
poset, it allows to interpret the volume of a region as
the number of points in it, from which one would recover the conformal
factor of the metric.

The reason for imposing the second condition in the definition of
faithful embedding  is not just that the small
lengths would not be meaningful, but also that the causal embedding with
the first conditions by themselves would be far from determining a unique
approximating spacetime: given any manifold with the right causal
structure
we would arrange the density to have a constant value by setting the
conformal factor appropriately; but in doing so, we would in general
introduce an unreasonable large curvature, or other small characteristic
lengths, in other words the causal embedding and the uniform density
condition alone would determine the continuum geometry leaving a room for 
   arbitrary variations on small scales (spacing between embedded points
or smaller) .

Now, let us go back to our starting question , i.e. To what extent a
causal set determines the properties of an associated manifold?

Having given the above definitions this question can be reformulated as
follow,

Given a manifold into which ${\cal P}$ can be faithfully embedded , how
unique
are the topology , differentiable structure and the metric?

It has been conjectured \cite{bs} , that
 the topology and the differentiable  structure are
unique, and the metric is determined up to "small variation".

This is the main  conjecture of the causal set approach, 
although it is not  stated in a
mathematical
rigorous way,
there are some evidence supporting it. For instance the study of the
dimension of causal sets  shows (almost) the dimensionality of the
associated manifold is unique, and there  some  arguments
supporting the uniqueness of the topology and the metric.

There is a very important point to note  here, the conditions imposed in
the definition of the faithful embedding are  very strong and in general,
 for a given causal set ${\cal P}$, there will be no manifold ${\cal
M}$ in
which ${\cal P}$ can
be faithfully embedded; in fact we expect almost all the causal sets not
to be faithfully embeddable anywhere, for example a crude estimation   
 shows that only a vanishingly small  fraction of causal sets with
a large $N$ number
of elements  can be faithfully embedded in $n$ dimensional Minkowski space,
i.e., $c 2^{n Nlog_2N -N^2/4 -(3/2 +log_2 e)N +log_2 N}$. 
	
The last conclusion may seem at first sight a disadvantage, however this turned out
 to be rather an advantage as we  will see in the coming sections,
   for the time being one is only interested
 to show that the causal set has a structure rich enough to imply all the
geometrical properties as we attribute to continuum spacetime, in other word
we are interested in the
uniqueness of the continuum approximation of causal set, when it exists.

\subsection{Dimension of causets}
One of the basic aspects  of the manifold is the dimension, so an obvious
first question is whether there is a good way to recognize the effective
continuum dimension of a causet.  

In general there is no very meaningful intrinsic definition of dimension
for a poset, however, it turns out
that the most useful definitions of dimension for posets are those in
which the dimension in some sense is not a property of the poset itself
only, but of the poset and some realization of it.    

First I will give two  definitions, one is of the conformal dimension 
 and  the combinatorial or linear
dimension, the former because of its direct physical meaning and relation
to what we will call later the physical dimension, and the
latter because of the many results known about it, although may turn to
have a
little to do with the causal one for dimensions higher than 3 as we shall
see. Than I will discuss the  "Physical dimension" and  the fractal
dimension (statistical) which should coincide with the physical one if the
causet is faithfully embeddable.

 \textbf{Linear dimension}

As  remarked earlier one of the natural realization of a poset is the
linear one, and we define the \emph{linear dimension} to be simply the
smallest
$n$ for which there exists a linear realization in $\Re^n$.This
dimension is
well defined since any poset has some linear realization for some $n$, for
example the binomial poset $B_n$ has a linear dimension $n$, and it is
easy to show that any poset with no more than  $n$ element can always
be realized as a subposet of $B_n$.

Moreover many upper bounds on the linear dimension have been set:

$ldim {\cal P} \le ldim {\cal P}' +1 $ where ${\cal P}'$ is
obtained from
${\cal P}$ by removing a
single element.

$ldim {\cal P} \le ldim ({\cal P}\setminus C) +2 $ where $C$ is a chain in
${\cal P}$.

$ldim {\cal P} \le width{\cal P}$.

$ldim{\cal P} \le (max{2, |\cal{\cal P} \setminus A|)} $ where $A$ is an
antichain
in ${\cal P}$. 

$ldim{\cal P} \le | {\cal P} |\setminus 2 $ for $ |P|\ge 4$.

$ldim {\cal P} \le 1+ width ({\cal P} \setminus M) $ where ${\cal M}$ is any 
set of maximal or minimal elements.

$ldim {\cal P} \le 1 + 2width({\cal P} \setminus A)$ where $A \not=
{\cal P}$ is antichain in ${\cal P}$. 

\textbf{Conformal dimension} 
The \emph{conformal dimension} of a causal set is the smallest $n$ for
which
there exists a causal embedding  in $n$-dimensional Minkowski space.

This defintion is concerned only with existing of causal embedding in
Minkowski space with no further condition, however, the study of the
conformal dimesnion is of a particular importance since
one of requirement of a  faithful embedding was that
each point would have suitable neighborhood which is approximately flat,
this neighborhood will contain a few points and due statistical
fluctuation these points need not be recognizably uniformly distributed,
hence
looking  like a small size causet embedded in a
Minkowski space time, not necessarly with uniform distribution.
So  suitable such subsets will contain the
information on the 
dimension of manifold in which the causet can be embedded .

The above definition although it may seem natural, need not be defined for
all posets, its definiteness lacks on some unproven (may be wrong!)
conjectures, for instance if one could prove that: \emph{adding one maximal
or
minimal point to a causal set can increase its conformal dimension at most
by
one}; than the conformal dimension would defined for any causet.

Note here that an equivalent version  for the above statement 
holds for linear dimension, the One -point
removal theorem, i.e, removing one point from a poset can decrease its
linear dimension at most by one.

This led  to the definition of special posets which encode all the
information about the linear dimension.

\textbf{Definition}: A poset is linearly \emph{d-irreducible} if it has a
linear
dimension $d$ and
the removal of any element reduces its dimension.The $n$-irreducible poset
with  minimum size is called \emph{$n$-pixie}.

It follows from this definition and the One-point theorem that each poset
with linear dimension $d$ contains a $d$-irreducible poset.
Thus the $d$-irreducible posets are $obstructions$ to have a linear
dimension less than $d$ \footnote{It is interesting to note here
the similarity between the role
of the
irreducible posets and the role of special  subgraphs which prevent a
graph from being planar , see \cite{mey} and reference therein
for similar problems in combinatorial
and Algebraic geometry.}.
By the one-point removal theorem the dimension of an $d$-irreducible poset
is reduced by one upon the removal of any of its elements.

At this point it is natural to ask to what extent the linear dimension is
related to
the conformal one?  

Although no  proof has been provided for the one-point removal theorem in
the
case of conformal dimension one might try to define conformally 
irreducible causal set,
along a
similar line and obtain  similar results 
when the
causal embedding exists, as for instance, a causal set with a causal
dimension $d$ must contain a $d$-irreducible causal set.This follow from
the fact that a weaker version of One-point removal theorem holds, since
if  a causal set were embeddable in $d$-dim Minkowski space we know that
by
removing elements  we would not increase the dimension , so by removing
elements whose removal does not reduce the conformal dimension, and when
no
such
elements remain the causal set is $d$-irreducible. In particular we have
the
following theorems;

\textbf{Theorem}\cite{mey}:\emph{Every 3-irreducible causal set is a
3-irreducible
poset and
conversely.}

\textbf{Theorem}\cite{mey}: \emph{A  causal set can be
embedded in two dimensional Minkowski space iff it has linear dimension at
most two.}

Naively one would hope that similar results hold in higher dimension but
there are counter examples, a simple example is shown in fig (1.2), where a
poset
which can only embedded linearly in four dimension or bigger but it
can be embedded
in three dimensional Minkowski space\footnote{Note here that we used the
fact that if a realization of a causet is found in terms of balls in $n$
dim Euclidean space ordered by inclusion, then it can be embedded in
$n+1$ dim
Minkowski space.}\cite{mey}.

\begin{figure}[t]
\epsfxsize=8cm
\centerline{\epsfbox{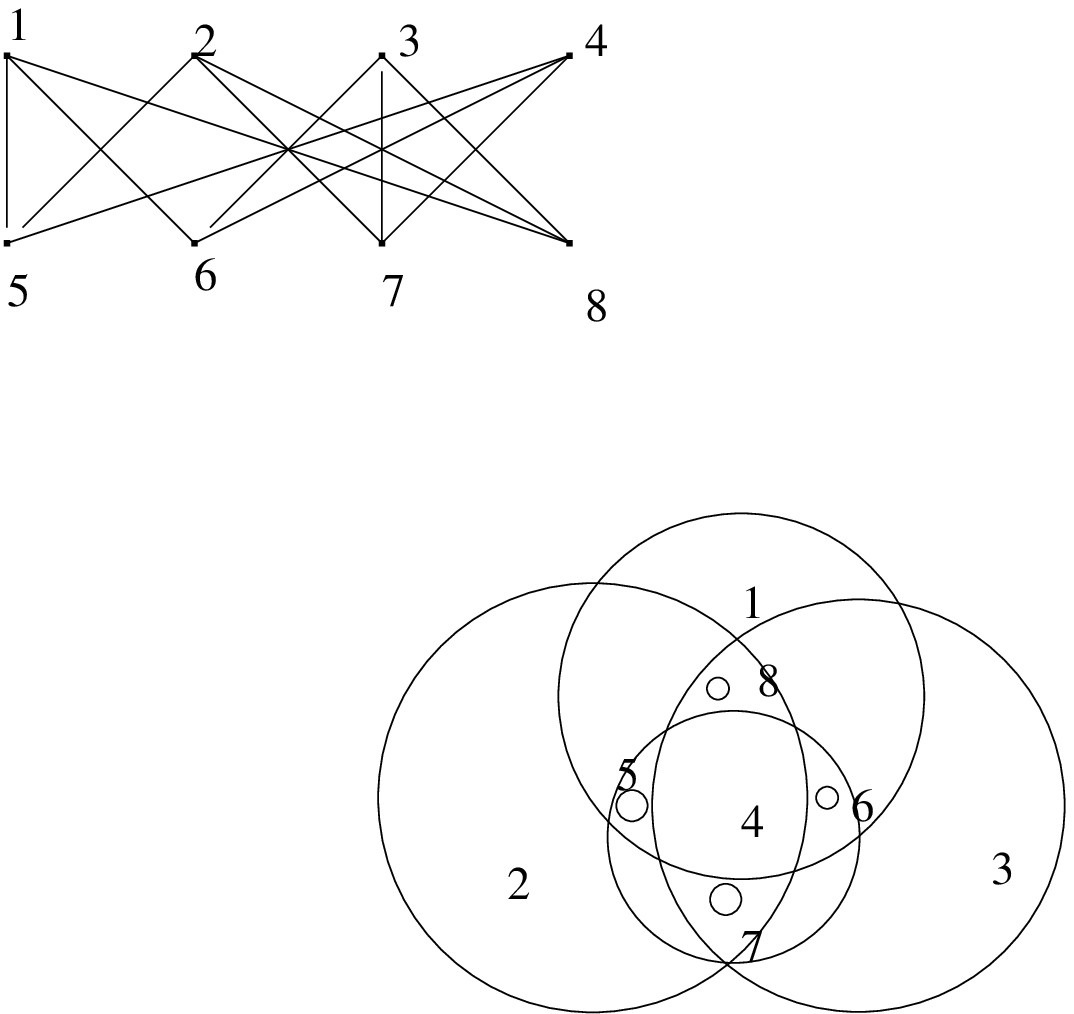}}
\caption{This diagram shows the explicit causal  embedding in 3-dimensional
Minkowski spacetime of a poset having  the above Hasse diagram, via its
realization in terms of balls in 2-dimension}
\end{figure}          

Now it is natural question  to ask  if there exist at all causets which can
be only in higher dimension $d\ge 4$, in other words,  are there
causets
capable of characterizing arbitrarily  great spacetime dimension?(Are
there $d$-irreducible causets with $d\ge 4$?) 

First it has been proven by Meyer \cite{mey}, that \emph{if} 
 a binomial $B_n$ 
poset  with $n\ge 6$ can be causally embedded in  a Minkowski space with
dimension $d$,
than $d$ is at least as large as some number, more precisely we have the
following theorem .

\textbf{Theorem}:\emph{The conformal dimension of binomial poset $B_n$ is
at
least as
large
as the minimal $d$ satisfying :}
$$
{n \choose d} \ge \sum_{i= d+1}^n {n \choose i}.
$$

In some sense this theorem doesn't prove any thing concerning the question
we asked, since it could be (and it seems likely to be [see cite{spro1} 
and references therein])
that there is no
causal embedding at all for such a binomial posets, however ,
it has been shown Brightwell and Winkler \cite{brig},  that the poset
$P_n$ made by
retaining from the
binomial poset
$B_n$ 
only  the relations embodied in the following rule:
$$
p_1\p p_2 \ria p_1\subseteq p_2  \qquad\mbox{and}\qquad  |p_1| =1
\qquad\mbox{or}\qquad |p_2| = n-1
$$

can be  embedded in  $n$ dimensional Minkowski space (or higher of
course).
and not in $n-1$ dimensional Minkowski space.

The last result shows that irreducible causet of arbitrary dimension do
exist,  they can be obtained simply by removing points from the above 
defined family of causets until they become embeddable in lower
dimension, their dimension
may decrease by two or more upon the removal of one element, however.

At the end of this section it may seem surprising that with our intuitive
definition of the conformal dimension we have not even able to proof its
existence
(Definiteness) of  causal sets,  moreover   it is very possible that the above
conjectures
may turn out to be wrong, and only some special posets have a definite
conformal dimension, in fact it has been suggested \cite{spro1} that , the
binomial
posets ($n\ge 6$) may not embeddable in any dimension\footnote{{\bf Note
added in the proof}: After this  
  thesis  was completed we learned that in fact there are causets that
won't embed in any Minkowski space \cite{t}} (however, if one allows
for highly curved spacetime the causal embedding could be possible
\cite{daug}).
The subject of causal embedding of posets is still under investigation
\cite{daug}.   
Finding the characteristic causets  similar to the family we mentioned
above (or at least their general properties) would, in
fact, shed a light on the construction of the dynamics of causal sets.

\subsection{The physical dimension and fractal dimension}
Recall that our starting point was that a faithful
embedding would give us almost all the geometrical information about the
associated manifold and in particular the dimension, however
 as remarked earlier we don't expect all posets to have a
faithful embedding -actually almost none will have- and any definition of
the dimension based on the dimension of the manifold in which the causal set
can be faithfully embedded would be far from  giving a well defined
mathematical entity.A more practical definition of dimension has been
proposed by Meyer , Bombelli, \cite{luca, mey} and is called \emph{fractal
dimension}\footnote{The word $fractal$ is to reflect the fact that the
dimension needs not
be an integer. }
or  \emph{Myrheim-Meyer dimension}, which  should reduce to the
 the \emph{physical dimension} defined to be the dimension of the
associated manifold when
it exists. 

The fractal  dimension is
statistical in character 
and here I give a very brief sketch of it.

The fractal dimension is defined in a $quasilocal $ way, being
associated
with different regions in the causal set rather than the whole set. 
 What one merely does here is to take  an
interval (small enough)   and count the number of  elements
it contains , and
define  a quantity using the interval for
which theoretical( statistical)
dependence on the  volume of the
interval  and the dimension
is known for causet  embedded in Minkowski space. Then by inverting the
relation governing the dependence one can in principle  obtain an
effective dimensionality for this interval. Now, if one can cover
the a large region of the causal set by intervals all roughly with
same size, and all yielding almost the same result for the dimension 
one can associate this dimension to the casual set, however if different
regions gave different results one would conclude that we were dealing
with a set which is not faithfully embeddable.
In general we expect  the resulting dimensions to agree, if the
causal set we consider is faithfully embeddable in a curved spacetime,
provided they are derived from intervals small enough that the
result is not significantly affected by curvature, however, these methods
are of course subjected to statistical fluctuations which should be small
for large intervals,  but it may happen that the interval
with 
the right size to offset statistical fluctuation becomes large not to be
negligible with respect to the radius of curvature of space-time.The above
mentioned
limitation on the applicability of these methods may be turned to an
advantage and, in fact, used to estimate the curvature\cite{ mey,
luca}

The following example illustrate how the idea of fractal dimension works.
  
For  $N$ points sprinkled uniformly in an interval in
Minkowski spacetime,
 the
expected number of elements to the future of a  point $x$ is $ V(J^+
\cap A)$\footnote{We work with unite density}, and the expected number of relations
in $A$ is
$$
<N_{rel}>  = \int_{A}  V(J^+ \cap A) dV = V(A)^2 f(n)
$$
where 

$$f(n) = \frac{\Gamma (n+1) \Gamma (n/2)}{4\Gamma (3n/2)}$$
Now, since the expected number of elements in $A$ is $ <N(A)>= V(A)$
, we may count the number of elements to approximate the volume , count the
number of relations to the right hand side of the above equation, plug them
in and invert $f(n)$ to obtain an approximation for the dimension of the
Minkowski space in which the causet can be faithfully embedded. 

 Numerical simulations  show that the Hausdorff dimension can be
effectively computed , and numerical results are available for points
sprinkled  in 2 ,3 ,4 Minkowski space and de Sitter space, with the computed
dimension
converging rapidly to the true one as the number of sprinkled points
becomes larger, the result being in general agreement with 
 error allowed by statistical fluctuation \cite{mey}. An   interesting 
result being for the case of "Kaluza -Klein cylinder" of 1+1-dimension.
In this example one sees clearly how the effective dimension falls gradually
from 2 down to 1 as the size of the interval in question increases , showing
how coarse-graining \footnote{ See next section for definition}
can
induce "dimensional reduction", and how such scale-dependent dimensionality
 or more generally topology becomes a perfectly well-defined concept in
the
context of causal set, more about this later.

\subsection{More topological information}

In the previous section we have revealed  that the causal
set contain
dimensionality information, now what about the 
more general
question,i.e, Does a causal set ${\cal P}$ encode the topology  of the
manifold it
is
sprinkled in?

This aspect  of the Kinematics of causal set is by far the
less developed, at least compared to the dimension. There have
been
many suggestions and interesting ideas  without being concretized, however.

Here I sketch very briefly some suggestive ideas of how one can hope to
extract
the "germs" of a notion of incidence, around which the topology is
essentially built, from the causal set.
The general idea is to try to
construct a finite topological space, or a finite simplicial complex, or
more  generally cell complex, independently of any mapping between the
causet and manifolds ,i.e,depending only on the structure of the causet
itself.

In general the topological spaces constructed from the causet have some
topological properties
which are not related to the topological spaces we may hope to get, as for
instance, if one  defines a simplicial complex from ${\cal P}$ by
associating with
every
element a vertex, and with every chain of length $k$ a $k$-simplex,
whose
vertices  are those associated with the $k+1$ elements in the chain.Thus,
if a j-chain is contained in a $k$-chain $(j < k)$ then the corresponding
$j$-simplex is a face of the $k$-simplex.The resulting simplicial complex
is called the \emph{order complex} of ${\cal P}$ \cite{luca}  and its
dimension
is the length
of the longest chain and hence the height of the causal set, and this
dimension  is of course not related to the dimension of the topological
space we were hoping to get.

On the other hand given a causal set sprinkled in a manifold ${\cal M}$, we can
define a cover of ${\cal M}$ by choosing a collection of Alexandrov
neighborhood
which cover it.Now a cover of a topological space defines a simplicial
complex, and one way to define is to construct the $nerve$\footnote{A
nerve of a finite cover is an abstract simplicial complex obtained by
associating a vertex to each open set and declaring that the opens sets
have non-empty intersection.} of the open cover, and we might hope that
the nerve of an appropriately constructed cover of ${\cal M}$ by Alexondrov
neighborhood would be a simplicial complex with the right topology to be a
triangulation of $\cal M$.The problem with this construction is that we
don't
know if such a good cover always exists in term of Alexondrov      
neighborhoods  whose intersection properties are expressible just in term
of $\cal{\cal P}$.

A more direct way to construct a simplicial complex using the sprinkled
causal set, is try to find an Lorentz analogue of the known
Dirichlet-Voronoi construction  for random lattice in Euclidean
space,
which yields a simplicial complex and an associated dual cell complex.
Or more general construction like the one used by Lee \cite{tdl} for
Euclidean
spaces
(based on arbitrary convex regions which differ
from each other by a translation and/or dilatation, and which is
equivalent
to Dirichlet-Voronoi when applied to spheres), this construction can be
repeated for stably causal space-time  with a sprinkling of points
satisfying the faithful embedding conditions, using a global time 
function $t: M \ria R$ (which is always defined for causally stable
spacetimes) to
isolate
a collection of Alexondrov neighborhood whose  endpoints  belong to the
same integral curve of  $t^a := g^{ab}\partial_b t$, to define simplices,
and the claim is that the collection of all these simplices forms a
triangulation of ${\cal M}$. see \cite{luca} for more discussion .

\section{Coarse-graining and emergence of structures}

The study of the conformal dimension, the faithful embedding, and the
topology of causet has so far revealed for us the following picture. 

Taking the causal set as whole may in general have no geometrical meaning
, in the sense of finding a Manifold which gives a continuum
approximation of the causet with all  the conditions of faithful
embedding full filed , and even without the requirement a faithful
embedding, the causet, for instance, may acquire no well defined causal
dimension.At this point we should remember that after all,   
 our theory we seek for must enable us to arrive to
the  a continuum  and 4-d dimensional picture of space time  we experience
down to the standard model scale or even to GUT's scale. 
 
To handle this problem  one may invoke two possibilities \footnote{
The way in which the continuum 4-d spacetime emerges may turn out
different,
although we believe it won't be very far from the second possibility 
given herein}

One possibility  is that in final theory, the dynamically
preferred causal sets are such that they do admit a faithful embedding,
and in this way one will have solved the problem for all scale not just
the large view
scale, however such a picture is hard to  believe for the following
reasons.

Although the discussion has 
so far been purely kinematical we expect that
the small
the microscopic structure to be characterized by quantum fluctuations, or
described by the interference or the superposition of many different 
causal sets, and we do not of course expect that, if a causal set is
faithfully embeddable , any small variation of its causal structure will
lead to another faithfully embeddable causet, since  one could be, e.g,
adding a link which creates one higher-dimensional pixie.
 
As a second reason, recall the success of many Kaluza-Klein type
theory, in which spacetime is a manifold, but its topology at Planck
scales is very different from the effective large-scale one.Such a
situation could not arise from a faithful embedding of a causal set ,
because of our requirement on the length scales defined by the geometry
unless the compactification radius is large enough
\cite{luca},
moreover as we will see below, if we want to put the causet theory as a
candidate for providing  a mechanism from which other effective
fields (not only gravitational) would appear in a purely causal fundamental theory,
the physically relevant
 causets should probably not all turn to
be  faithfully embeddable .

The second possibility , motivated by the above discussion, is to look
for a procedure in which one can
associate, in a consistent manner, a manifold to non faithfully
embeddable causal set.To do this recall that our arguments  indicate that
the small-scale structure of causal sets which is preventing the causal set
from being faithfully embeddable, so if one can smooth out these
structure  such that the new resulting causal set become embeddable in
some manifold, this manifold would be large scale approximation of the
causal set, this process in general is called the $coarse-graining$ and is
in the core of statistical and quantum field theory. So far there
has been no systematic way of defining a coarse-graining that would
fit automatically our expectations, in fact a systematic procedure 
would  emerge naturally from a Path Integral (Sum of
histories)
formulation
of the dynamics of causet\footnote{It should be noted here that even in
Quantum Field Theory there is no \emph{general} way of derving the
croase-grained dynamics.}, such a formulation has not yet
been constructed and we will postpone  the discussion  of some suggestive
idea to the coming chapter, and try here to give some tentative ways 
which have been proposed in \cite{luca}.

One way that one can think of is to identify elements in some set into
equivalent classes, and quotienning.Doing out this equivalence relation,
inducing somehow a structure among equivalence classes (In Kaluza-Klein
type theories the group action defines a natural equivalence relation),
however  such equivalence relation must not introduce
any inconsistencies in the relationships among classes of elements, a
possible consistent definition is the following.
Consider a collection of Alexandrov sets $A_i= A(p_i,q_i) \in {\cal P}$ having all
roughly equal volume $V$ such that $\cup_i A_i = {\cal P}$, and induce on it
the  partial order $ A_i \p A_j$ iff $p_i \p p_j $ and $ q_j \p  q_j$.This
makes ${A_i}$ into  a causal set , which we call a \emph{cover
coarse-graining} of ${\cal P}$, and it of course preserves only future
of ${\cal P}$ with
characteristic volume scale larger than $V$, however  a potential problem 
associated
with this definition is that, we might pick few $A_i$ 's which are strongly
boosted with respect to the others (i.e they look very long, along null
directions), and would have few relations, of a kind that may make the
resulting causet complicated in an undesired way.

Other way to define  coarse-graining is via the notion of subset .
A \emph{subset coarse-graining } of a causal set ${\cal P}$ will be a
${\cal P}' \subset {\cal P}$,
with the induced partial order, satisfying a condition intended to ensure that
it represents a large-scale view of ${\cal P}$, to do this we require
that there
exists  a parameter $p\in [0,1]$ such that the fraction of all
$n$-element
Alexandrov sets  of ${\cal P}$ which contain $k$ element of
${\cal P}'$ is
approximately $ {n \choose k} p^k (1-p)^{n-k}$ .One can think of
${\cal P}'$ as having
been obtained by picking at random a fraction $p$ of the elements of
${\cal P}$,
which makes ${\cal P}'$ appear sprinkled with uniform density in
${\cal P}$, however
with this
definition we  may be forcing too much randomness into ${\cal P}'$
which may
leave it (strictly speaking) 
as non-embeddable as ${\cal P}$ , e.g , because we are left
with high-dimensional
pixie somewhere

Having given the above tentative definitions for coarse-graining let us
see
how this process may  effect the embedding of a causal set.
As regards to the subset coarse-graining , if a causal set ${\cal P}$
has a conformal
dimension $n$, it is easy to show that the resulting subset
coarse-graining ${\cal P}'$ will have conformal dimension $n' \le n $.

A similar result for the cover coarse-graining is hard to establish, however if
one could choose in each set of the cover an element $r_i \in A_i$, such
that $r_i \p r_j$ iff $A_i\p A_j$, the cover set ${A_i}$  would be
equivalent to the subset coarse-graining ${r_i}$ , and the above result 
for the dimension would hold between ${\cal P}$ and ${A_i}$ , but it
seems
unlikely that such a points exist in general, and the situation could  even
be worse, since
it
cannot be excluded that if the cover is "bad" in the
sense described above, new higher-dimensional pixies are created by the
links of the "long" Alexondrov sets, thus increasing the dimension.  
 
Although we have so far been unable to show that the two proposed way for
coarse-gaining will wash out all unwanted feature\footnote{ Here the
unwanted feature are meant to be the future which prevent the causet from
being faithfully embeddable , otherwise they are very welcome} and do not
bring any new undesired features, we can give a  qualitative picture for 
the effect of a \emph{would be}   systematic coarse-graining.

After some degree of coarse-graining, an initially non-embeddable causal
set can acquire a well-defined physical dimension, which might however
correspond to a Kaluza-Klein type manifold , and need not be the
"macroscopic" dimensionality of the causal set, obtained by further
coarse-graining.

Let us now see how new effective fields (not only gravitational) may
emerge from a purely causal theory through the coarse-graining
mechanism.

We start by a causal set which may have no geometrical meaning , then
after performing a coarse-graining , we  obtain a new causet $P'$
which is
faithfully embeddable in some manifold (this manifold might 
have non-trivial topology , either globally, because of a possible
Kaluza-Klein nature, or  because of the presence of localized
structures, e.g handles).

Now the new set ${\cal P}'$
will obviously contain less information, and just by looking at ${\cal
P}'$ , one
cannot tell how many extra elements there were, and how they were
related.If one uses only ${\cal P}'$ to define the dynamics (say an 
amplitude
associated to some history )  the resulting amplitude of a given history will
in
general be
different from the amplitude we would get using ${\cal P}$, thus
leading to a
different selection rules of dynamically preferred causal sets, which are
the
ones that determine the classical limit;
and since the right
amplitude should be calculated using ${\cal P}$ the extra
information, or more precisely the relevant information, lost by
the coarse-graining process are needed in order to continue using the
same
procedure with ${\cal P}'$.
The question now is to find a way to preserve the information on the
original set in the coarse-grained one. It has been proposed \cite{luca} 
that one
can
do this by attaching numbers to various elements of the structure in
${\cal P}'$,
which tell us where the additional elements were and how they were linked.
As an illustration, consider the following examples. If ${\cal P}'$ is
a cover
coarse-gaining, to each element $p_i= A_i \in {\cal P}'$ we can
associate the
number of elements in the original $ A(p_i,q_j) \in {\cal P} $ , to each
pair
$A_i, A_j$ the number of elements in $A(p_i, q_j) \cap A(p_j, q_j)$,...,
to each collection $A_i, A_j, ....., A_k$ the number of elements in
$A(p_i, q_j ) \cap A(p_j, q_j)\cap ...\cap
  A(p_k,q_k)$.This possibility is suggestive, since we expect the
$A_i^,$ s with non-empty intersections to be nearby; in particular if we
defined a simplicial complex from ${\cal P}'$, a collection of $k$
nearby
$A_i^,$ s would define a k-simplex in the complex. A possible way in which
the  effective fields may arise is to associate to each $k$-skeleton of
the complex, which then, by its nature, would correspond to a different
kind of tensor field.

To conclude this chapter let me summarize the  general picture.

As regards the main conjecture of causets, if we put the various results
we have obtained, we can proof the following results.

\textbf{Theorem}\cite{luca}: \emph{If we randomly sprinkle points in any
finite region of a
Minkowski space of arbitrary dimension $n$ with increasing density, after
a finite number of steps we will end up with a set of points whose causal
relations define a causal set $P$ with conformal dimension $n$}.

Now the argument used in showing the above theorem [Bombelli \cite{luca}]
uses only compact
regions defined by $n$-pixies and it is natural to assume that this result
continue to hold for general
space-time in which all length scales defined by the metric are bounded
below by some constant, so we can apply locally the above
theorem.

Now using the above theorem we can go a step in proving the uniqueness of
a faithful embedding.

\textbf{theorem}\cite{luca}:\emph{Given two faithful embedding of the same
causal
set, than the two
associated manifold must have the  same dimension.}

This can be proven as follow,

Let $f: {\cal P}\ria ({\cal M},g)$ be a fixed faithful embedding of dimension
$n$, now
$f({\cal P})$ can be looked as a sprinkling  in ${\cal M}$,
in which
case the previous
theorem can be applied . Therefore, ${\cal P}$ must contain $n$-pixies as
subsets.
Let also $f' : {\cal P}\ria {\cal M'}$ be other faithful embedding ,  since
${\cal P}$
contains $n$-pixies , $M'$ must have at least dimension $n$. 
Interchanging
the role of  ${\cal M}$ and  ${\cal M'}$ establishes the equality between the two dimension.

It is also possible to argue that the two manifold are globally
approximately isometric \cite{luca}.
 
Notice here that even if we assume at the end that we were able to
prove
the uniqueness of the faithful embedding when it exists, along the
above
lines  of argument, we would not thereby be able  to characterize it
for practical purposes in a
pure
causal
set terms, and we would be only at the stage of the continuum theory of
space. To understand in pure causet term we  would need a constructive
proof, most probably constructing simplicial complex out of causets
 which are a triangulation of a manifold, and, perhaps , dealing with
the question of how many different causal sets give different
triangulation of the same manifold.

   As regard to the emergence of the picture of space time we experience
today , we have seen that the causets at the fundamental level may
exhibit
no geometrical meaning,  being purely causal, only after
coarse-graining the causet may acquire  continuum
geometrical approximation,  although not necessary a 4-dim
continuum
geometry we experience down to standard model scale, which may appear only
after further coarse-graining , 
in meanwhile , some intermediate
topological
structures  may manifest themselves in terms of geons , foam-like
structure
, wormholes, or internal manifold a la  Kaluza -Klein, and the
coarse-graining process will make fundamentally different structures
appear
similar. After each stage of the averaging process , the causal extra 
information of the original causet lost by the averaging process and not
manifested as geometrical structure , may be encoded as fields living on
this geometrical background.

\section{ The Issue of Locality}

One of the properties that the continuum physics  emerging from 
 causet dynamics must have, is locality i.e the action contributed by
causets
corresponding to a given Lorentzian manifold $(M,g)$ , should turn out to
be an integral of a locally defined quantity in $M$.
Now, as mentioned earlier for faithful embedding in a manifold $M$, 
points
which appear to be near neighbors in one frame will be boosted in other
frames, which makes the notion of nearest neighborhood drastically
different 
from the the one in ordinary  lattices, points in Lorentzian
manifold
have  a metric neighborhood  which converges to the light cone rather to
the point itself. This makes the recovery of locality rather a
nontrivial task . In the case of regular Lorentz, square say (which breaks
Lorentz
invariance) or Euclidean 
lattice , locality is achieved by having only near neighbor interactions,
in the continuum limit (lattice spacing going to zero) such interaction
will become local.

Now, the question of locality has not yet been settled , however, as
preliminary indication towards the recovery of locality was the the study
"scalar field" living in a fixed ( or " background") causal set (such
additional degrees of freedom might or might not be necessary to
incorporate
"matter") sprinkled in two dimension \cite{daug, rab}. The main idea goes
as
follow , 
one produces several
sprinklings of an Alalxondroff neighborhood in 1+1 Minkowski space.

A discretized version of the of the Laplacian operator is applied to
different scalar fields (generic enough ), and then averages of the
resulting actions for each field   is taken.
A match of the average values to the true continuum actions is the
criterion under which the conclusion that locality is recoverable. 
The numerical results for the above program showed that the discrete
version agree quite well with the continuum values. The agreement is to
within a single standard error for all  fields  which are not
rapidly varying and for which boundary terms are not expected to be
present
,however if the fields are rapidly varying or some boundary terms are
expected
to be present, the results are decidedly not too good, and this is should
in
fact be expected. These results could be an indication that the locality
can be recovered , regardless of the necessity of the presence of such
degree
of freedom to incorporate "matter", and thry also illustrate how 
locality may be
emerge via a route different from that of the near neighbors.

\chapter{Dynamics of Causal Set }

As mentioned in the introduction   any procedure
to quantize gravity is associated with technical and conceptual (interpretive)
 questions which  go beyond the question of 
quantizing gravity in the usual sense \cite{i, s}, and even the
study of quantum field theory in fixed background  addresses deep
questions which are quantum in character at the end , and  supported the
idea that General relativity and quantum mechanics cannot coexist,
however, it is widely believed that those questions will find their answer only upon 
  a much deeper understanding 
of the laws which  govern the dynamics of the
fundamental substance of spacetime.

Now, one of the fundamental concepts of "traditional "
dynamics is the
Hamiltonian which generates the time evolution, but  because time itself  
is discrete in the causet  approach one cannot hope to write down a
Hamiltonian.
Indeed it is not
even clear what configuration space could mean for causets, and therefore
unclear what Hilbert space such a Hamiltonian could act on,  in fact this
approach based mainly on, Algebra of operators to be interpreted physically in
terms of measurement, and Hilbert space, uses heavily  the notion
of spacelike Hypersurface. Apart from the fact that this approach is very
questionable  technically and conceptually \cite{s} , causet  does not
induce any
thing natural on a maximal antichain, which can then be evolved.
The only available framework to work with thus appears to be that of the
sum-over-histories, since it is by nature a spacetime approach, and it
seems more natural to write quantum dynamical theory for causets in the
sum-over-histories  because of their essential covariance, as we will see
below.

\section{Old Proposals For Quantum Amplitude}

In looking for the correct amplitude, there are a couple of requirements
to
guide us. Most obviously, the requirement that the dominant contribution
must in suitable circumstances come from those causets which , upon coarse
graining, give rise to
the continuum geometry of classical gravity.
A related requirement is that the effective action contributed by the
family of all causets corresponding to a given Lorentzian manifold
$({\cal M},g)$, should turn out to be dominated by the Hilbert-Einstein
action.

Before I sketch some proposals let me note a very important point here,
the classical dynamics for individual causets will probably not be defined
at all, since the amplitude function can have no actual derivative because
the causal set cannot be varied continuously , and the dynamics will not
be able to be viewed as having "arisen via quantization of some classical
structure" -just as one would expect for truly fundamental theory
\footnote{J.Wheeler put it in this way "Surely the lord did not on Day One
create geometry and on Day two proceed to ` quantize it`. The quantum
principle , rather, came on Day One and out of it something was built on
day Two which on the first inception looks like geometry but which on
closer examination is at the same time simpler and more sophisticated.}. 
  
To give a sum over histories formulation for causets one would associate to
each causet $C$ an amplitude $A(C)$; and dynamics would be contained in
the amplitude-function $A(.)$ , together with the combining rules
telling how to use the amplitude to construct meaningful probabilities.

No form for the amplitude has yet been obtained , although recently Sorkin
and Rideout have proposed  a model for the classical stochastic dynamics of
causet which may be considered as the first real step towards an
understanding of the quantum dynamics which  I leave its discussion to
the next section, and try here to give  briefly sketch of some old 
proposals.

These proposal are all based on associating the standard form of amplitude
to the a given causet, by defining an action using one of the many
structures which can be defined
on causets and the amplitude is obtained by the standard way (exponentiation the
action), for instance Bombelli \cite{luca} has considered the Multiloop
action defined by as,

$$
S_m(C) := ln M
$$

Where $M$ is the number of distinct multiloops in $C$.

Bombelli argued that the action satisfies approximate additivity and hence
one may hope to recover locality in the continuum approximation, moreover,
he added an analogy between the this proposal and the 2-dimensional Ising
model whose partition function can be written in a Multiloop form, although
the multiloops here are not exactly the same as the one for
causets. Now, near  some critical temperature  this 2-dimensional Ising
 model is known
to be equivalent to local
scalar
field theory with definite mass, and one may hope to recover locality in a
similar fashion.

Apart from these speculation this proposal has not been developed to any
extent.

Another due to Geroch, is that $S(C) =2\pi Z(C)$ where $Z(C)$ is
a functional on the causal set $C$ such that it is integer valued when $C$
is faithfully embeddable into a manifold and not integer valued when $C$
is not .Then

$$
\sum_c e^{i S(C)}
$$
will have constructively adding contributions for faithfully embeddable
causets.
Moreover, if the spectrum of $Z(C)$ is such that there exists $C^*$
satisfying

$$
Z(C^*) =Z(C) + \frac{1}{2}
$$
for every non-embeddable $C$, then non-embeddable causets will cancel.
Apart from the fact that an explicit expression for an action  
satisfying the above requirements  has not been given, 
this last proposal does not fit with our expectation that not all the
causets which dominate the sum-over-histories  be faithfully embeddable in
a manifold, however it could serve as coarse-grained dynamics.

As I remarked earlier in above proposal and many other \cite{luca, mey,
daug} no calculation has been done to test the validity of the actions,
because of  technical difficulties due mainly to fact that one doesn't
know how to carry sums over all causets [see \cite{daug} for some
proposed
techniques]. To investigate the possible consequences of the causet
dynamics  Sorkin \cite{spro1}  has considered the following
amplitude $A= exp (i \beta N_{rel})$, $N_{rel}$ being the number of
relation as defined in the previous section. To prevent 
the amplitude-sums from diverging, one can take the total number of
causal set elements to be a fixed integer $N$.
Now, if we were to go
to the corresponding "statistical mechanics" problem by continuing $\beta$
to imaginary values, then we would be dealing with a random causal set of
$N$ elements, and with probability-weight given by the "Boltzmann factor"
$exp (- \beta N_{rel})$. It happens that just this problem has been
studied in connection with a certain lattice -gas model.
The first result of interest is that, in the
"thermodynamic limit" $N\ria \infty$ , at least two, and probably 
an infinite number, of phase transitions occur as $N_{rel}/N^2$ is 
varied (corresponding to varying $\beta N^2$ in the "canonical
ensemble"). For small values of this parameter, the most probable causal
sets possess only two "layers", and the phase transitions mark thresholds
at which successively greater numbers of layers begin to contribute.
In some very general sense the causal set is thus becoming more 
manifold-like with each transition. 

Another  interesting phenomena accompanying the 2-level to 3-level 
phase transition
is the spontaneous breaking of time-reversal symmetry. In the 2-level
phase
the most probable configurations look similar to their  T-reversals, but
the initial 3-level phase the causet of high-probability
have very unequal numbers of elements in their top and bottom layers. The
claim was not of course that this was the root of the cosmological
time-asymmetry, but it does show the possibility that something of
the sort could ultimately emerge from a better understating of causal set
dynamics.

\section{Classical Stochastic Dynamics for Causal Sets}
The previous proposed quantum dynamics for causets were mainly based on
some guesses and analogies with standard dynamics, and they could not be
developed further due to the technical difficulties we mentioned, and it
seems that a deeper understanding of what one may  first mean by
  dynamics for causets is needed.

Recently Sorkin and Rideout have a proposed a model of defining the
evolution of causet as stochastic process described in terms of the
probability  of forming designated causets \cite{SR}. That is, the
dynamical law
will
be a rule which assigns  probabilities to suitable classes of causets.
Though the dynamics proposed there deals only with classical probabilities, 
can help us to get used to a relatively unfamiliar type of dynamical
formulation, and would  guide us
towards physically suitable conditions to place on the theory.

Recall that from the sum over histories point of view, quantum mechanics is
understood as a modified stochastic dynamics characterized by a
non-classical probability-calculus in which alternatives interfere. In this
sense the resulting theories from this dynamics should provide a
relatively accessible "half way house" to full quantum gravity .

After reviewing the main aspects of this dynamics we outline a possible
implications of it. The proofs of the results quoted
below and more detail can be found in \cite{SR,  rid}.
\subsection{Sequential Growth}
The dynamics proposed can be regarded as a process of "cosmological
accretion". At each step of this process an element of the causal set
comes into being as the "offspring" of a definite set of the existing
elements. The phenomenological passage of time is taken to be
manifestation
of
this continuing growth of the causet. Thus, the process is not thought to
be
as happening "in time"  but rather as "constituting time", which means in
 a practical sense that there is no meaningful order of birth of the
elements other than that implied by the order relation $\p$.
However in order to define the dynamics one "has" to introduce an element
of "gauge" into the description of the growth process, namely 
 the births are treated as if they happened in a definite order with
respect to
some fictious "external time", represented by a natural labeling of the
elements , that is, labeling by integers $0,1, 2,...$ which are
compatible with causal order $\p$. The gauge invariance is captured by the
statement that the labels carry no physical meaning (this will be called
"discrete general covariance").

It is helpful to visualize the growth of the causet in terms of paths in a
poset $\cal P$ of finite causets. 

\begin{figure}[h]
\epsfxsize= 14cm
\centerline{\epsfbox{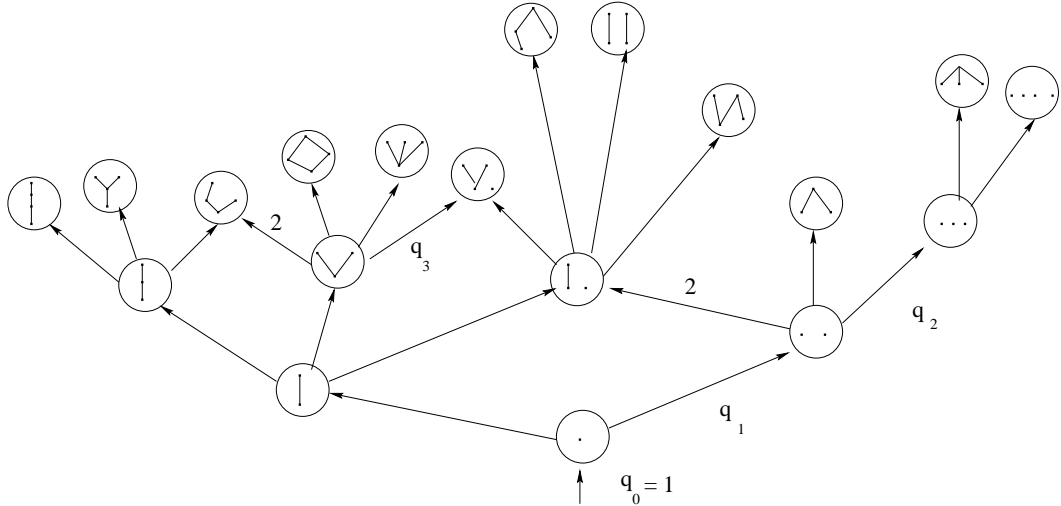}}
\caption{The poset of finite causets }
\end{figure}                       

The growth will be a sort of Markov
process
taking place in $\cal P$. Each finite causet is one element of this
poset.

A \emph{child} of causet $C$ is a causet obtained by accreting a single
element
to the causet, and $C$ is called the \emph{parent} in $\cal P$, the
causets
obtained in this way from $C$ will be called collectively a \emph{
family}. Drawing
$\cal P$ as a Hasse diagram of Hasse diagrams, we get Fig (2.1).

Any natural labeling of a causet $C\in \cal P$ determines uniquely a path
in $\cal P$
beginning at the empty causet and ending at $C$. As we mentioned before we
want 
the
physics to be independent of labeling, so different paths in $\cal P$ leading
to the
same causet should be regarded as representing the same (partial) universe, the
distinction between them being  "pure gauge".

The child formed by adjoining an element which is to the future of every element
of the parent will be called the \emph{timid child}. The child formed by
adjoining an element which is space like to every other element will be called 
the
\emph{gregarious child}. A child which is not timid will be called a
\emph{bold}
child.

Each parent-child relationship in $\cal P$ describes a transition $C\ria
C'$, from one
causet to another induced by the birth of a new element. The past of the
new
element will be referred as the \emph{precursor set} of the transition

The set of causets with $n$ elements will be denote by ${\cal C}_n$, and the
set
of all transition from ${\cal C}_n$ to ${\cal C}_{n+1}$ will be called
\emph{stage
n}.

The dynamics for the stochastic growth  is defined by giving, for each
$n$-element causet $C$, the \emph{transition probability } from it to each
of its possible children. Were  not we to impose some restrictions on the
transition probabilities we would be left with
 freedom of associating a free parameter to each possible transition, 
however, as we mentioned above one has to impose path independence
condition in order to restore gauge invariance, the path independence
condition can be stated in term of probability by requiring that, the net
probability of forming any particular $n$-element causet $C$ is independent
of the order of birth we attribute to its elements. Beside this
condition there are other two conditions which seem natural to
impose.

-\textbf{Bell causality condition} :The physical idea behind
this condition is that events occurring in some part of a causet  $C$ 
 should be influenced only by the portion of $C$ lying to their past (hence
the name bell causality). In this way, the order relation constituting
$C$
will be causal in the dynamical sense.

This condition can be stated in the following way: the ratio of transition
probabilities leading to two possible children of a given causet depend
only on the triad consisting of the two corresponding precursor set and
their union. 

\begin{figure}[t]
\epsfxsize=5cm
\centerline{\epsfbox{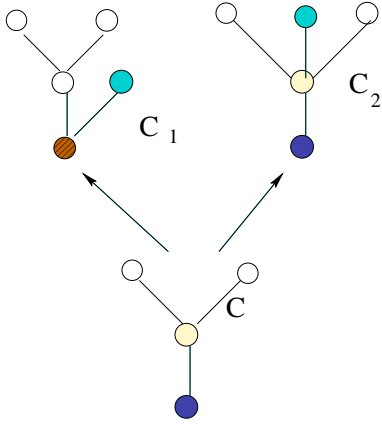}}
\caption{Illustrating Bell causality}
\end{figure}          

Thus, let $  C \ria C_1 $ designate a transition from $C\in
{\cal C}_n$ to $ C_1\in {\cal C}_{n+1}$,  and similarly $  C   \ria
C_2$. Then,
the
Bell
causality condition can be expressed as the equality of two ratios:
$$
\frac{prob(C\ria C_1)}{prob(C\ria C_2)} = \frac{prob(B\ria
B_1)}{prob(B\ria
B_2)} 
$$ 
where $B\in {\cal C}_m, m\le n$, is the union of the precursor set of $C\ria
C_1$
with the precursor set of $C\ria C_2$, $B_1 \in {\cal C}_{m+1}$ is  $B$  with
an
element added in the same manner as in the transition $ C\ria C_1$, and
$B_2\in {\cal C}_{m+1}$  is $B$ with an element added in the same manner as in
the transition $ C\ria C_2$ , Fig (2.2), Fig(2.3) show an example.

\begin{figure}[h]
\epsfxsize= 5cm
\centerline{\epsfbox{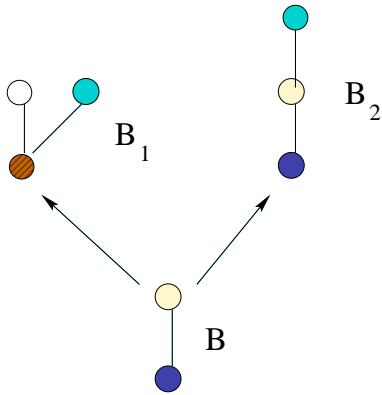}}
\caption{Spectators do not affect relative probability}
\end{figure}

- \textbf{Markov sum rule}

As with any Markov process, one must require that the sum of the full set
of transition probabilities issuing from  a given  causet be unity. 
However, this  depends in a subtle manner on the extent to which causets
elements are
regarded as "distinguishable". What is done here is to identify distinct
transitions with distinct precursor sets of parent, which amount to
treating causet elements as distinguishable.

Now, with these requirement on the dynamics one can show \cite{SR,
rid} :

-The number of free parameters are reduced to \emph{ at most one
parameter per family}

-\emph{ The probability to add a completely disconnected element depends
only on the cardinality of the parent causal set}. 

If the causets are regarded as entire universe, then a gregarious child
transition
corresponds to the spawning of a new, completely disconnected universe,
and
the probability for this to occur does not depend on the internal structure
of the existing universe, but only on its size, which seems eminently
reasonable.

Let $q_n$ be the probability of such a gregarious child transition at stage
$n$ than,

- The probability $\alpha_n$ for an arbitrary transition from ${\cal C}_n$ to
${\cal C}_{n+1}$  is given by:

\beq
 \alpha_n =\sum_{k=0}^{m} (-1)^k {m \choose
k} \frac{q_n}{q_{\varpi-k}}
\eeq

where $m$ stands for the number of maximal elements in the precursor set
and $\varpi$ for the size of the entire precursor set. 

The above form for the transition probability exhibits its causal nature
particularly clearly: except for the overall normalization factor $q_n$,
$\alpha_n$ depends only on invariants of the associated precursor set .

Now, since the $\alpha_n$ are classical probabilities , each must lie
between 0 and 1, and this in turn would restricts the possible values of
the $q_n$ . It turns out that it suffices to impose only one inequality
per
stage, more precisely it suffices that  $q_n> 0$ for all $n$ and
$\alpha_n\ge 0$ for the timid transition from the $n$-antichain.
Moreover if we define the following quantities
$$
t_n = \sum_{k=0}^{n} (-)^{n-k} { n\choose k} \frac{1}{q_k}
$$
than the full set of inequalities restricting the $q_n$ will be
satisfied iff $t_n \ge 0$ for all $n$ with $t_0 =1$ , since $q_0=1$.

One can invert the above equation and recover the $q_n$ in terms of the
$t_n$,
\beq
\frac{1}{q_n} =\sum_{k=0}^{n} {n \choose k} t_k
\eeq

Thus , the $t_n$ may be  treated as free parameter subjected only 
to $t_n \ge 0$ and $t_0=1$ , if this is done, the
remaining probabilities $\alpha_n$ can be re-expressed more simply in
terms of the $t_n$ as.
\beq
 \alpha_n = \frac{\sum_{l=m}^{\varpi} { \varpi -m \choose \varpi -l }
t_l}{\sum_{j=0}^{n} {n \choose j} t_j}
\eeq

Equation (2.2) implies that

$$
q_0\equiv 1 \ge q_1 \ge q_2 \ge q_3 \ge ....
$$
So if we think of the $q_n$ as the basic parameters or "coupling
constants" of our sequential growth  dynamics, then it is as if the
universe
had a free choice of one parameter at each stage of the process, however
the choice is not completely free since the allowable values of $q_n$ at
every stage are limited by the choices already made. But if we think of
$t_n$ as the basic parameters, one has free choice at each stage.
Unlike the $q_n$ , the $t_n$ cannot be identified with any dynamical
transition probability.
Rather they can be realized as ratios of two such probabilities, namely as
the ratio $x_n/q_n$ , where $x_n$ is the transition probability from an
antichain of $ n$ elements to the  timid child of that antichain.

For every  \emph{labeled } causet $\tilde{C}$ of size $N$, there is an
associated net probability of formation $P(\tilde{C})$ which                         
 is the product of the transition  probabilities $\alpha_i $
of the individual births described by the labeling:
$$
P(C) = \displaystyle\prod_{i=0}^{N-1} \alpha_i 
$$

Which is independent of the labeling ($\alpha_i$ being solutions
of the dynamics, satisfying the general covariance condition), this can be
brought out more clearly by defining,
 
\beq
\lambda(\varpi , m) = \sum_{k=0}^{\varpi-m} { \varpi -m\choose k}
t_{k+m}
\eeq

 Whence

\beq
P(\tilde{C}) =\frac{ \displaystyle \prod_{i=0}^{N-1} \lambda (\varpi_i,
m_i)}{\displaystyle\prod_{j=0}^{N-1}
\lambda(j, 0)} = P(C)=\frac{ \displaystyle\prod_{x\in C} \lambda(\varpi(x),
m(x))}{\displaystyle\prod_{j=0}^{|C|-1} \lambda(j, 0)}
\eeq

where $\varpi (x) =|past x| $ and $m(x) = |maximal (past x)|$.

The net probability of arriving at particular $C\in P$ is not just $P(C)$
but
$$
Prob_N(C) =W(C) P(C)
$$
where $N=|C|$ and $W(C)$ is the total number of paths through ${\cal P}$
that
arrive at $C$, each link being taken with its proper multiplicity.

Now the above expression is manifestly "causal" and "covariant", however
, it
has no direct physical meaning in the sense of carrying a full invariant
meaning. A statement like "when the causet had  a given number of element
it
was a chain"   is itself meaningless before a certain birth order is
chosen.

This, also, is an aspect  of the gauge problem, but not the one that
functions as a constraint on the transition probabilities that define the
this dynamics. Rather it limits the physically meaningful $question$ that
one can ask of the dynamics.

As an example of a truly covariant question, let us take "Does the
two-chain ever occur as partial stem of $C$?".
Recall that a partial stem is  a finite subset of $C$ which contains its own
past. So the above question is equivalent  to asking whether or not $C$ is
an
antichain.

The answer to this question  will be a probability, $P$, which is natural
to identify as
$$
P=\lim_{N \to \infty} Prob_N(X_N),
$$
where $X_N$ is the event that "at stage $N$", $C$ possesses a partial stem
which is a two-chain.

 \subsection{ Sample Cosmologies}

Although  the dynamics we discussed is in its pure classical stage, it is
interesting  
 to explore and understand possible  physical consequences
of
various choices for $t_n$. One of the most interesting question is of
course which dynamics (choice of $t_n$), if any,  would give us general
relativity, or what are the conditions that one has to impose on the
choices so that the resulting spaces from this dynamics resemble the
Minkowski space or de Sitter (may be after coarse-graining).  

To start with, let us consider the case of limited number of past links

$$
t_i =0, i > n_0
$$
With this values of $t_i$ it is easy to see that $\alpha_n$'s   vanish if
$m>n$. Hence, no element can be born with more than $n_0$ past
links. This means in particular that any realistic choicer of parameters
will have $t_n > 0$ for all $n$, since an element  of a causal set
faithfully embeddable in Minkowski space would have an infinite number of
past links.

 A known example for randomly growing causet
is what was
called in \cite{SR} "transitive percolation". It is an especial simple
instance
of sequential growth dynamics, in which each new element forges a causal
bond independently with each existing  element with probability  $p$,
where $p\in [0,1]$ is a fixed parameter of the model.
This model can be solved independently of the requirement stated for the
dynamics and the transition probability $\alpha_n$  is given by
$$
\alpha_n = p^m(1-p)^{n-\varpi}
$$
Where we used the same notation as before.

With the above expression one can show that the Transitive percolation
dynamics satisfy all the requirements of the general dynamics.
In fact, it corresponds to the choice $t_n=t^n$, where $t=\frac{p}{1-p}$.

From a physical point of view, transitive percolation has some appealing
feature, both as model for relatively small region of spacetime and as
a cosmological model for spacetime as whole, where the universe cycles
endlessly through phases of expansion, stasis, and contraction back down to a
single element.

Other thing interesting that one learns from percolation model is that, the
 causets generated by the dynamics, do not at all resemble the 3-tier,
generic causets of Keilman and Rothshild mentioned in the previous chapter,
but rather they have the potential to reproduce a spacetime or a part of one. 

Nevertheless, the dynamics of transitive percolation cannot be viable
as a theory
of quantum gravity, since at least two reasons, one obvious reason is
that it
lacks quantum interference, being stochastic only in the pure classical
sense,  the other reason
is that the future of any element of the causet is completely
independent of anything "spacelike" to that element. Therefore, the only 
spacetimes which a causal set generated by transitive percolation
could hope to resemble would be homogeneous, such as Minkowski or de
Sitter spacetimes, but due to the periodic recollapse allude to earlier
in transitive percolation, one could at best hope to reproduce a small
portion of such  homogeneous spacetimes.

On the other hand, in computer simulations of transitive percolation,
two independent (and coarse graining invariant) dimension estimators have
tended to agree. These dimension indicators vary with $n$ and one must
rescale $p$ if one wishes to hold the spacetime constant. To do any
better, one would scale $p$ so that it decreased with increasing $n$
\cite{SR} and reference therein. This suggests that the physically
interesting choices of $t_n$ could be that which decrease with $n$ and by
the percolation example faster than exponentially in $n$. An reasonable
and simple choice could be $1/n!$, which we will discuss below.

\subsection{Possible Cosmological Implications}

It is an old idea that the universe might start, and keep bouncing,
getting bigger with each bounce. The explanation of why it is so big now
would then be that there have been many previous bounces. In general
relativity with bouncing universe, what happens in a particular cycle is
independent of what happened in earlier cycle, except insofar as they set
the initial conditions for it. People have speculated that the couplings
constant themselves might change, but in classical general relativity
(insofar as it allows a bounce at all), that cannot occur.  

Now let us pick up a particular cycle of expansion and possible
contraction and call it "the current era" for short. In the causal set
context, in the current era the first event is by definition a single
element $e_0$ which is to the future of everything not in the current era
and to the past of every thing in it. Now, in percolation if we compute
relative probabilities for events of the current era, it is enough to
truncate the causet at $e_0$. This follows directly from the definition of 
percolation, so the dynamics
in  the current era  is completely independent of the past eras. 
We will call this property \emph{locality in time}. 

Now what happens in the other dynamics? 

First let us note the following. The probability of transitions are given
by eqt(2.3)
$$
\alpha_n= \frac{\lambda(\varpi_n, m)}{\lambda(n,0)} 
$$
Where 
$$
\lambda(\varpi, m) = \sum_{k=0}^{\varpi-m} { \varpi -m \choose k}
t_{k+m}  
$$

Now , recall that $\varpi$ stands for the for the precursor set defined
as
the past of the new elements. If we denote by $N$ the number of element in
the past of $e_0$,
 
 than
$\lambda(\varpi,m)$ can be written as
$$
\lambda(\varpi,m) =\sum_{k=0}^{\varpi'+N-m} { \varpi'+N -m\choose k}
t_{k+m}                             
$$
Where $\varpi'$ stands for the precursor set in the current era. But $m$ 
must also stand for the current era, since it is the number of element of
the precursor set and they must all belong to the current era. Hence we
conclude that in the current era one can use the same formula for
$\alpha_n$ , except we replace $\varpi$ by $\varpi +N$. So the question of
locality of dynamics is turned now to the question of what happens to the
$\lambda$'s when we make this substitution. It is easy to see that the
dynamics will be local only iff all $\lambda$'s just change by  common
$N$-dependent factor, so the dynamics of the current era will perfectly
independent of $N$. This of course happens for the peculation model. Does
it happen for other models? The answer is no, because we have the
following
simple uniqueness theorem.

\textbf{Theorem}
The only dynamics  which satisfies the time locality condition is the
percolation.

\textbf{Proof}

The question is to prove that the following property holds only for
percolation model
\beq
\lambda(\varpi+N,m)= r_N \lambda (\varpi, m)
\eeq
for arbitrary $m$, $\varpi$ and $N$\footnote{The  collapse is a random 
event so the equality must hold for
 arbitrary $N$}.
where $\lambda (\varpi, m)$ is defined by 
$$
\lambda(\varpi, m)= \sum_{k=0}^{\varpi-m} { \varpi -m \choose k} f(k+m) 
$$
Since the aim is to prove that this could happen only when
$f(m+k)=\alpha t^{k+m}$,   $\alpha $, $t$,  are free parameters, one
can argue as follow.

Because equality (2.6) must hold for any $\varpi$ and $m$ it must hold in
particular for the transition where $\varpi=m$ 
, or one may pick up any special, but generic enough,
case to establish the result.

So we have
\beq
\lambda(N+m,m)=r_N \lambda(m,m)
\eeq

On the other hand we have for arbitrary $\varpi$
$$
\lambda(\varpi+N,m)=\lambda(\varpi +1+N-1,m)= \lambda(\varpi'
+N-1,m)=
$$ 
$$
r_{N-1}
\lambda(\varpi +1,m)= r_1 r_{N-1} \lambda(\varpi ,m)=r_N \lambda(\varpi,m)
$$
from which it follows that,
$$
\forall\  N, \ \
r_N=r_1 r_{N-1} \Ria r_N=r_1^N \equiv r^N
$$
 
Having established this we go back to eqt(2), which can  now be written
as,
$$
\lambda (N+m,m) =r^N \lambda(m,m)
$$

By writing $r$ as $r\equiv t+1$ and
putting all together we get.
$$
\forall\ N,\ \ \
\lambda (m,m)\sum_{l=0}^{N} {N \choose l} t^l=\sum_{k=0}^{N} {
N \choose k} f(k+m) 
$$
From  which it is not difficult to establish that,
$$
\lambda(m,m)=f(m), \ and \ f(k+m)= t^k f(m)
$$ 
Now, since $f(k+m)$ is symmetric by definition under the exchange of $k$
and $ m$ we conclude that, $f(m)$ must be of the form , $\alpha t^m$.
Q.E.D

Thus there can be no function $f$ that satisfies the required criteria
expect the one corresponding to transitive percolation.

Having established this, let us go ask about the effect of bouncing  on
the other dynamics. If we imagine that the causal set has bounced a very
large number of times, so $N$ can be taken to be very large namely
$$
N \ria \infty
$$

Let now us look at the behavior of the asymptotic dynamics . To do
this  
it is enough to look at the asymptotic behavior of $\lambda (\varpi +N,m)$
in the limit $N \ria \infty$. Here we try to investigate the asymptotic
dynamics of two models.

We first consider $ t_n= t^{n} \displaystyle
\frac{\Gamma (p+1)\Gamma(n+1)}{\Gamma(n+p+1)}$

$p$ and $t$ are two positive couplings constant.

For $p=0$ it corresponds to percolation, $p$ integer to
integrating the
percolation result $p$ times and than multipying it by $\frac{p!}{t^p}$.

For arbitrary $p$  the exact
result is.

$$
\lambda(n,m,p) =t^{m}\frac{\Gamma (p+1)\Gamma(m+1)}{\Gamma(m+p+1)}
F([-n, m+1],[m+p+1], -t])
$$                                                        
where $F$ is the generalized hypergeometric function.

The asymptotic form for given $p$ and $N \ria \infty$ is  
\beq
\lambda_{eff}(n,m,p)\equiv \lambda(\varpi+N,m)=
\frac{t^{m-p}
p!(1+t)^{n+p}}{N^p}(1+O(\frac{\varpi-m}{N})).
\eeq

Where $n\equiv N+\varpi -m$.

The above results shows that after many bouncing the model will converge
 percolation, more precisely the period dominated by percolation
gets bigger with each bounce, and the dependence on $p$ disappears (being 
only a common factor); in some sense after many bouncing the model forgets
completely this coupling and behave just like a percolation.

Let us now have a look on the more interesting case $t_p= t^p/{p!}$.

\beq
\lambda (\varpi,m)=\sum_{k=0}^{n} {n \choose k}
\frac{t^{k+m}}{(m+k)!}
\eeq

 Now, this series can be written in term of Kummer function 
$M(\alpha,\beta,x)$
 \footnote{M.Abramowitz and I.Stegun, Handbook of Mathematical
Functions,
chapter 13.}, which is one of the independent solutions of the following
differential equation
$$
x M'' +(\beta -x) M' -\alpha M=0
$$

The exact form of $\lambda$ is given by
\beq
\lambda(N+\varpi,m)= \frac{t^m}{m!} M(-n,m+1, -t)
\eeq

For $N \ria \infty$ , which  also means  $ n \ria \infty$ since $\varpi
\ge m$, $\lambda$ has the following asymptotic form
\begin{eqnarray}
\lambda_{eff} &=& lim_{n \to \infty} \frac{t^m}{m!}M(-n, m+1, -t)
\nonumber \\ 
&=&
 \frac{t^m}{\sqrt{\pi}} e^{-t/2}
(\varpi +N-\frac{m}{2})^{-\frac{2m+1}{4}} e^{2 \sqrt{f}} 
\end{eqnarray}
where $f$ is given by
$$
f= (\varpi+N-\frac{m}{2}) t                                       
$$

Now, if we further assume that $N \gg \varpi $, so we are at the early stage at
the era 
, expand in power of $\frac{2\varpi-m}{\sqrt{N}}$ and neglect
$\frac{2\varpi-m}{N}$ in confront of $\frac{2\varpi-m}{\sqrt{N}}$, 
 keeping the first no
trivial
dependence on $\varpi$  and dropping irrelevant terms we get

\beq
\lambda_{eff}= (\frac{t}{\sqrt{N}})^m 
\exp{\left( (\frac{2\varpi-m}{2})\sqrt{
\frac{t}{N}}\right)}
\eeq

We note first that the resulting dynamics is local in the sense we
described above. Other interesting thing to note here, is the model is
described by two effective  couplings $t$ and
$1/{\sqrt{N}}$,
the later
being as an
initial conditions set by the previous eras and is very small and
getting smaller with each bounce. This example
shows  how the  bouncing could effect the dynamics.   

The most relevant question to ask here is, if any of
this possible dynamics produce (under coarse graining) a life like
universe, in the limit $m, \varpi \gg 1$ i.e.  whether the
behavior of the dynamics is  of a  given cosmology (Minkowski, de
Sitter, etc
). For example perhaps one could ask  in this limit $m, \varpi \gg 1$, 
whether
$\lambda$ was
maximized where the relation between $m$ and $\varpi$ was like that of
the spacetime in question. This point needs a further
exploration .


\chapter{Black Hole Entropy as Links}

One of the most  remarkable  developments in theoretical physics that have
occurred in
the past thirty years, was undoubtedly the discovery of the close
relationship between certain laws of black hole physics and the ordinary
laws of the thermodynamics. Today, well into its third decade of the
development, black hole thermodynamics remains intellectually
stimulating and puzzling at once. It appears that these laws of black hole
mechanics and the laws of thermodynamics are two major pieces of a puzzle
that fit together so perfectly that there can be little doubt that this
"fit" is of  deep significance.
The existence of this close relationship between
these laws seem to be guiding us towards a deeper understanding of the
fundamental nature of spacetime, as well as  understanding of
some aspects of the nature of thermodynamics itself \cite{Wald, sorsmbh}.

\section{Black Hole Thermodynamics}

In this section I briefly review some aspects of the
thermodynamics of black holes.

It was first pointed out by Bekenstein \cite{beken1, beken2} that a close
relationship might
exist between certain laws satisfied by black holes in classical general
relativity and the ordinary laws of thermodynamics.

Bekenstein noted that the area theorem of classical general relativity, 
stating that the area of a black hole can never decrease in any process,
is closely analogous to the statement of the ordinary second law of
thermodynamics, which states that the total entropy of closed system never
decrease in any process. Moreover, Bekenstein proposed that the area of a
black hole (times a constant of order unity in Planck units) should be
interpreted as its physical entropy. The above
proposal
was  confirmed by Bardeen, Carter and Hawking  \cite{bhc}, they proved
that in general
relativity ,
the surface gravity, $\kappa$, of a stationary black hole must be constant
over the event horizon. Which is analogue to the zeroth law of
thermodynamics, which states that the temperature, $T$, must be uniform
over a body in thermal equilibrium. The analogue of the first law of
thermodynamics was also proved. In the vacuum case, this law states that
the difference in mass, $M$, area, $A$, and angular momentum, $J$,
of
two nearby stationary black holes must be related by
$$
\delta M=\frac{1}{8 \pi}\kappa \delta A + \Omega\delta J
$$

where $\Omega$ denotes the angular velocity of the event horizon. Which
is analogue to the first law of ordinary thermodynamics, which states that
the differences in energy, $E$, entropy, and other state parameters of two
nearby thermal equilibrium states of a system are given by
$$
\delta E= T\delta S+"work terms".
$$
Now these analogies suggested the following identifications,
$E\longleftrightarrow M$, $T \longleftrightarrow \alpha\kappa$, and
$S\longleftrightarrow A/{8\pi \alpha}$, where $\alpha $ is some
undetermined constant. A hint that this relationship might be of a
physical significance arises from the fact that $E$ and $M$ represent the
same physical quantity, the total energy of the system. However, in
classical general relativity, the physical temperature of a black hole is
absolute zero, so there can be no physical relationship between $T$ and
$\kappa$. Consequently, it also would be inconsistent to assume a physical
relationship between $S$ and $A$.  Hawking's discovery of
quantum particles creation in the presence of black hole, was a
breakthrough
in the understanding of the laws of black hole mechanics, and showed that
the analogy between these laws and that of thermodynamics was not merely a
mathematical similarity but rather it has real physical significance
\cite{hawking1, hawking2}.
Hawking calculation showed that due to quantum effects, a black hole
radiates to
infinity all the species of particles with a perfect blackbody spectrum,
at temperature 
$$
T=\frac{\kappa}{2\pi}= \frac{1}{8 \pi M}
$$
Thus, $\frac{\kappa}{2\pi}$ truly is the physical temperature of the black
hole.
An other piece of evidence for this physical connection came from 
a derivation of Hawking and Gibbons of the area law using the
Euclidean
approach to quantum
gravity, by evaluating the partition function, in its zero loop
approximation for an Euclidean action, for Schwarzschild metric, namely they
considered, the partition function defined by \cite{hag}
\beq
Z= Tr e^{-\beta {\cal H}} = \int {\cal D}g e^{-S_E}
\eeq

where $S_E$ denotes the "Euclidean action", and the integral is taken
over all Euclidean paths which are periodic in Euclidean time with period
$\beta=1/T$.
An argument using canonical ensemble then establishes that 
\beq
S= ln Z +\beta E
\eeq
Now by evaluating (3.1) in the zero loop approximation by simply
evaluating
$S_E$ , Remarkably, Hawking and Gibbons found that the entropy derived
using (3.2), in this approximation is given precisely by

\beq
S=A/4
\eeq
which is with the right  factor to complete the analogy.

However, the above derivation has some disturbing aspects. As it can be
seen from the relation of the temperature and the mass (energy) for the
Schwarzschild black hole, the temperature varies inversely with its mass
energy, and hence the Schwarzschild black hole has a negative heat
capacity. This result should not be surprising on physical grounds, since
an ordinary self-gravitating  star in Newtonian gravity also has a negative
capacity; if one remove energy from a star, it contracts and heats up. 
As in the case of an ordinary star, this negative does not imply any
fundamental difficulty in describing the thermodynamics of black holes,
since the micro-canonical ensemble still should be well defined for a
finite system containing a black hole, and black hole can exist in stable,
thermal equilibrium in a sufficiently small box with walls that perfectly
reflect radiation. However, the negative heat capacity  implies that a
Schwarzschild B.H  cannot exist in a stable thermal  with  ordinary heat
bath at fixed temperature\footnote{The instability arises because ,
fluctuations cause a black
hole to absorb an extra amount of thermal radiation from its
 hotter environment, causing the B.H to grow without bound .}.
But such an equilibrium should  be  necessary in order to justify the use of
the canonical ensemble for black hole in the above derivation. Technically
this can be addressed to the following fact that in order for the integral
to converge for the canonical partition function , the entropy must  be
concave function of the energy , which is not the case for S.B.H. Thus ,
there appears to be a logical inconsistency in the above Euclidean path
integral calculation , since the result seem to invalidate the method used
to derive it. So, before the partition function can be used as a probe of
black hole thermodynamics, it is necessary to stabilize the B.H.. There have
been several ideas to over come this problem , most notably the
proposal by York based upon the micro-canonical ensemble .
 Nevertheless, the essentially classical nature of the
Euclidean path integral derivation of the formula $S=A/4$ remains
rather mysterious \cite{bken3}.

The final picture of the thermodynamic nature of the laws of the B.H
mechanics can be summarized as follow:

Consider a black hole formed by gravitational collapse, which settles
down to a stationary final state. By the zeroth law of B.H.M, the surface
gravity $\kappa$, of the stationary black hole final state will be
constant over its event horizon. Consider a \emph{quantum} filed
propagating in
this background spacetime, which is initially in any (nonsingular)
state.\emph{Then at asymptotically late times, particles of this field
will be radiated to infinity as though the black hole were a perfect
blackbody at the Hawking temperature $\kappa/{2 \pi}$}.Thus, a stationary
black hole truly is  a state of thermal equilibrium, and the above
temperature is the physical temperature of a black hole.
It should be noted that this result relies only on the analysis of quantum
fields in the region exterior to the black hole. In particular, the
details of the gravitational field equations play no role, and the result
holds in any metric theory of gravity obeying the zeroth law.
Moreover, the final state (equilibrium ) doesn't depend on the detail of the
collapse or the detail of the interior and  will be described by only few
parameter as any
thermodynamical system i.e. energy , angular momentum, electric charge
etc...and its entropy is given by (3.3). The description of the final
state of
the B.H by only these few parameters is known as the No-Hair theorem.

\section{The Generalized Second Law of Thermodynamics (GSL)}

The  line of the discussion of the previous section strongly suggested
that $A/4$
must be regarded as the physical entropy of the B.H. However this seems at
first to be in conflict with the  quantum particles creation
process, since the mass  of the black hole must decrease in the process
if energy is to be conserved\footnote{This violation can occur because the
expected stress-energy tensor of the quantum field violates the null
energy condition at the horizon of the black hole, and this condition is
necessary in order the area increasing theorem [ref Wald].} and hence
the second law of B.H.M is violated. On the other hand, in the presence of
B.H one can take matter and dump it into the B.H in which case -at least,
according to classical gravity-it will disappear into the singularity
within the black hole. In this manner, the total entropy of matter in the
 universe can be decreased, violating the second law of
thermodynamics. Note, however, that when the total entropy, $S_m$, of the
matter outside  of B.H is decreased by dumping matter into a B.H, the
area
will tend to increase. Similarly, when the area is decreased during the
particle creation process, thermal matter is created out-side the black
hole, so
$S_m$ increases. Thus, although $S_m$ and $A$ each can decrease
individually, it is possible that the \emph {generalized entropy}, $S_t$,
defined by
\beq
S_t= S_m +\frac{A}{4}
\eeq
 never decreases.

It  is now widely believed that $\delta S_t\ge 0$ in all process. So far
the
increase of the generalized entropy has passed all the tests -\emph{Gedamken
experiments}  and
there are more general arguments -at least in the cases which can be treated as
small perturbation of S.B.H-  in support of the increasing of $S_t$. 
The increasing nature of the total entropy  has became
known as the \emph{generalized second law} GSL.
If we accept its validity, GSL  would then have a very natural
interpretation: it simply would be the ordinary second law applied to a
system containing a B.H and $A/4$ is no more than the true
physical entropy if the B.H.
 
Indeed, in the absence of a complete quantum theory of gravity, it is
hard
to imagine how a more convincing  case could be made for the merger of the
laws of B.H.M with the laws of thermodynamics. Nevertheless, there
remain
many puzzling
aspects to this merger. 
Prominent among them are the following: 

(1)The physical origin of the
entropy $S_{bh}$ i.e. What statistical mechanics
behind black hole thermodynamics. Can the origin of $S_{bh}$ be understood
in essentially the same manner as in the thermodynamics of conventional
systems, as suggested by the apparently perfect merger of black hole
mechanics with thermodynamics, or is there some entirely new phenomenon at
work here?

2) What mechanism insures that the generalized entropy grows in any
situation i.e. Why does the second law continue to hold \footnote{There is also the question known as the
information lost puzzle which I will not discuss it here , however I 
should 
mention that this question is of a fundamental importance, since it seems 
to interplay between two fundamental aspects of the physical world ,
namely the structure
of space-time "the meaning of singularity" and the interpretation of
quantum mechanics "measurement" \cite{hartle}, and it is generally
believed that all
the puzzles of the black hole are not independent and will be solved once
we really solve one of
them.}

\subsection{On the Origin of the B.H Entropy }
Usually , the laws of the thermodynamics  are not believed  to be fundamental
laws in their own right, but rather to be laws which arise from the
fundamental "microscopic dynamics" of a sufficiently complicated system when 
one
passes to a macroscopic description of it. The great power and utility of
the laws
 thermodynamics stems mainly from the that fact the basic form of the
laws
does not depend upon the detail of the underlying microscopic dynamics of
particular system and , thus they have a "universal " validity .

Now  although we have mentioned earlier the black hole behaves exactly as
any
thermodynamical system,  any direct statistical way to count for its
entropy or to explain the validity of the GSL, encounter many problems
which makes the black hole different in many respects \cite{
sorsmbh,Wald}. 

In the first place, it is at least peculiar that the number of black hole states
would be proportional to $e^{Area}$ rather that $e^{Volume}$ as for other
thermodynamics system. This peculiarity becomes more troubling if one
consider the
example of Oppenheimer-Snyder spacetime, in which a Freidman universe of
\emph{arbitrary} size is joined onto the interior of a Schwarzschild black hole of
arbitrary mass. The existence of such a solution leads to the conclusion
that the
number of possible \emph{interior} states for a black hole is really
infinite.

Another important issue that arises in the context
of black hole involves
ergodic behavior. The increasing character of entropy of thermodynamical
system usually may be 
 predicted on
the
assumption of ergodicity generally stated :\emph{ generic dynamical orbits sample
the entire energy shell,
spending "equal times in equal volumes"}. However gross violation of such
ergodic
behavior occur in classical black holes, the course of events inside a
collapsing star leads classically to a singularity, and it  is not at all obvious
that this is consistent with ergodic exploration of all available states. 
 But even the ergodicity is restored  there would remain
a
problem with internal equilibrium, which is 
necessary   in order to deduce the entropy from the
value of few
macroscopic parameters. And since mere knowledge of the external appearance
of the black hole tells a little about its interior, and realistic black
hole would seem to be far from internal equilibrium , this makes the black
hole very different from standard thermodynamical system.

Another problem with any direct counting of states of B.H which would lead
to the split of the entropy the B.H and that of its surrounding matter
can be only deduced upon an assumption of weak coupling, which doubly wrong 
in the case of black hole.
The coupling from outside to inside is not weak but very strong , while
the reverse coupling is very weak (although not as nonexistent). Indeed ,
this last observation points up the fact that conditions in the interior
should be irrelevant, almost by definition , to what goes on outside. And
since the second law, as ordinarily formulated for B.H., makes no reference
to conditions inside, it seems unpleasant
to have to include the entropy of matter that has fallen into the black
hole in counting up total entropy, since neither the entropy nor any
other property of this matter can in principle be measured by an
observer outside the black hole\cite{sorsmbh}.

An apparent interesting difference between the B.H thermodynamics and
ordinary
thermodynamics is that the laws of the latter are statistical in nature
-valid only with high probability- while the laws of B.H are rigorous
theorem in differential geometry. However, it appears overwhelmingly likely
that
the those laws will emerge, from the underlying statistical mechanics of
 a fundamental microscopic theory gravity, and those laws of black hole
should hold exactly only in some thermodynamic limit.

A more striking difference is that no derivation of the second
law can even get started without applying some form of coarse
graining. However, for black hole we cannot in principle measure internal
states unless we go inside the B.H in which case we still would
not be able to report our results to observers remaining outside the B.H.
Thus black hole affords an obvious objective  way to coarse-grain,
neglect
whatever is inside the horizon  and making the notion of entropy 
more fundamental in the context of black hole.

Although there are many ways to get the entropy the statistical aspect is
not
exposed. For instance, general relativity is a field theory and it
describes
infinitely many degrees of freedom to the spacetime metric, and in the
 statistical
mechanics of field, statistical entropy first appears at one-loop
level. If the gravitation degree of freedom are treated  classically,
no sensible thermodynamics should be possible\footnote{This situation
also
arises for the electromagnetic field in a box , where a classical
treatment
run in to the Ultraviolet catastrophe}

\subsection{Entanglement Entropy}

The density operator in quantum mechanics or the phase  density in
classical mechanics represents our knowledge about the system. This knowledge
is more or less complete: clearly our information is maximum when we can
make predictions with full certainty , and its larger when the system is
in pure state than when it is in a statistical mixture. Moreover, this
system is better known the number of possible micro-states is small or
when the probability for one of them is close to unity than when there are
a large number of possible micro-states with all approximately the same
probability.
A macro-state (statistical state) is the set of possible microstates $
|n>$ each with its
own probability $ q_n$ for its occurrence. The probabilities $ q_n$ are
positive and normalized
$$
\sum_n q_n=1
$$
The various micro-states $|n>$ are arbitrary vectors of Hilbert space
$\cal{H}$.
The density operator is defined as follow
\beq
\ro = \sum _n |n> q_n <n|
\eeq
The evolution of the density matrix is via arbitrary linear law which
preserves the positivity property of $\rho$ and ,$Tr\ro$=1.

The statistical entropy  associated with the density operator $\ro$ is
defined by

\beq
S(\rho) =-k  Tr (\rho ln \rho)= -k\sum_n p_n ln p_n
\eeq

This statistical entropy measure the unavailable information (missing)
about the system, which one might acquire by knowing the system better
through microscopic measurement. Now  as mentioned earlier the black 
hole and due to the presence of the event horizon one has no access to
the inside region,  affords an objective way of coarse-graining, namely 
neglect whatever is inside the horizon, an entropy based on this type
of coarse -graining is known as the \emph{Entanglement Entropy}.

Entanglement was introduce first to understand the Unruh effect as
resulting from ignoring the states beyond the Rindler horizon, however
,its
first use as a possible source for the B.H. entropy was proposed by
Bombelli, Koul, Lee and Sorkin (BKLS) \cite{sbk}. The
observation that was made by BLKS was that the exterior region of the
black hole   
has a well defined autonomous dynamics, -no information is fed into it front
the inside horizon- one can expect a second law to apply to an entropy defined
exclusively in it. To that end they introduced $\rho^{ext}$, the
reduced
density matrix corresponding to the observables available on a spacelike
hypersurface $\Sigma$, extending from the horizon to spatial infinity.

Defined in that way $\rho^{ext}$, will undergo a well defined
behavior,
namely, expressed in Shrodinger picture, $\rho^{ext}(t_1)$ associate $x$ with a
hypersurface $\Sigma(t_1)$ will develop into unique state $\rho^{ext}(t_2)$
associated with a hypersurface $\Sigma(t_2)$ if the later lies strictly in the
future of $\Sigma(t_1)$, i.g $\Sigma(t_2)$ will be in the domain of dependence
of $\Sigma(t_1)$ but not vice versa: Evolution of the external states is well
defined into the future but not the past, however the evolution is not
unitary (only the overall evolution need be unitary). Moreover the entropy
$S_{ext} =-tr (\rho_{ext}ln \rho_{ext})$ will in general be nonzero even the
overall state is pure, and need not remain constant. The way in which the
above
entropy is defined can be thought as an extreme type of coarse graining in
which one neglects not only the correlation between internal and external
states but even the existence of the internal region itself.  
That this can lead to any well-defined evolution at all for $\rho^{ext}$
is
due
to one-way character of the future horizon , which allows the spacetime region
\emph{outside} of it develop \emph{autonomously}.

Now, the above type of entropy when calculated turned out to be divergent
due the to the entanglement between values of the quantum field just
outside and just outside the horizon, and if no cutoff were introduced
the entropy would diverge. However  if a cutoff is introduced the result
would comes out to be proportional to the area of the horizon with
proportionality constant quadratic in the cutoff. To the same conclusion
arrived Sredincki \cite{sred} who rediscovered the idea of BLKS and
pointed out that
the global vacuum states of a
scalar field in flat spacetime, when restricted to the exterior region
of an imaginary sphere, is in a mixed state there. The density matrix of
this mixed states arise from tracing out  those parts of the global states
that reside inside the sphere; its entropy is evidently related to the
unknown information about the sphere's interior. This entropy is
non vanishing only because the exterior states is correlated with the
interior one. In the sphere case also the quantum entanglement entropy
comes out to be proportional to the sphere's surface area with a
coefficient which diverges quadratically in the high frequency cutoff.
The entanglement has been lately  considered by several authors  with
the same conclusion. Kabat and Strassler further show that the density
operator in question is thermal irrespective of the nature of the
field \cite{ks}.

In trying to identify the entanglement entropy as the black hole entropy one
ends up with the conclusion that its dependent on the number of species
of
fields that exists in nature , since each filed must make its
contribution
to the entanglement entropy . Yet eqt(3.3)  says nothing about the number
of
species
!. To overcome this problem Sorkin and t'Hooft suggested that different
species contribute, but the contributions of the actual species in nature
exactly add up to $A/4$. The point of view here is that the list of 
elementary
particles species is \emph{prearranged} to chime with gravitational physics.
Bekenstein
estimate such a contribution for the list of elementary particle we know and 
argued
that it is not inconceivable that due to the gravitational and other 
interactions, $S_{ext}$ ends up $ A/4$.
Another different proposal for the resolution of the species problem was 
offered by
Jacobson. If the Hilbert-Einstein action were all induced 
it thus would make sense to identify the entanglement entropy and the B.H
entropy.

However, it should be noted here that the above resolution depends on
black
hole entropy being all the entanglement one. For instance BLKS argued
that
such a
contribution must be present and need not be the only one. And Callan and
Wilczek claimed that it is only a correction to the tree level part of
$ S_{bh} $. 

Regardless of the fact that the entanglement is the full or a part of
the black hole, if on accepts its presence a Ultra-violet cutoff should
be present to prevent the entropy from diverging and hence
make any sense of it. BLKS also suggested that  the physical entropy may
turn to be finite due to the quantum fluctuation of the horizon, and
Sorkin argued further that this may change things dramatically, by
cutting off the entangled field modes at a rather long wavelength, so
that they become a relatively small correction to the original B.H entropy.
 
On the
other hand Bekenstein argued that the entanglement entropy is operationally
finite (at least in flat spacetime) the claim was that , it is untrue, in
general, that one knows
nothing about the interior sates. For example knowing the size of the
internal region one can set a bound  on the energy  which one has to
trace
out and hence providing cutoff and making the entropy operationally
finite.

\section{Counting Links}
As we have seen earlier one of the most interesting aspects of the B.H
thermodynamic is its
universality, for instance the basic form of the laws appears to
be independent  of the detail of the precise Lagrangian or Hamiltonian of
the underlying theory of gravity -like any thermodynamical system. For
instance the zeroth laws of B.H
thermodynamics can be derived for Stationary-axisymetric black without any
reference to the field equations, and for the first and the second law
can be derived only upon the assumption that the field equation was
derived
from a diffeomorphism covariant Lagrangian. More precisly  
 the entropy will be given by an integral of a local
geometrical expression over the black hole Horizon, and it seems that the
black hole laws are not tied to any specific model-in manner similar to
any thermodynamical system- and all what one has
to have is an event horizon (a region of spacetime where no thing can
escape). The presence of the event horizon prevents one from measuring
the internal configuration (acquiring information ) unless one goes inside
in which case one still would not be able to report his results to
observers remaining outside the black hole, hence making the emergence
of the statistical description rather automatic for black holes, leading
one to coarse grain the interior region of the black hole, and therefore
to a notion of entropy. We have also seen that the entropy for quantum
fields
defined by tracing out the states inside some region of space time turn
out to be proportional to the area of the boundary if an Ultra-violet
cutoff is introduced, however from the discussionof the previous section,
it
seems rather unnatural to attribute the B.H entropy ( or at least all of
it) to the entropy restored  in the quanta of matter field (
gauge fields other than the gravitational) unless the Hilbert-Einstein
action is all induced neither the
emergence of the B.H
laws  nor some recent derivations of the black hole entropy suggest
this. The point we want to make here is that the B.H entropy could be 
better understood as pure
gravitational-spacetime - namely, 
\emph{as a response of having an event horizon which hides information
about
a
region of space time, and it is value measure the amount of missing
information about the region of spacetime inside the black hole.}

The question that arises here is how to measure these information. In the
continuum
picture there seems no way to do this, and it is not even obvious how to
formulate the question, and as we shall see later the quantity we
calculate would diverge if no ultra-violet cut-off were present in the same
way that the entanglement entropy for quantum fields diverges.

In the causal set context such question cannot be formulated either in
the current stage of development of the theory, and
 deriving the entropy from the first principle is not possible without
having the causets dynamics in hand, however, recall that the underlying
picture of spacetime in the causal set context is discret and hence one
may hope to capture  the black hole entropy, to a good approximation, not
by counting states directly 
but 
rather by counting discrete
elements of the causal set itself. A hint for this comes from the case of
box of
gas, where one starts by a density operator (counting states, taking into
account quantum numbers,,etc) but in the classical limite and for a pefect
gas one ends up merely counting the number of molecules to leading order.
Now in the black
hole context we expect that the entropy can be understood in as
entanglment in a
sufficiently generelized sense, and we may try to estimate the leading
behaviour of this entanglment by counting discret elements (entanglement
between discrete elements). In trying to look for those elements we recall
that the black hole entropy is mainly a measure of the area of its horizon
in Planck units, so we have just to come up with this measure in the
causal set picture\footnote{Note here that
it is actually very difficult to ge the area of cross section of a null
surface by counting, unlike what
a Euclidean intuition might suggest; for instance, no one knows a simple
measure of spacelike distance in causets that works in general.}.  
We also
think heuristically of "information" flowing
along links, and we may hope to eastimate  how much information is
getting entangled by the causet growth process using links. Now links also
are the building blocks of the causet, remember that knowing the links
between
the elements is equivalent to the knowledge of the whole causal
set\footnote{It is enough to know links, and the other relations are
implied by the axioms of the poset.}, and it
seems natural that by counting the links between elements that lie outside
and inside the
horizon one would account for the missing information about the region
inside  the balk hole.
With this interpretation the black hole
entropy is a type of entanglement between causet elements in-outside the
horizon, in the presence of the horizon one does not know which points
outside the horizon are linked to  the ones inside the horizon . 
 
The aim of the rest of this chapter is to show that by counting links
 with suitable  causality conditions on the element in-out defined with
respect to the
horizon and  a given hypersurface  on which we seek to evaluate the
entropy, will come out proportional to the area of the horizon in
causal set units.

\textbf{Our program}

Recall that  in the causal set appraoch the underlying structure of
spacetime is a causal set and
the manifold picture arise only as  large scale approximation (most
probably after 
coarse graining) of
the causal set, via faithful embedding.

Our goal is to consider a causet obtained via random sprinkling in a
black
hole background with density $l_c$, so this causal set is faithful
embeddable in this back ground by definition, and try to count the number
of links with certain causality ( maximality and minimality)  conditions
which will specify below.

Before I outline the program and give detail account of the derivation
of the different results, let me recall the definition of link and the
maximality
and minimality in the context of a faithfully embedded causet.

Let first  $C$ be any causet and   $x$ and $y$ two element in this
causet.
The interval  defined by two elements of the causet is defined by 
$$
J(x,y) :\{ p\in C | x\p p\p y\}
$$
Now if $J(x,y)$ contains no element (except $x$ and $y$) we say there is
\emph{link} between $x$ and $y$.

A point $p$ in is said to be maximal (minimal) in $C$ if there is no
element in it is future (past).

Now if the causet is embeddable (not necessarily faithfully) in spacetime
with Lorentzian metric the above definition can be translated to the
following.

$$
x \p y \Ria x \in J^-(y)
$$

If $J^+(x)\cap J^-(y)$ is empty (contains no element except $x$ and $y$)
then there is a link 
between $x$ and $y$.
 
If $J^+(p)$ $ (J^-(p))$ contains no elements (except $p$ itself) then $p$
is said to be
maximal (minimal). 

Now if the causet is faithfully embedded in a some
manifold, we talk  about the probability of having a point in a given
volume
or the probability of having a link between two points ..etc.

First recall  that the probability that there will be exactly $n$
embedded
points in a given Alexondrov volume is given by , $\frac{(\rho
V)^n}{n!} \ e^{-\rho V}$. The probability of having no point is $e^{-\rho
V}$ and that of having exactly one point is an infinitesimal volume element
$dV$ is just $\rho dV$ . For  fixed  $x$  the probability of having a link
with another point $y$ ($x\in J^-(y)$ say) , is given by
$$
P( x\p y)  = e^{-\rho V(A(x,y))} \rho dV(y)
$$
Now , let $\chi (x,y) $ denotes the variable whose value is $1$ if there
is
a link between these points and $0$ otherwise. The expectation  value
is
 $$
< \chi (x,y)> = 0 \cdot P(\chi =0)+1\cdot P(\chi =1) = P( \chi=1)
$$
So we conclude that the expected number of links with $x$ is given by
$$
<n_l(x)>= \int_{J^+(x)} e^{-\rho V(A(x,y)} \rho dV(y)  
$$

Now the quantity we are aiming to calculate is the expected number of
links
between points defined in specific regions in space time and satisfying
 certain max and min conditions.
The Max and  Min and link condition are just statement about a specific
volume
being empty :The probability that a volume contains no point.
If we denote this volume by $V$ this will be $e^{-\rho V}$ .

The probability of a having  point $x$ in infinitesimal volume  
$dV(x)$
 and point $y$ in $ dV(y)$ is $\rho^2 dV(x)dV(y)$

Than the probability of having a link is

$$
 P (x,y)=e^{-\rho V(x,y)} ({\cal{D}} ) \rho dV(x) \rho dV(y)
$$

Where ${\cal{D}}$ stands for the fact that $x$ and $y$ are subjected to
specific Max and Min and  belong to specific regions, and $x\p y$.

Now it is easy to deduce  that the expected number of
links between the points satisfying the condition of the domain $\cal{D}$
is 
\beq
<n> = \int_{\cal{D}} e^{-\rho V(x,y)} \ \rho^2 dV(x) dV(y)
\eeq

This is our main equation which we will use in the rest of this chapter.

We will set $\rho =1$ ($l_c =1$).

\begin{figure}[h]
\epsfxsize=8cm
\centerline{\epsfbox{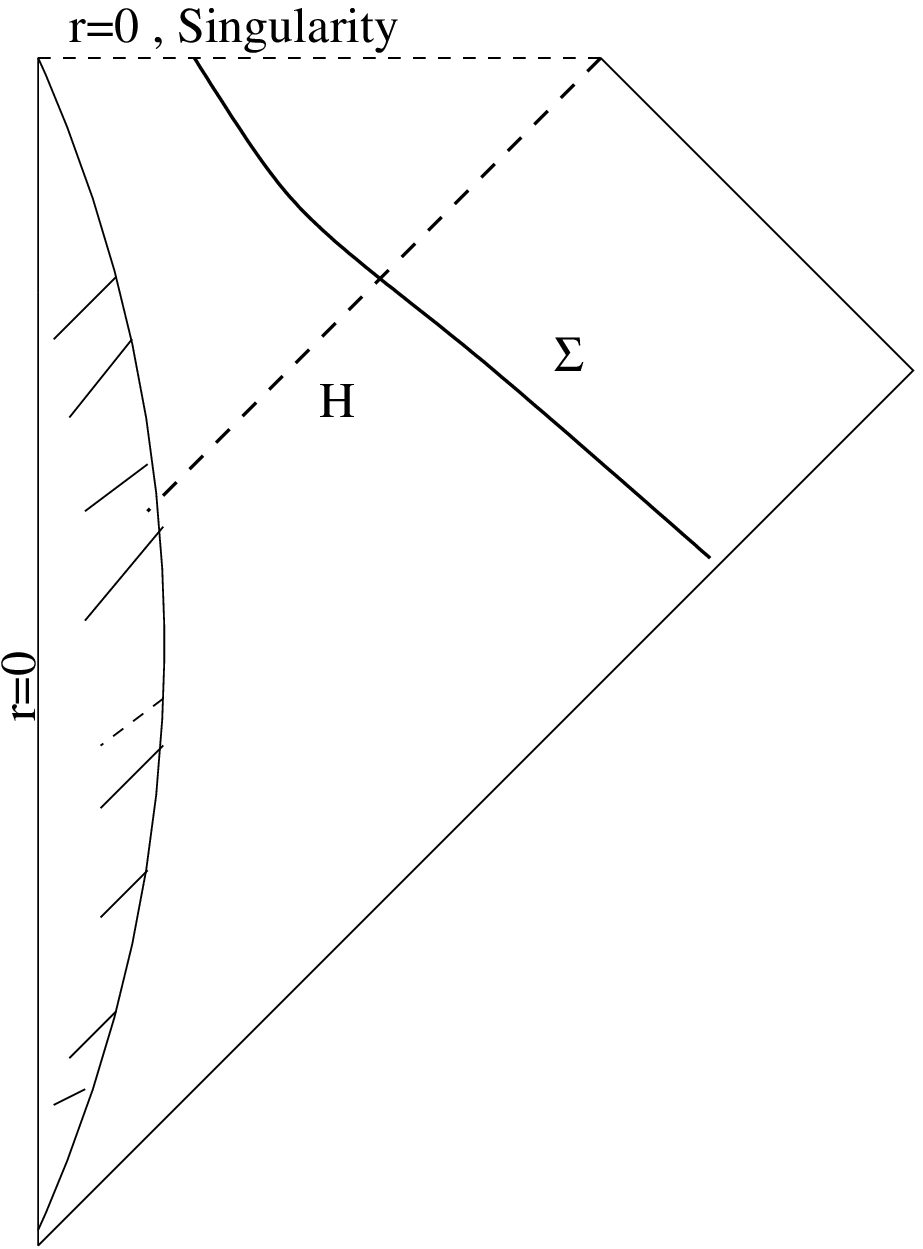}}
\caption{}
\end{figure}

Now, consider a collapsing spherically symmetric start, which produces the
spacetime given by Fig(3.1), chose a hypersurface $\Sigma$ which has a
well
defined time evolution,  which could be null or space like,
and
intersect the horizon

consider a point in the region $ J^-(H)
\cap
J^-(\Sigma)$ and other in the region $J^+(H)\cap J^+(\Sigma)$ and such
that $y\in J^+(x)$. Now were we to count the number of links between pairs
$(x,y)$  with no further conditions, the result could be shown to be
infinite for spacelike and for a
null surface could be shown to go like $ (2M)^4$ . The infinity can easily
be
understood as coming form pairs of points which are null related (they
have a zero
volume) and there are infinity of them. Now the only way to make this
result
finite is to suppress those contributions. A natural a choice for doing
this is to impose maximally condition on $x$ and minimality condition
on $y$. The minimality  and maximality conditions will be chosen as
vvfollow:
we will demand $x$ to be maximal in $ J^-(\Sigma)$ and $y$ be minimal in
$J^+(\Sigma)\cap J^+(H)$ .
The maximality condition is  natural a choice, since we would like to
account for information missing (links) in a given hypersurface $\Sigma$,
and if
one did not impose such a condition, one would be counting information
which were already acquired   
in  preceding hypersurface .
The fact that we do not impose  a similar condition on $y$ is because this
condition would give zero for a null case, but the result must agree 
for null or spacelike if both intersect the horizon in the same time,
moreover for stationary black  the  results should  agree in all cases.

The above two Max and Min conditions are the ones which will adopt to
derive the area law.

Now, to carry the calculation in  4-d Schwarzschild black one has to
evaluate the volumes to ensure the link and maximality  
and minimality (Max and Min for short) conditions, however 
due to the complexity of the null geodesic ( the non radial ones), these
volumes are very hard to get and may turn out to have a very complicated
form making the calculation unmanagable\footnote{To our knowledge
Alexandrov neighborhood has never been calculated for the Schwarzschild
case 
(let alone imposing Max
and
Min conditions), even in a simple spacetime like de Sitter the volume has
only an infinite expansion in term of the curvature and can be only useful
when the curvature is small. The Alexandrov volume in general geometry can
be
expanded in term of the curvature, valide for small curvature, and since
we are mainly intrested in the case of large mass and expect that our
counting is controled by  near horzion geometry, such an expansion
will be a very good approximation, although it remains to impose Max and
Min conditions which turned to be
 no trivial task even in the flat case.}.

To establish our result we will take the following road. 

1) -We  first consider a dimensional reduction of the 4-d Schwarzschild
obtained by
identifying each two sphere $S^2$ to a point, the result is 2-d black hole
, and it is different from the standard 2-d black hole only in one respect
being not a solution  of 2-d Einstein equation. Than we will count the
expected number of links with a modified Max and Min conditions. These
conditions are slight modification of the condition which will adopt in
the
4-dimensonal case and do not change the argument we want to make at the
end, they are introduced only because of their convenience to get an
exact result, and are not arbitrary, moreover we will see that
these two conditions give  the same result.  

This calculation will be done using a null hypersurface and the result
will be shown to be a constant. This result is expected if we want the
counting in 4-dimensional to give some thing  proportional to the area of
the horizon.

2)-In the second step we will consider a black hole produced by a
collapsing spherical null  shell of matter, the spacetime produced has
two
regions one flat with an event horizon and other Schawrzschild with event and
apparent horizon. We repeat the same counting using the same Max and Min
conditions in the 2-dimensional reduction of this spacetime using null 
hypersurface lying completely in the flat region and intersecting of 
course  horizon,  and  restrict the counting only to the points which 
lie in the flat region. Now, the result will be shown to equal to that
 of Schwarzschild case up to a negative correction vanishing when the
collapse is
pushed to future infinity. Putting the first step and
the second step
together we conclude  that in order to establish the area law for the
Schwarzschild
case it is enough to establish using the flat region and pushing the
collapse in 4-d to future infinity, since the agreement between the two
calculations in 2-d can be transfered simply to an agreement about the
proportionality  to the area of the horizon between the Schwarzschild and the
null shell  case, more precisely, 
if the expected number of links counted in the flat region with collapse
pushed to infinity gave $<n> = c \pi $T$^2$ , where $T$ is the time in
which the hypersurface intersect the horizon, the counting in Schwarzschild
case would give $<n> =c\pi$ $(2M)^2$, where $M$ is the mass of the black
hole.

3) In the third step we will use a space-like hypersurface and show for
the null shell case
with the collapse
pushed to infinity that the result must be a constant,  and how the apparent 
divergence coming from null related pairs are cured
by the our Min, Min and link conditions. This calculation will be done
using
the Max and Min conditions which we will adopt in 4-d. A necessary
consistency check for  our calculation is to show that the result is
the
same for null and Spacelike case. Even though we won't prove this here,
we will give an argument  which leads to  expect that the
two cases should agree\footnote{Indeed a check in 2-d should be enough, 
and this will appear in \cite{dour}.}.  
 
4)The four and the last step will be doing the calculation in 4-d using a
space-like hypersurface
for the collapsing null shell pushing the collapse to infinity and showing
that the result is proportional to the area of the horizon.

\subsection{2-dimensional Schwarzschild spacetime}

Consider the two dimensional Schwarzschild spacetime obtained from the
realistic 4-dimensional black hole space time 
 by identifying each 2-sphere $S^2$ to a point, 
  the resulting two dimensional
spacetime has exactly the same causal structure as the S-sector of the
4-dimensional. 

For simplicity we have ignored the presence of the collapse, this of
course will not
change the argument since the detail of the collapse  should be 
irrelevant 
as mentioned earlier, or  we can choose the hypersurface to intersect
the horizon far from the collapse and the result will not be affected by
the presence of collapse. 
 Moreover, we could consider an eternal black hole 
where not
only the region of spacetime which is present in a realistic collapse is considered, and the result will turn out the
same. So only the portion showed in Fig(3.2) will be considered for
calculation.

\begin{figure}[h]
\epsfxsize=8cm
\centerline{\epsfbox{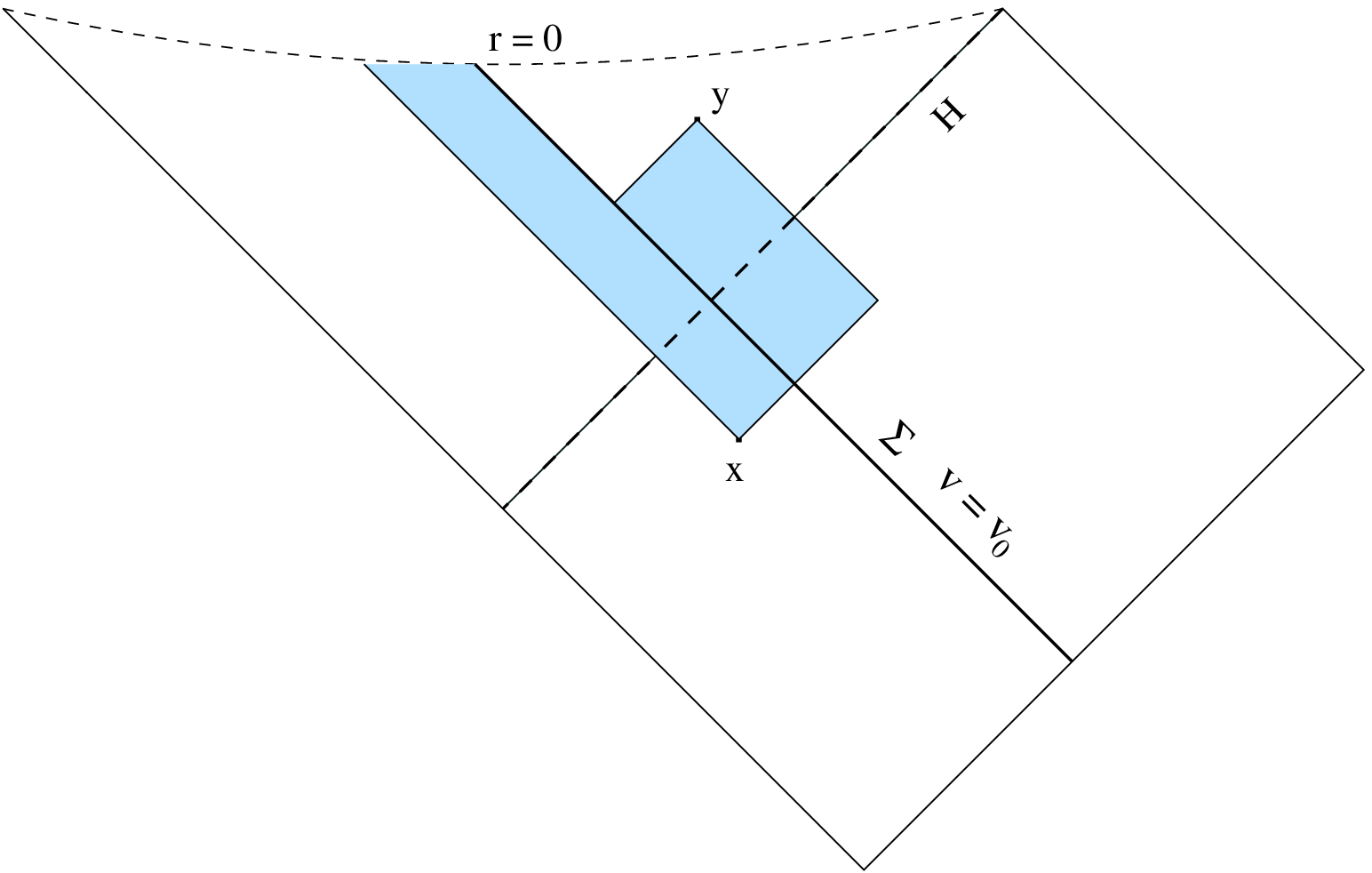}}
\caption{}
\end{figure}

The metric describing this spacetime is given by

\beq
ds^2 = -\frac{16M^3}{r} e^{-r/{2M}} dudv
\eeq

where we use Kruskal coordinate and $ (u, v)$, $r$ is defined implicitly
via the following equation
\beq
uv= (1-\frac{r}{2M}) e^{r/{2M}}
\eeq
Since we are dealing with macroscopic black hole, $2M \gg 1$ in causal set
units, which is expected to be around Planck scale.

Now let $\Sigma$ be a null hypersurface with equation $v=v_0$, and $(x,y)$
be a pair of points satisfying the following conditions
Fig (3.2). 
\beq
 \left \{\begin{array}{c}
x \in J^-(\Sigma) \cap J^-(H)\\
y\in J^+(\Sigma)\cap J^+(H)\\
 x \ \ \mbox{max in} \ J^-(\Sigma)\\
y \ \ \mbox{min in} \ J^+(\Sigma)\cap J^+(H)\\

\end{array}\right.
\eeq
Note that here there is no need to impose the link and the causal relation
between $x$ and $y$ since condition (3.10) already imply them.

The volume needed to ensure those conditions can easily be calculated
 and is given by
$$
V = r(\frac{v_x}{v_0})+r_{0y}^2+r_{xy}^2-r_{yy}^2-r_{x0}^2
$$
Where we used the following notation
\beq
u_i v_j = (1-\frac{r_{ij}}{2M}) e^{r_{ij}/{2M}}
\eeq

Now using (3.7) the expected number of links can be written as

\beq
<n> =(16M^3)^2 \int_{0}^{v_0} dv_x \int_{-\infty}^{0} du_x
\int_{v_0}^{\infty} dv_y\int_{0}^{1/v_y} du_y
\frac{e^{-r_{xx}/{2M}}}{r_{xx}}
\frac{e^{-r_{yy}/{2M}}}{r_{yy}}\ e^{-V}
\eeq
Now , due to the complicated implicit dependence of the volume on $v_x,
u_x, v_y$ and $u_y$ it is very hard to get an exact asymptotic formula for
this integral but it is not difficult to show that it is finite and is
bounded between the two following value.
\beq
 \frac{\pi^2}{6} \le <n> \le  \frac{\pi^2}{3}a 
\eeq

Where $a\equiv 2M \gg 1$.

Of course this bound it is not useful, however we will see later that the
lower bound must be the exact leading order of the integral. Now, since
our
aim
is to justify  the restriction to the flat region in four dimension , we
 can modify the Max and Min condition
slightly to make the calculation possible in both Schwarzschild region and
flat region.

To do that let us impose the following conditions.
\beq
 \left \{\begin{array}{c}
x \in J^-(\Sigma) \cap J^-(H)\\
y\in J^+(\Sigma)\cap J^+(H)\\
 x \ \ \mbox{max in} \ J^-(\Sigma)\cap J^-(H)\\
y\ \  \mbox{min in} \ J^+(H)\\
\end{array}\right.
\eeq
\begin{figure}[h]
\epsfxsize=8cm
\centerline{\epsfbox{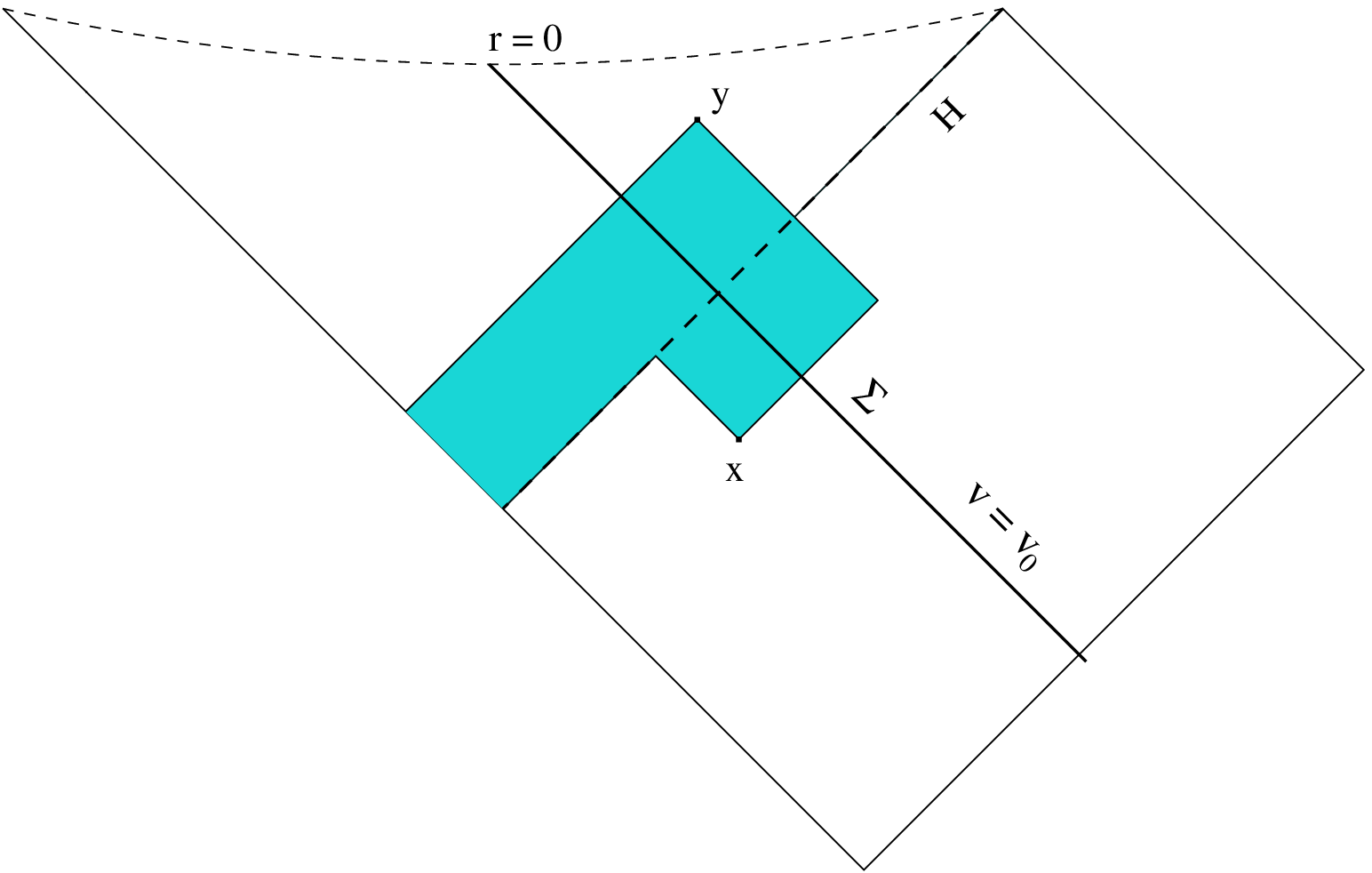}}
\caption{}
\end{figure}

The volume needed to ensure these conditions is evaluated in Appandix A. 
and is
given by,  Fig(3.3)
\beq
V= (2M)^2+r_{xy}^2-r_{xx}^2-r_{yy}^2
\eeq
 
The expected number of links becomes now
\beq
<n> = (16M^3)^2 \int_{0}^{v_0} dv_x \int_{-\infty}^{0} du_x
\int_{v_0}^{\infty} dv_y \int_{0}^{1/v_y} du_y 
\frac{e^{-r_{xx}/{2M}}}{r_{xx}} \frac{e^{-r_{yy}/{2M}}}{r_{yy}}  e^{-V}
\eeq
Changing the  variables
$$
(u_x, v_x, u_y, v_y) \ria (r_{xx}, r_{x0}, r_{xy}, r_{yy})
$$
using (3.11)  it is easy to evaluate the corresponding Jacobian
$$
J= \frac{4}{(16M)^3} \frac{r_{xy}}{r_{xy}-2M} \frac{r_{x0}}{r_{x0}-2M} \
r_{xx} \ e^{r_{xx}/{2M}} \ r_{yy}\ e^{r_{yy}/{2M}}
$$
where we implemented the notation ( $r_{x0} \equiv r(u_x v_0)$).

Let 
$$
x = r_{xy} , \ y = r_{x0} \ z= r_{xx}
$$

Now the boundary of the integral can be easily deduced form (3.11).

We end up with 
\beq
<n> =4 (e^{-a^2} \int_{0}^{a} e^{r_{yy}^2} dr_{yy}) I(a)
\eeq

Where $ a= 2 M \gg 1$ for a macroscopic B.H.
and
\beq
I(a) =\int_{a}^{\infty} dx \frac{x}{x-a} e^{-x^2} \int_{a}^{x} dy
\frac{y}{y-a} \int_{a}^{y} e^{z^2} dz
\eeq

In the first place it is interesting to note that the expected number of
links does not
depend on $v_0$ , reflecting  the stationarity of spacetime (black
hole). In fact any dependence on $v_0$ (at least the leading contribution)
would kill any hope for this calculation to produce any thing of physical
significance, for stationary black holes.

Now, $I(a)$ is shown in Appandix.B to have the following asymptotic
expansion,
$$
I(a) =\frac{\pi^2}{12} a + O(1/a)
$$
It is also easy to show that 
$$
e^{-a^2} \int_{0}^{a} e^{r_{yy}^2} dr_{yy} = \frac{1}{2a}+ O(1/a^2)
$$
Now putting all together we get,

\beq
<n> =\frac{\pi^2}{6} +O(\frac{1}{a})
\eeq

Although the conditions we imposed 
are not the ones which we will be adopted, this results shows that such
counting, were it done in four dimension would turn out to be
$v_0$-independent and proportional to the area of the horizon namely $
(2M)^2$.

Before we  move to the collapsing null shell let us see where the dominant
contributions
comes from. To that it is enough to consider only the integral denoted by
$I(a)$ .

\begin{eqnarray*}
 I& =& \underbrace{\int_{a}^{\infty} \frac{x}{x-a} e^{-x^2} dx
\int_{a}^{x}dy\int_{a}^{y} e^{z^2} dz}_{I_1} \\
& & \mbox{} +\underbrace{a\int_{a}^{\infty}\frac{x}{x-a} e^{-x^2} dx
\int_{a}^{x}
\frac{1}{y-a} \int_{a}^{y} e^{z^2} dz}_{I_2}
\end{eqnarray*}

Let us first concentrate on $I_1$.

It is easy to see that $I_1$ can be written as
\begin{eqnarray*}
I_1&=& \int_{a}^{\infty} \frac{x}{x-a} e^{-x^2} dx \int_{a}^{x} dz
 e^{z^2} \int_{z}^{x} dy = \int_{a}^{\infty} \frac{x}{x-a} e^{-x^2} dx
\int_{a}^{x} dz
 e^{z^2} (x-z)\\
& & \mbox{} = \int_{a}^{\infty} \frac{x}{x-a} e^{-x^2} dx
\left[\frac{x}{2z}e^{z^2}-\frac{1}{2}e^{z^2}\right]_{z=a}^{z=a}
+\frac{1}{a}(\mbox{the first term})\\
& & \mbox{} =-\frac{1}{2a} e^{a^2} \int_{a}^{\infty} x e^{-x^2}
+\frac{1}{a}(\mbox{the first term})\\
& & \mbox{}= -\frac{1}{4a} +O(1/a^2)
\end{eqnarray*} 
          
So we that the term $a$ in the expansion of $I$ must have  come from
$I_2$.

Now 
$$
I_2 =a\int_{a}^{\infty}\frac{x}{x-a} e^{-x^2} dx
\int_{a}^{x}
\frac{1}{y-a} \int_{a}^{y} e^{z^2} dz
 $$
The region $y \gg a$ cannot give $a$, because in this region $I_1$ is
bigger in absolute value than $I_2$ as can be easily seen. So the dominant
came from $y\ria 0$ and hence from $u_x$ near zero, in fact as can be 
easily seen the term $\frac{1}{y-a}$ diverges at $ y=0$ and this
what makes it different from $I_1$. Although this argument
doesn't say any thing about the pints $y$, the points $y$ which dominate
the contribute are the ones sitting near the horizon (null related to the
$x$'s ), since other points are exponentially suppressed due to the link
 and max
conditions we are imposing on them. This result shows
clearly that this type of counting is controlled by the near horizon
geometry; for instance  pairs of points $(x,y)$ sitting  arbitrarily near
to the
hypersurface, with $y$  arbitrarily near the horizon, leading to
arbitrarily small volume, do not give any significant contribution if $x$ is
far from the horizon  namely with coordinate $|u_x|\gg 1$ ($y\gg a$).

\subsection{2-dimensional collapsing null matter}

We now consider a spherically collapsing null shell of matter with stress
energy tensor
given by
\beq
T_{vv} = \frac{M \delta(v-b)}{4\pi r^2}
\eeq
the other components are identically zero.
Penrose diagram for the spacetime (after dimensional reduction $S^2
\ria $ point) is shown in Fig(3.4).

\begin{figure}[h]
\epsfxsize=8cm
\centerline{\epsfbox{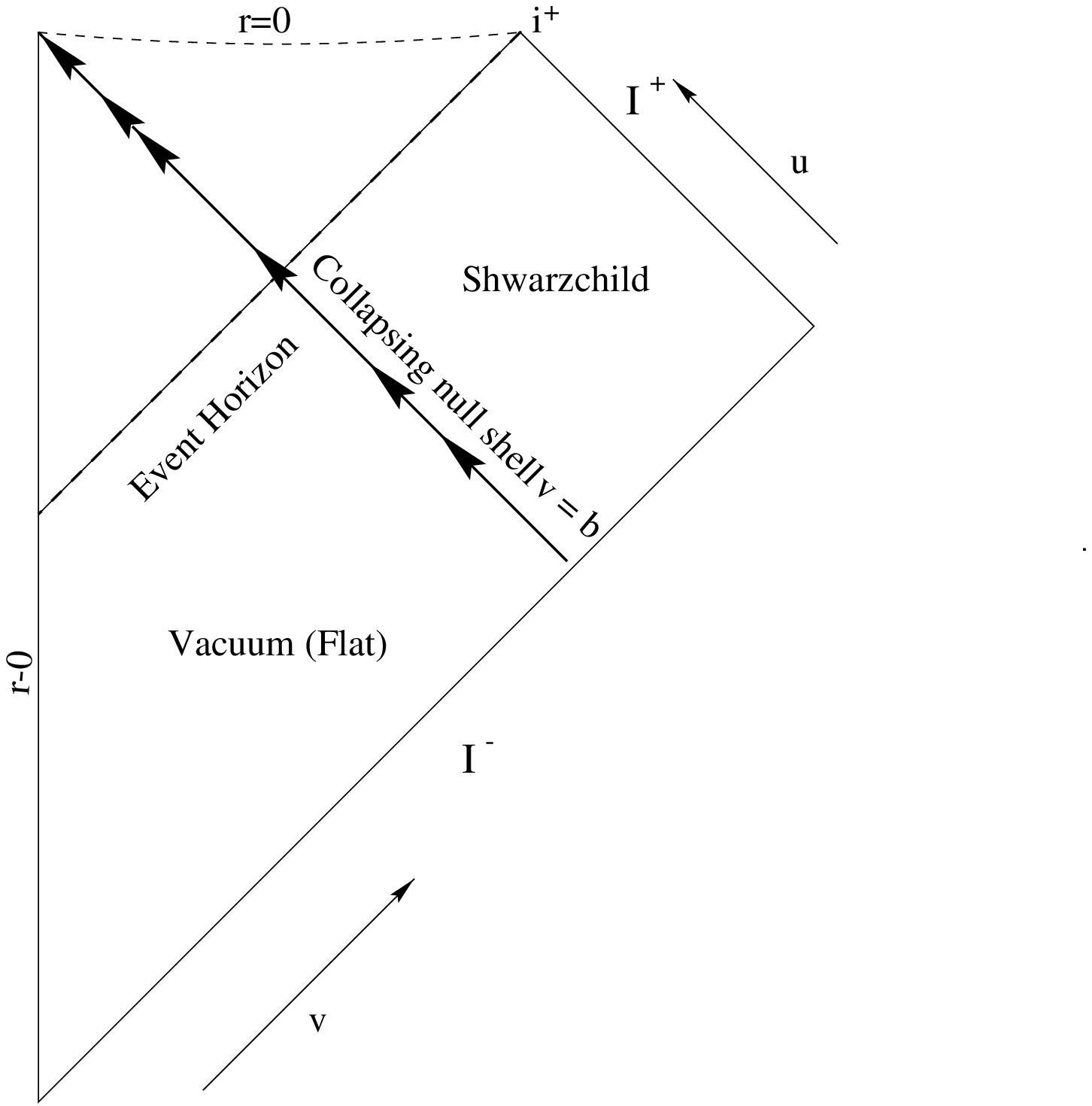}}
\caption{}
\end{figure}

Let us chose a null hypersurface $\Sigma$ parameterized by $v=a<b$,
Fig(3.5).
Here $a$ is of course different from $ a$ in the Schwarzschild
case.

\begin{figure}[h]
\epsfxsize=8cm
\centerline{\epsfbox{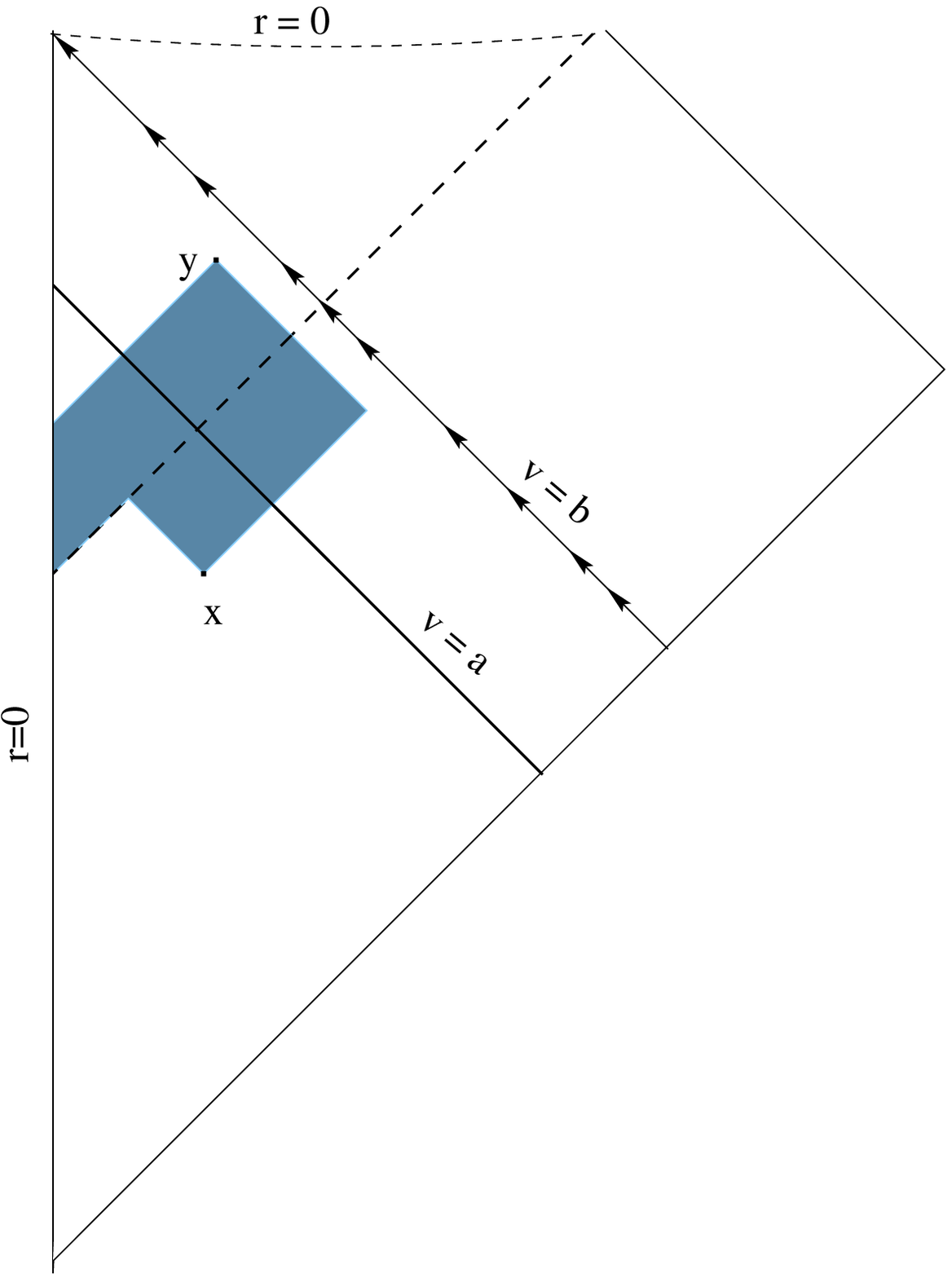}}
\caption{}
\end{figure}                

We will assume that $b\gg a\gg1$ ,the assumption that $b \gg a$ is natural
if we want to restrict the counting to the flat region only. The second
assumption  we take it for the time being as technical, however it will be
justified when we deal with the four dimensional case.

Imposing the same Min and Max conditions we imposed in the Schwarzschild
case
we get for $<n>$ the following expression
\beq
<n> = \int_{a}^{b} dv_y \int_{0}^{v_y} du_y \int_{\infty}^{0} du_x
\int_{0}^{a} dv_x e^{-V}
\eeq
where
$$
V= u_y v_y -u_x(v_y-v_x)-u_y^2/2
$$
It is easy to perform the integration over $v_x$ and $u_x$, and we end up
with
$$
<n> = \int_{a}^{b} dv_y \int_{0}^{v_y} du_y \ e^{-u_yv_y +u_y^2/2} \ ln
\left(\frac{v_y}{v_y-a}\right)
$$
Changing the variables from, $(u_y, v_y)$ to $ (x=v_y, y=v_y- u_y)$ we
get
$$
<n> =\int_{a}^{b} dx \ ln\left(\frac{x}{x-a} \right) e^{-x^2/2}
\int_{0}^{x} dy \ e^{y^2/2}
$$

Since $a\le x\le b$, and assuming that $a \gg 1$ , it is easy show that
$$
e^{-x^2/2} \int_{0}^{x} e^{y^2/2} dy = \frac{1}{x} +O(\frac{1}{x^2})
$$
It follows that
$$
<n> =\int_{a}^{b} \frac{1}{x} \ ln\left( \frac{x}{x-a}\right) dx
+\mbox{higher orders}
$$
By higher orders here, we mean terms which are $\le \frac{1}{a}$ the first
term.
Now it is also easy to show that
$$
\int_{a}^{b} \frac{1}{x} ln\left(\frac{x}{x-a}\right) =\frac{\pi^2}{6}
-\l(a/b)
$$
where
$$
l(a/b) =\sum_{k=1}^{\infty} (a/b)^k/k^2
$$
This series is of course convergent and is vanishing in the limit $b\ria
\infty$.
The presence of the negative contribution $l(a/b)$ is expected, since
when $\Sigma$ in near the Schwarzschild region (near the collapse), one
should not neglect the links coming from the Schwarzschild which would
give a significant contribution .

Now,  this result if of interest, by assuming that the collapse is pushed
away, or even to future
infinity , one can restrict oneself to the flat region and the result
would
turn out to be the same as the Schwarzschild case, vice-versa , in the
collapsing null matter, we can restrict the calculation only in the
Schwarzschild region given that the Hypersurface chosen intersects the
horizon away from the collapse (in its future). This agreement means that
if the calculation were done in  four
dimension the result will be proportional to the area of the
horizon , and  the proportionality constant will be the same, in the
Schwarzschild case and the flat case, so to establish the area law for the
Schwarzschild case it is enough to do calculation in the flat region in the
for the collapsing null shell pushing the collapse to future infinity.

Before I move to the 2-d spacelike hypersurface, let me note the following,
By  imposing the Max and Min which
will be adopted for the 4-dimensional, the expected number of links using
only the flat region turned out to be (after performing two integral),
$$
<n> = \int_{0}^{\infty} ln(\frac{a+y}{y}) e^{-(y+a)^2/2} dy \int_{y}^{a+y}
e^{x^2/2} dx
$$
and it  can easily be shown that:
$$
<n> =\frac{\pi^2}{6}+ O(1/a)
$$ 
This result  have two implications. It implies that the lower
bound in (3.13) is exactly the leading order of the integral. The other
possible implication, is that this type of counting could be insensible to
some
conditions involving the region inside the black hole. This should not be
surprising, since  some conditions inside the black could be expected  to
be
irrelevant as far as the conditions have the same behavior near the
horizon \cite{dour}.

 \subsection{Space-like case}

\begin{figure}[h]
\epsfxsize=6cm
\centerline{\epsfbox{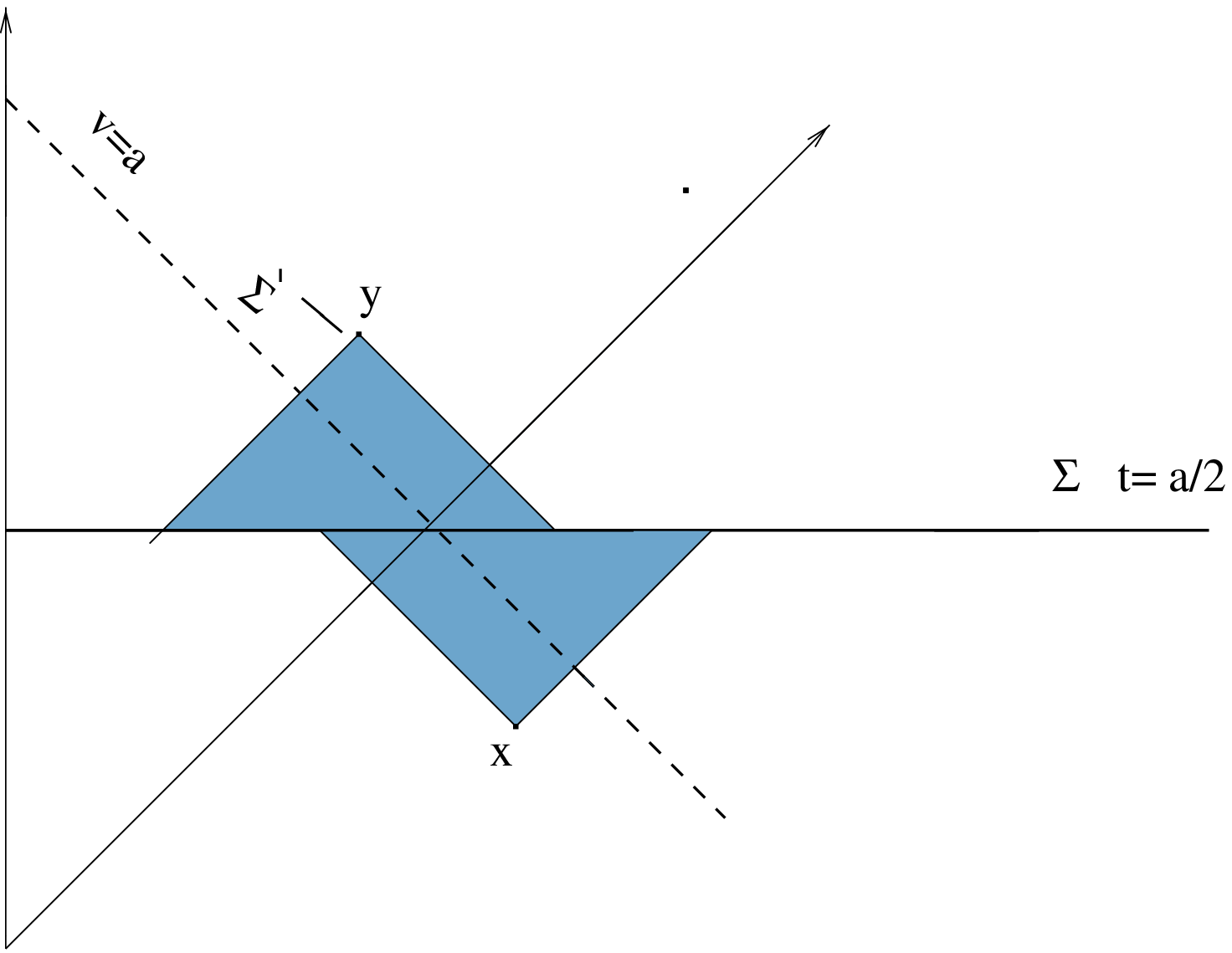}}
\caption{}
\end{figure}                               
\begin{figure}[h]
\epsfxsize=6cm
\centerline{\epsfbox{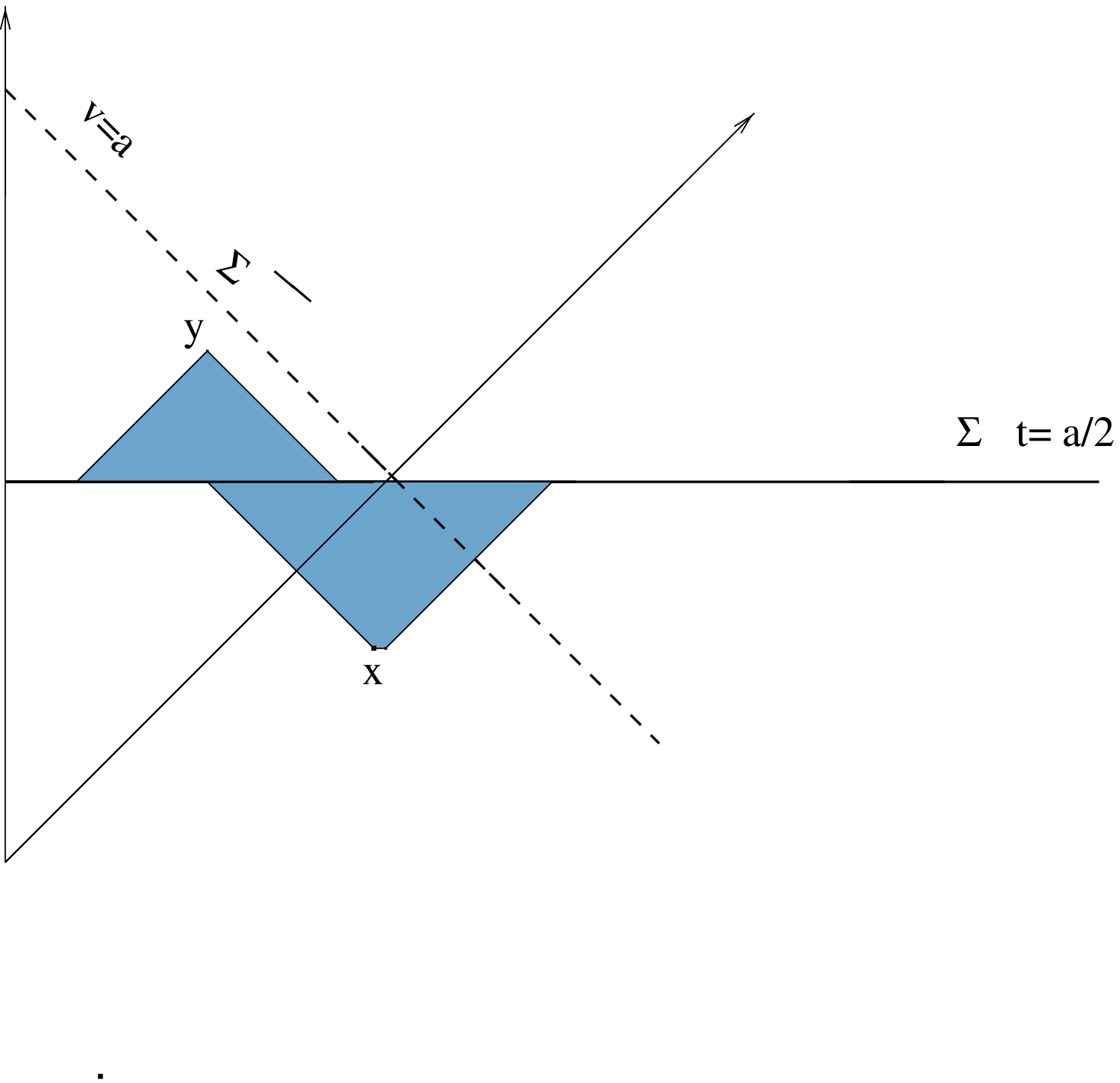}}
\caption{}
\end{figure}                               

This section we show that doing the calculation in 2-d reduction of 4-d  
flat using
spacelike hypersurface and imposing our Max and Min condition which we 
will
adopt is 4-d   the expected number of links is constant, we also show how
apparent divergences are cured by the Max and Min and link.

Pushing the collapse to infinity the space time produced is just half of
the
of the Minkowski spacetime. Consider a space like hypersurface $\Sigma $
with equation $t=a/2$ and consider a pair of points $(x,y)$ satisfying the
following conditions.

 \beq
 \left \{\begin{array}{c}
x \in J^-(\Sigma) \cap J^-(H)\\
y\in J^+(\Sigma)\cap J^+(H)\\
 x \ \ \mbox{max in} \ J^-(\Sigma)\\
y\ \  \mbox{min in} \ J^+(\Sigma)\cap J^+(H)\\
\end{array}\right.
\eeq

Using null coordinate  $(u,v)$ we have.

\beq
 \left \{\begin{array}{c}
 v_x\le v_y \\
u_x+v_x \le a ,\ \ u_x\le 0 \\
u_y+v_y \ge a, \ \ v_y \ge u_y \\
\end{array}\right.
\eeq

In the time like case one has to distinguish different cases for each we
have a different volume corresponding to the Max and Min conditions and
link. 

Let us first introduce  a null hypersurface $\Sigma '$ with $v=a$ .
Now as can be seen Fig(3.6) is  qualitatively the same as the
contribution 
we have already evaluated for the null case, so it will just give a
constant. Fig(3.7) will not give any significant contribution, because of
the
Max and Min condition. Now the only possible divergence or some thing
different from a constant can only come from contribution of Fig(3.8) and
Fig(3.9), and it easy to see that both contributions have the same possible source
of  apparent divergence, so we will
evaluate here just one of them, the other can be shown to be a constant

Now since the apparent divergence  due to the pairs with  $u_y \ria 0$
and $ t_x= a/2$, we will
count
just contributions coming from
$ u_y\le a$  (the other contribution $u_y \ge a$ will of course be
exponentially suppressed with respect to the one coming form $u_y\le a$.

Now it is easy to read from Fig(3.8) the  rang of $u_x,v_x,u_y,v_y$.

\begin{figure}[h]
\epsfxsize=6cm
\centerline{\epsfbox{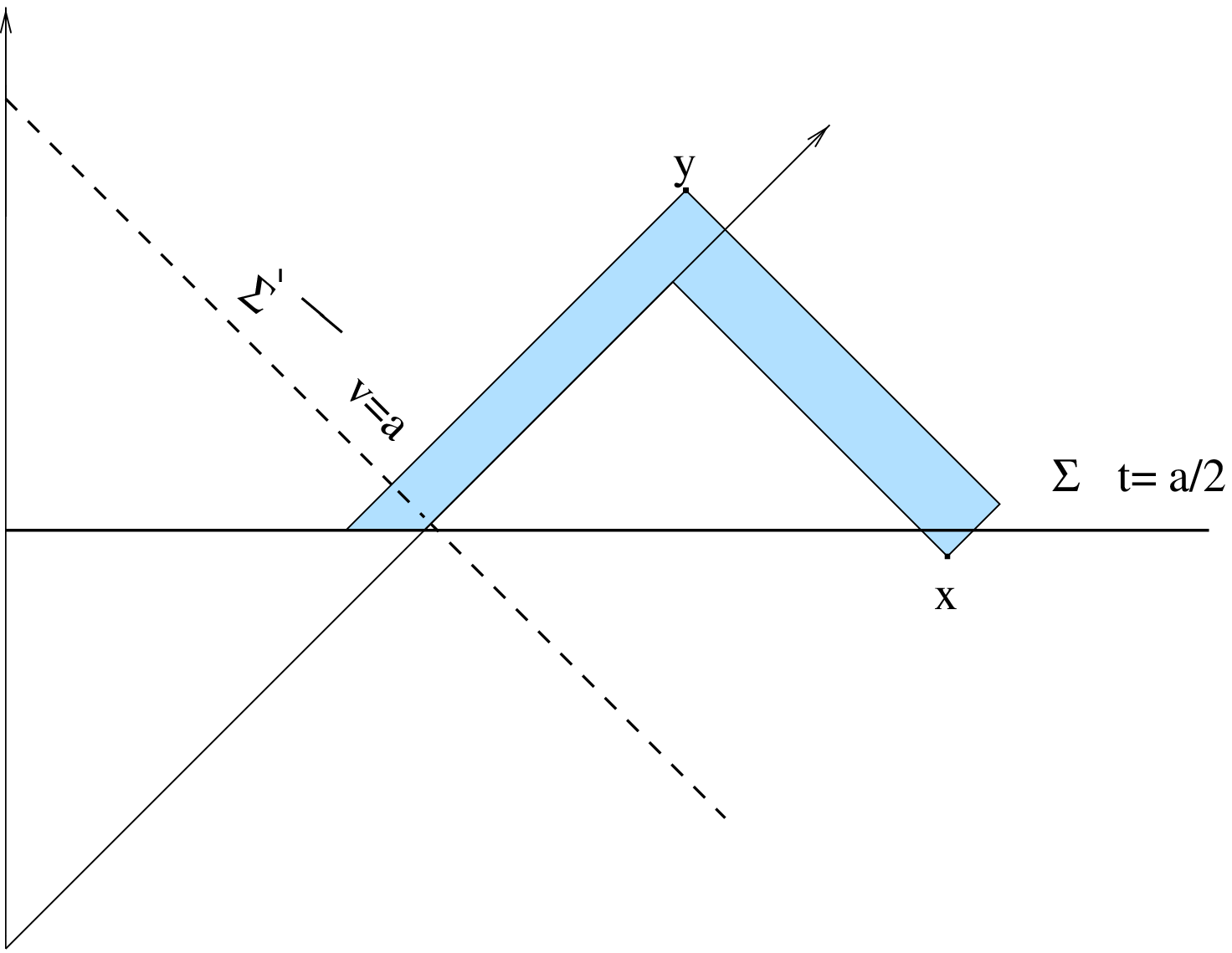}}
\caption{}
\end{figure}                    

\begin{figure}[h]
\epsfxsize=6cm
\centerline{\epsfbox{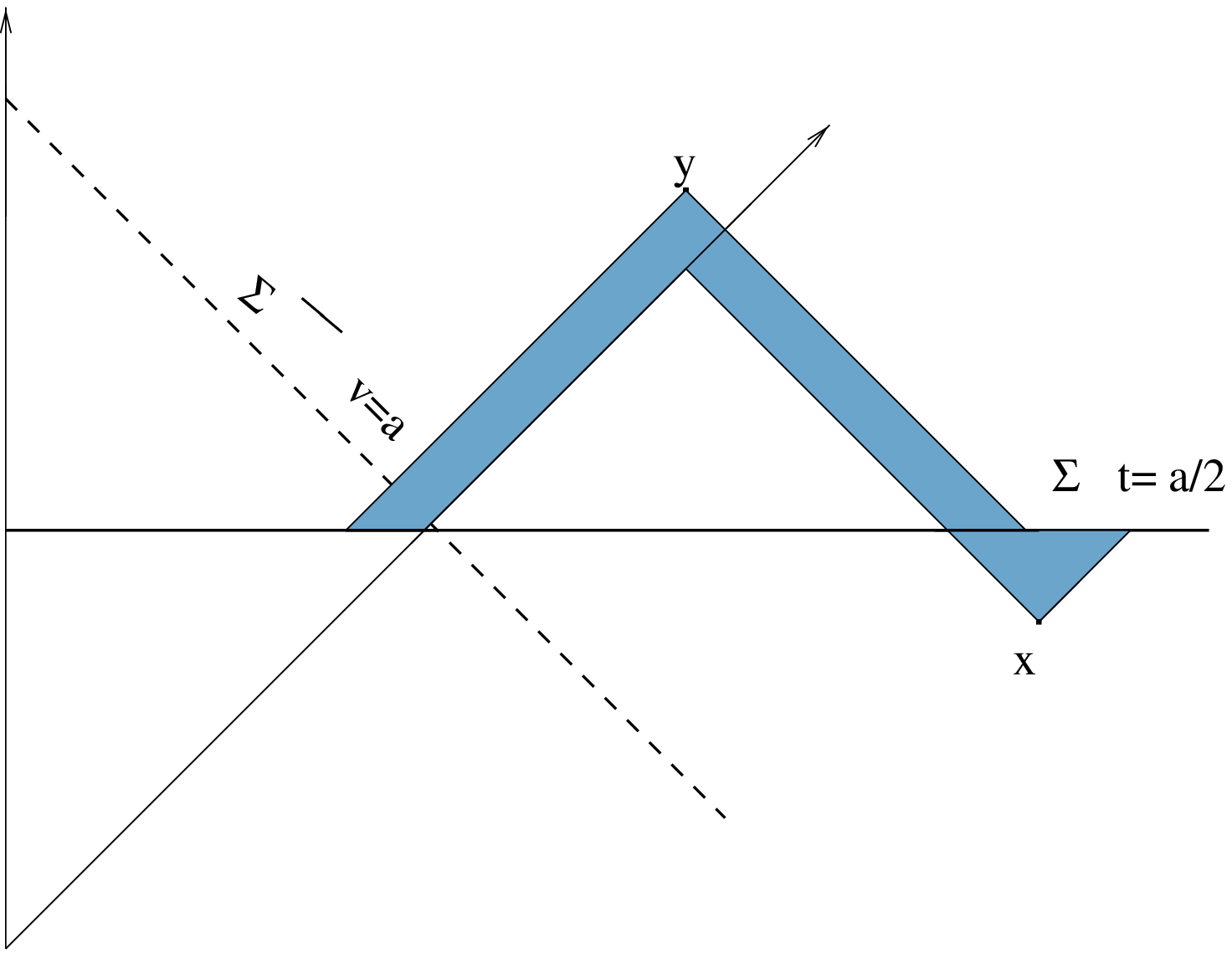}}
\caption{}
\end{figure}                               
\beq
 \left \{\begin{array}{c}
 a \le v_x\le v_y \\
u_x+v_x \le a ,\ \ u_x\le 0 \\
u_y+v_y \ge a, \ \ a\le v_y\ge u_y \le a \\
 v_y+u_x\ge a \ \ 
\end{array}\right.
\eeq

The volume needed which will use is the sum of $V_1$ and $V_2$.
\beq
V= v_y(u_y-u_x) +u_x v_x-u_y a +u_y^2/2
\eeq

the expected number of links
\beq
<n>_a = \int_{0}^{a}du_y\int_{-\infty}^{0}du_x\int_{a}^{a-u_x} dv_x
\int_{a-u_x}^{\infty} dv_y e^{-V}
\eeq
It is easy to perform the integration over $v_x$ and $v_y$ and we get
\beq
<n>_a=\int_{0}^{a} dy \int_{0}^{\infty} dx \frac{1}{x(y +x)}
e^{-xy-y^2/2}(1-e^{-x^2}) =I(a)
\eeq
Where we made a simple change of variables $ u_x =-x$ and $ u_y =y$.

To show that $<n>_a$ has the following form 
$$
<n>_a = (1) +O(1/a)
$$

It is enough to show that $I(\infty)$ is finite.
 
The only source of divergence of this integral could be from $x\ria
\infty$ and $ y\ria 0$, so let us study this region.

We first change the variable from $ x,y$ to $ x, z-x+y$, $I$ becomes
\beq
I = \int_{0}^{\infty} dx \int_{x}^{\infty} dz 
\frac{1}{x(z)}
e^{-z^2/2+x^2/2}(1-e^{-x^2})                                          
\eeq
 Let us split the $I$ into two contributions.

$$
( \int_{0}^{\lambda} +\int_{\lambda}^{\infty}) dx 
\frac{e^{x^2/2}-e^{-x^2/2}}{x} \int_{x}^{\infty} 
\frac{e^{-z^2/2}}{z}
$$
Where $\lambda >1$.
To prove the convergence of this integral it is enough to prove the
convergence of the following integral
$$
I_{\lambda} = \int_{\lambda}^{\infty} \frac{e^{x^2/2}}{x}
\int_{x}^{\infty} \frac{e^{-z^2/2}}{z} dz
$$
It is easy to see that
$$
I_{\lambda} \le \int_{\lambda}^{\infty} \frac{e^{x^2/2}}{x^2}
dx\int_{x}^{\infty} e^{-z^2/2} dz
$$
We have also
$$
\int_{x}^{\infty} e^{-z^2/2} dz =
\sqrt{\frac{\pi}{2}}(1-\Phi(\frac{x}{\sqrt{2}}))
$$
$$
\Ria I_{\lambda} \le \sqrt{\frac{\pi}{2}}\int_{\lambda}^{\infty}
\frac{e^{x^2/2}}{x^2} (1-\Phi(\frac{x}{\sqrt{2}})) dx \le 
$$
$$
\sqrt{\frac{\pi}{2\lambda}}\frac{1}{\lambda} \int_{\lambda}^{\infty}
\frac{e^{x^2/2}}{x^{1/2}} (1-\Phi(\frac{x}{\sqrt{2}})) dx \le    
$$
But 
$$
\int_{\lambda}^{\infty}
\frac{e^{x^2/2}}{x^{1/2}} (1-\Phi(\frac{x}{\sqrt{2}})) dx 
$$                                             
 is finite, since,
$$
\int_{0}^{\infty}
\frac{e^{x^2/2}}{x^{1/2}} (1-\Phi(\frac{x}{\sqrt{2}})) dx
=\frac{1}{2^{5/4}} \Gamma(1/4) \footnote{ Gradshteyn / Ryzhik , Table of
Integrals , Series and 
Products, page 649. }                             
$$

The conclusion of this calculation is that once the Max and Min 
conditions are suitably chosen there can be no divergence and in two
dimension the expected number of links is just of order 1, which means
that
in four dimension the result would turn to be proportional to the area
of the horizon. 

Before we move to the 4-dimensional case let us see why one would expect
the expected number of links to be the same for the null and spacelike
hypersurface. Since the counting is based on causal relations and is
manifestly Lorentz invariant (in flat case), by Lorentz invariance all
space like  spacelike planes must give the same answer, but "in the limit
of
tilting, a spacelike plane becomes null".  This gives a reason to expect
that that null Sigma
would produce the same result as spacelike Sigma.

\subsection{4-dimensional case}

Now as we have demonstrated in the previous section that 
in order to establish the area law one can consider the collapsing
null-shell of matter case and pushing the collapse to future . 
 
In 4-dimension if we push the collapse to future infinity, we will
mainly
be dealing with a cone, and  counting the number of links between
couple
of points $(x,y)$, with the following conditions.
\beq
\left \{\begin{array}{c}

x \in J^-(H),\  x \ \mbox{max in} J^-(\Sigma) \\
y \in J^+(H)\cap J^+(\Sigma), \ y \ \mbox{min in}  J^+(H)\cap
J^+(\Sigma)\\
x \in J^-(y)\\
\end{array}\right.
\eeq

Let us use the spherical coordinates and let 
$$
x= (t_1, r_1, \theta_1, \phi_1), \  y=(t_2, r_2, \theta_2, \phi_2)
$$
The above conditions can be written in terms of these coordinates as

\beq
\left \{\begin{array}{c}
x \in J^-(H)\cap J^-(\Sigma) \Ria  t_1\le T/2, t_1\le r_1\\
y \in J^+(H)\cap J^+(\Sigma) \Ria t_2 \ge T/2, r_2\le t_2\\ 
x \in J^-(y) \Ria (t_1-t_2)^2-(\vec{r_1}-\vec{r_2})^2 \ge 0.\\
\end{array}\right.
\eeq

In the 4-dimensional case, the form of the volume needed to ensure
the Max and Min conditions  depends on the relative position of the points
out-in , more precisely depend on the how   the light
cones of $x$ and $y$ intersect respect to the hypersurface.
It turns out [see Appendix.A] that one has to distinguish effectively 
three categories

a)  $t_1+ t_2 +\Delta r\le T$

b)$t_1+t_2 -\Delta r \le T $

c) $t_1+t_2 -\Delta r\le T \le t_1+t_2 +\Delta r\le T$    

 In Appendix.A we develop a  technique for evaluating those volumes, and
evaluate two of them.

For the case (a) the volumes turns out to be particularly simple and is
given by the sum of two cones indicated in.

$$
V_a = \frac{\pi}{3} (t_1 -\frac{a}{2})^4 +\frac{\pi}{3} (t_2
-\frac{a}{2})^4 
$$

Let $ x= t_1-a/2 $  and $ y= t_2-a/2$.

The expected number of links coming from this type of contributions, can
be written as
\beq
<n>_a = \int_{\cal{D}}  e^{-\frac{\pi}{3}(x^4+y^4)} dV_x dV_y
\eeq

where 
\beq
\cal{D} := \left\{ \begin{array}{c}
 0 \le\theta_i\le \pi, \ \ \  0\le\phi_i\le 2\pi \\
r_1\ge x+a/2, \ \ \ r_2\le y+a/2 \\
x\le 0,\ \ \ y\ge 0 \\

\Delta r+x+y\le 0, \ \ \ \Delta r^2-(x-y)^2 \le 0\\
\end{array} \right.
\eeq

This integral is evaluated in Appendix.B and it is shown that it has the
following
asymptotic expansion

\beq
<n>_a = \frac{\pi^3a^2}{16}( c+ (1/a))
 \eeq

Where $c$ is given by the equation of appendix
\beq
c=\int_{0}^{\infty}\int_{0}^{x}  (x-y)^4
e^{-\frac{\pi}{3} (x^4 +y^4)} \ dy
 \eeq

The other contribution namely $b$ and $ c$ are much harder to evaluate due
to the complicated algebraic form of the volume however they will give
some
thing similar to the $a$ except for the proportionality constant, this can
be seen easily from  the two dimensional case where similar contribution
give constant, and the only source for the infinity which may come from
points null where $x$ is near  $\Sigma$ and point $y$ sitting near the
horizon is cured by the Max and Min conditions. Now, since the expression
of the volumes for the other cases does not at all suggest that similar
exact calculation can be done, one may replace these volumes by an
effective
volumes obtained by expanding the original ones  around their zeros
(around the horizon and
 the values for which the pairs are linked) and keeping only the leading
order. And to avoid having
contibutions from pairs which do not orginaly contribute (coming from
regions far from the horizon), one may add other conevnient volume which
would
suppress these contributions [Work in this direction is in progress]. 

Now, if we denote by $\gamma_c$ the sum of the different coefficient
similar to $c$ and we recover the causal set units we can write for $<n>$
\beq
<n> = \gamma_c \frac{A(\Sigma\cap H)}{l_c^2}\left( 1+
(\frac{1}{\sqrt{A}})\right)
\eeq

Where $A(\Sigma\cap H)$ is the area of the cross section in which the
horizon $H$ intersect the hypersurface.

If we atribute all the B.H entropy  or part of it to $<n>$ we conclude
that the causal
set scale must be of the order of Planck scale.

Let us now come to our assumption about $a$. Equation (3.33) shows that
the
number links is proportional to the area  of the horizon $A$ , up to
corrections which goes like $\frac{1}{\sqrt{A}}$, and as we have
seen  that the dominant contribution comes from the pair of elements 
sitting near the horizon, so if the area of the horizon is small in causal set units , the pairs
contributing to the expected number of links will be few and as any
statistical calculation, this counting will be subjected to statistical
fluctuations. 
Those fluctuations go like $1/\sqrt{<n>}$, hence we may conclude
that the subleading term  $1/\sqrt{A}$ in the expression of $<n>$ is 
not too meaningful. 
 Those fluctuations should  of course be extremely small if $a\gg1$
or when
one
deals with
macroscopic black hole and this counting becomes very accurate, in the
case of Planck black hole such a counting
cannot of course carry over, or even to expect that the entropy will be
given by B.H formula. 
 
This observation  offers other possible interpretation
of this counting, the expected number of links could be understood as
a measure of the roughness of the horizon due to the discreteness of the
causal set. In counting the links, one could be effectively counting
"horizon elements" of the sort.  

\subsection{Concluding Remarks}
We have shown that  the expected number of links with
natural causality conditions may provide a source of the black hole
entropy, and the causal set scale must be of the order of Planck scale  
beyond which one would expect the continuum description of spacetime to
fail and one has only to deal with causal set . Other interesting
conclusion, is that these counting is controlled by the near horizon
geometry this seems to be in agreement with general idea that the entropy
is coming from near horizon states (as Carlip's \cite{carlip} calculation and
String
calculation\footnote{It is worth noting here that calculation given in
\cite{suss} is different from
the recent D-brane calculation and suggestes very strongly that
the black hole entropy is located relative to the horizon, whereas D-brane
calculation is done in flat spacetime where there is no horizon and seems (so
far) to be dealing with balck as whole.} \cite{suss}), and has the
advantage
that it in principle can be applied
in many cases, for instance this counting suggests an entropy for 2+1 black
hole also, proportional to the circumference of the horizon.

The  interpretation of the black hole entropy
as entanglement in-outside the horizon between  causal set elements which
we started with could be switched to other possible interpretation, namely
that by this counting one is measuring the roughness of the horizon due to
the discreteness of the causal set. 

It should be noted here that this type
of counting is not fundamental by itself, or statistical derivation of the
B.H entropy,  more fundamental derivation and interpretation can only
emerge from a better understanding of the causal set dynamics, 
as for instance the question of the entropy
increasing cannot be addressed here, however, if this type of counting
turns out to count the number of "entangled states"  defined
by tracing out   some a well defined quantum states inside the horizon, 
  than the validity of
GSL could follow from the proof of reference \cite{stp}. What one
is merely doing here could be, as we mentioned earlier, similar to counting
the number of molecules
in a box of  gas, which would turn up to a logarithmic
factor the entropy of the gas. This result is by itself of interest,
since in this way the black hole has revealed for us the  discreteness
"atomicity" nature
of spacetime, just as the quest for the statistical
mechanics
of a box of
gas tought us something important about the nature of ordinary matter on
atomic scales, revealing the existence of atoms and their size.


\appendix
\chapter*{Appendix 1}
\chapter{Evaluation of some volumes}
\section{2- dimensional Schwarzshild spacetime}

In this section we evaluate the volume needed to insure the Max and Min
 and link conditions defined by eqt(3.14) .

\beq
V= 16 M^3 \int_{u_x}^{u_y} \int_{v_x}^{v_y} \frac{e^{-r/{2M}}}{r} du dv+
 16 M^3\int_{0}^{u_y} \int_{0}^{v_x} \frac{e^{-r/{2M}}}{r} du dv
\eeq

where $u$ and $v$ are defined implicitly via the following equation
\beq
uv= (1- r/{2M})e^{r/(2M)}
\eeq
We  first evaluate the first part,
$$
V_1\equiv 16 M^3 \int_{u_x}^{u_y} \int_{v_x}^{v_y} \frac{e^{-r/{2M}}}{r}
du dv
$$
Let us  change the variables,
$$
(u,v)\ria (u,r(uv)),\Ria dv = -\frac{r}{(2M)^2u } e^{r/2M} dr
$$
$$
 \Ria V_1  = -4 M \int_{u_x}^{u_y}\int_{r(uv_x)}^{r(uv_y)} \frac{du}{u}
dr= -4 M \int_{u_x}^{u_y} ( \frac{r(uv_y)}{u} -\frac{r(u v_x)}{u}) du
$$
We will evaluate the first term .

Let 
$$
 I_1\equiv 8 M \int_{u_x}^{u_y}  \frac{r(uv_y)}{u} du
$$

$$
I_1\equiv 8 M \int_{u_x}^{u_y}  \frac{r(uv_x)}{u} du 
$$

Moving to a new variable,
$$ 
u \ria r(u v_y)\equiv r_y, \Ria du = - \frac{r_y}{(2M)^2} e^{r_y/{2M}}
dr_y
$$

$$
\Ria I_1 = \frac{1}{M} \int_{r_{xy}}^{r_{yy}} \frac{r_y^2}{1-r_y/{2M}}
dr_y= r_{yy}^2 -r_{xy}^2 +ln(u_y/u_x)
$$

Where we have used eqt(2).

Now, $I_2$  can be evaluated in similar fashion, and we get
for it ,
$$
I_2 = \frac{1}{M} \int_{r_{xx}}^{r_{yx}} \frac{r_x^2}{1-r_x/{2M}}
dr_y= r_{yx}^2 -r_{xx}^2 +ln(u_y/u_x)
$$                                   
$$
\Ria V_1= r_{yx}^2-r_{xx}^2 + r_{xy}^2-r_{yy}^2
$$
To get the second part of the volume $V$ we just substitute  $v_y$ and
$u_x$ by $0$
and  use the fact that $r(0)= 2M$ .
$$
V_2 = (2M)^2 -r_{yx}^2
$$
$$
\Ria V= V_1 +V_2 = (2M)^2 +r_{xy}^2-r_{xx}^2 -r_{yy}^2 
$$

\section{4-dimensional Flat Spacetime}

This section we develope a technique for calculating volume needed to
ensure maximality and minimality conditions  in flat space time.

We will first need the volume of the intersection of two balls (3-d solid 
sphere), so we start deriving it.

Let $ S_1$ and $ S_2$ be two balls with radius $R_1$ , $R_2$, and centers
$\vec{r_1}$ , $\vec{r_2}$ respectively,  
and let $ S_1\cap S_2$ denote their
intersection. Fig(A.1).
\begin{figure}[h]
\epsfxsize=8cm
\centerline{\epsfbox{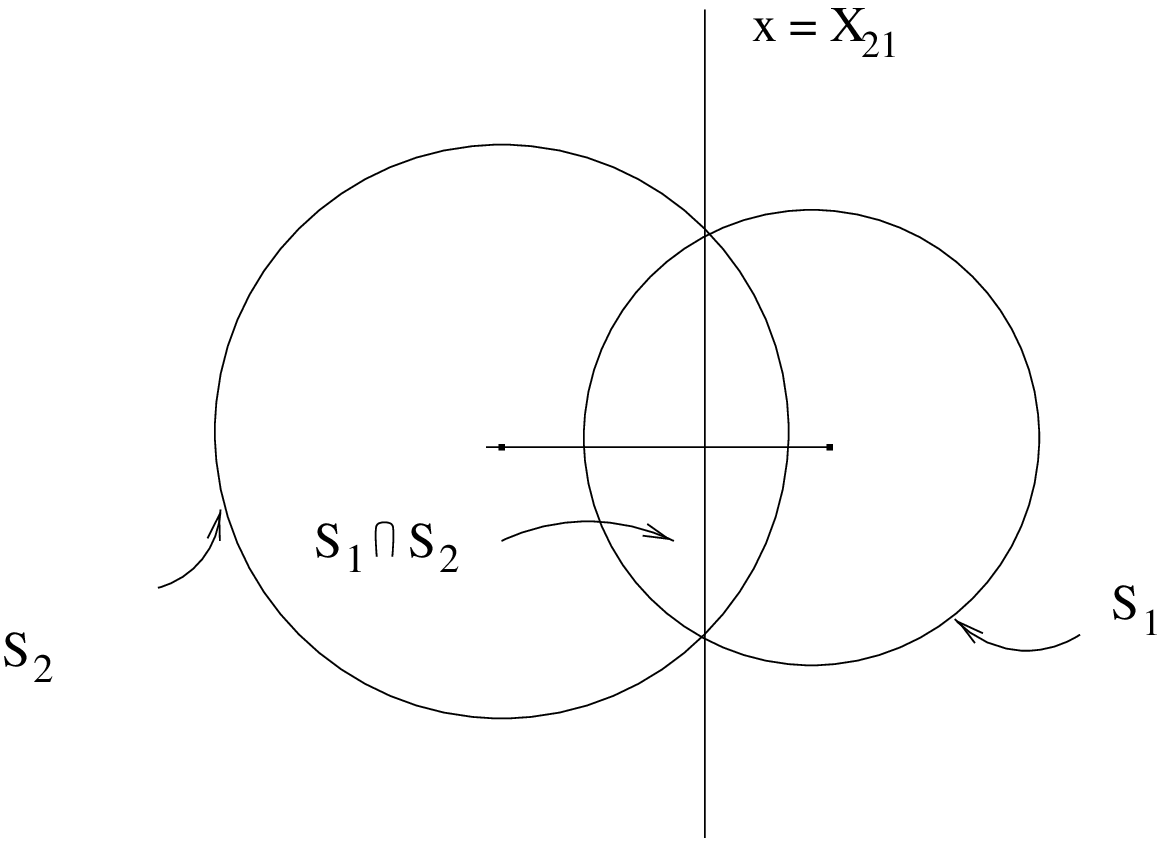}}
\caption{}
\end{figure}                   

First, it is always possible to chose the coordinate system (3-d) such
that the equations of the balls take the form.
$$
(x-a)^2+y^2+z^2\le R_1^2
$$
$$
x^2+y^2+z^2\le R_2^2
$$

Now let $x= X_{21}$ denote the equation of the plane that divide the
intersection of the two balls to $S_1$'s  and $S_2$'s
contribution. Then it is easy to show that
\beq
X_{21}= \frac{R_2^2-R_1^2+a^2}{2a}
\eeq
 Thus the volume of the intersection can  be written as
\beq
V(S_1\cap S_2)=\pi \int_{a-R_1}^{X_{21}}
(R_1^2-(x-a)^2)dx+\pi\int_{X_{21}}^{R_2}(R_2^2-x^2)dx
\eeq
It is easy to evaluate the integrals, and the final result is
$$
V(S_1\cap S_2) =\pi \left( \frac{2R_1^3}{3}+ \frac{2R_2^3}{3}-R_2^2 X_{21}
-R_1^2 (a-X_{21})+(a-X_{21})^3/3+X_{21}^3/3 \right).
$$
Written in invariant form
\beq
V(S_1\cap S_2)= \pi\left( \frac{2R_1^3}{3}+ \frac{2R_2^3}{3}-R_1^2 X_{12}
-R_2^2 (X_{21})+(X_{12})^3/3+X_{21}^3/3 \right).              
\eeq
Where we have used the following notation.
$$
X_{21}= \frac{R_2^2-R_1^2+{\Delta r}^2}{2 \Delta r}, \ \ \
X_{12} = \frac{R_1^2-R_2^2+{\Delta r}^2}{2 \Delta r}
$$

$$  
 {\Delta r}^2\equiv (\vec{r_1} -\vec{r_2})^2. 
$$

\subsection{Volume of the Union of Two Alexandroff Neighborhoods}
Other volume which we will need is the  
  the volume of the union of two Alexandroff neighborhoods.

Consider three points in 4-dimensional Minkowskin space, $p_0$ , $p_1$,
$p_2$, such that,
$$
p_2\in J^+(p_1),\ \ \  p_2\in J^+(p_0),\ \  p_1\not\in J^{\mp}( p_0)
$$ 
The aim is to calculate the
the following volume,  Fig(A.2).

\beq
V( J^+(p_0)\cap J^-(p_2) \cup \cap J^-(p_2)\cap J^+(p_1))
\eeq
\begin{figure}[h]
\epsfxsize=8cm
\centerline{\epsfbox{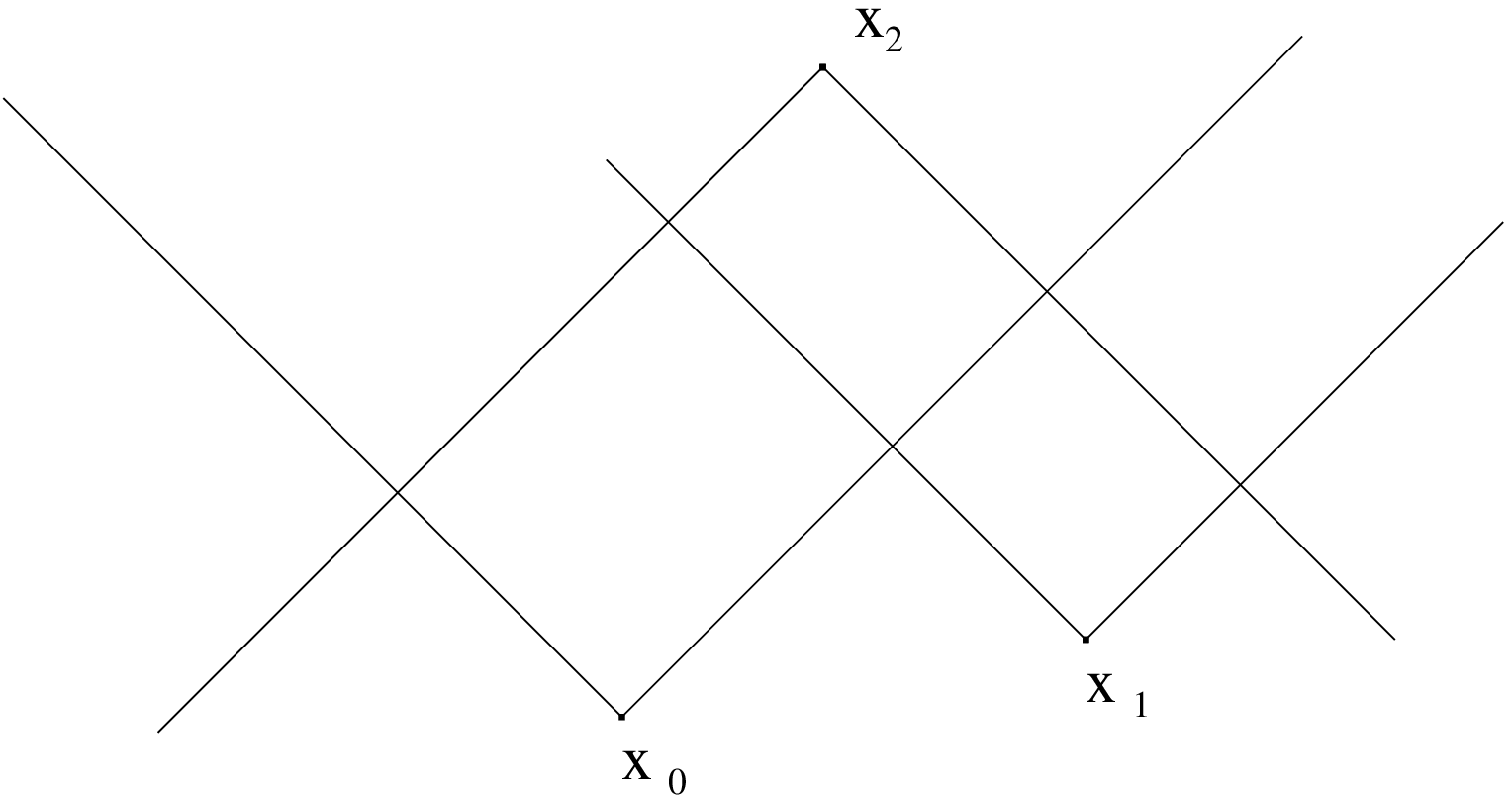}}
\caption{}
\end{figure}                    

To do this , we use the freedom, boosts, rotations, translations, and 
reduce Fig (A.2) to
Fig(A.3), namely we choose a coordinate system such that the coordinates
of 
of the points, $p_0$, $p_1$,and  $p_2$ are given by
$$
p_1=(\tau_1, a, 0,0), \ \ \ p_2 =(\tau_2,0,0,0),  , p_0\equiv 0= (0,0,0,0)
$$
The equations of the  solid light  cones are given by
$$
(x-a)^2+y^2+z^2 \le (t-\tau_1)^2, \ \ \ x^2+y^2+x^3 \le (\tau_2-t)^2, 
$$
$$  
x^2+y^2+x^3 \le t^2
$$

\begin{figure}[h]
\epsfxsize=10cm
\centerline{\epsfbox{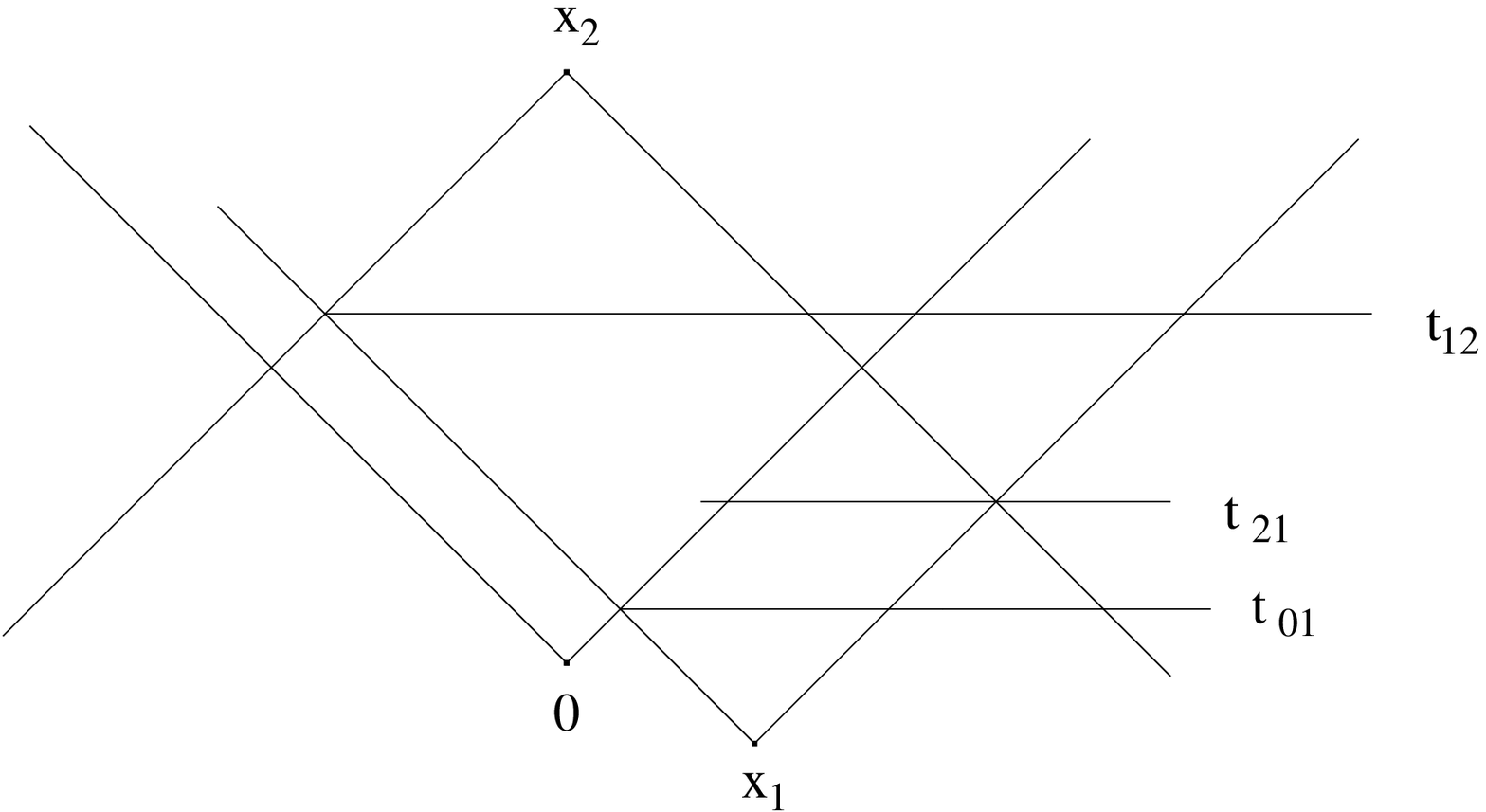}}
\caption{}
\end{figure}                    

In order to evaluate this volume it is necessary to find the times in which
the cones
intersect. Those times can be easily deduced from the light cone
equations.
$$
t_{12} =\frac{\tau_1+\tau_2+a}{2}, \ \ \ t_{21} =\frac{\tau_1+\tau_2-a}{2},
$$
$$
t_{01} = \frac{\tau_1+a}{2}, \ \ \ t_{02} =\frac{\tau_2}{2}.
$$

Fig (A.3) illustrate the evolution of the slicing in time of the volume.

Now using fig(A.4), it is not hard to establish that
\begin{eqnarray*} 
V(A_{02}\cup A_{12}) & = &  V_A(0,2) +\int_{\tau_1}^{t_{01}}
V(S_1)+\int_{t_{01}}^{t_{21}} V (\overline{S_0\cap S_1})
+\int_{t_{21}}^{\tau_2/2} \overline{S_0\cap S_2\cap S_1} \\
& = &  V_A(0,2) +\int_{\tau_1}^{t_{21}}V(S_1)+
\int_{t_{21}}^{\tau_2/2}
V(S_2\cap S_1) -\int_{t_{01}}^{\tau_2/2} V(S_0\cap S_1)
\end{eqnarray*}

\begin{figure}[h]
\epsfxsize=8cm
\centerline{\epsfbox{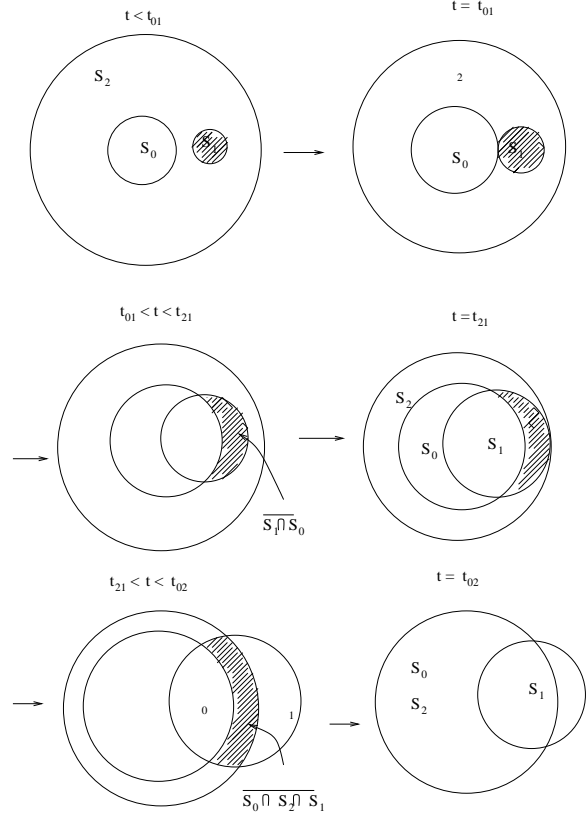}}
\caption{The shaded parts are the one needed to be included beside the
Alsexondroff neighborhood $A(0,2)$. In this figure we chosen $t_{01} \le
t_{21}$, however the form of the integral at the end does not depend on
this assumption.} 
\end{figure}                    

Where, 
\begin{eqnarray}
V(S_2\cap S_1)& = &\pi ( 2/3(t-\tau_1)^3+2/3 (\tau_2-t)^3-(t-\tau_1)^2
X_{12}-(\tau_2-t)^2 X_{21}  \nonumber \\
&  & +X_{12}^3/3+X_{21}^3/3)
\end{eqnarray}
\begin{eqnarray}
V(S_0\cap S_1) &=&\pi( 2/3 t^3 +2/3 (t-\tau_1)^3 -t^2 X_{10} -(t-\tau_1)^2
X_{01} \nonumber \\
 &  & +X_{10}^3/3 +X_{01}^3/3) 
\end{eqnarray}
Where,
\beq
X_{21} = \frac{\tau_2^2-\tau_1^2 -2t(\tau_2-\tau_1)+a^2}{2a}, \ \ \
X_{12}=\frac{\tau_1^2-\tau_2^2 -2t(\tau_1-\tau_2) + a^2}{2a}
\eeq

\beq 
X_{10} = \frac{a^2-\tau_1^2+2t\tau_1}{2a}, \  \  \ X_{01} =
\frac{\tau_1+a^2-2t\tau_1}{2a}
\eeq
Where we have use eqt(A.3) and eqt(A.5).
The above integral can easily be calculated and we give only the final
answer,
\beq
\int_{\tau1}^{t_{21}} V(S_1) dt=\frac{1}{48}(a+\tau_1-\tau_2)^4
\eeq
\beq
\int_{t_{21}}^{\tau_2/2} V(S_2 \cap S_1) dt
=\frac{1}{24}(a-\tau_1)(2\tau_2 a-
3\tau_1 a-\tau_1^2)(a+\tau_1-\tau_2)^2/a
\eeq

\beq
\int_{t_{10}}^{\tau_2/2} V(S_0\cap S_1) dt= \frac{1}{48a}(a^2-\tau_1
a+\tau_2 a-2 \tau_1^2)(\tau_2-\tau_1 -a)^3
\eeq

And we  get for $V$
\beq
V=V_A(2,0)+\frac{1}{24a}(a+\tau_2)(a-\tau1)^2 (a+\tau_1-\tau_2)^2
\eeq
Let us no write the volume in a Lorentzian invariant form using
the three independent invariant that can be formed from $\tau_1$
, $\tau_2$
and $a$.
These three  can be chosen in many ways, however it turns out that the
most convenient one are the following
\beq
 B= \tau_1^2 -a^2 =t_1^2 -r_1^2 
\eeq
\beq
C=\tau_2^2 =t_2^2-r_2^2
\eeq
\beq
S= t_1 t_2 -r_1 r_2 \sin \theta_1 \sin \theta_2 \cos (\phi_1-\phi_2) 
\eeq

Using the above equations it is easy to deduce that,

\beq
\tau_1 = \frac{S}{\sqrt{C}}
\eeq
\beq
a= B+\frac{S^2}{4 C}
\eeq

After a lengthy algebraic manipulation, $V$ takes the following
form,
\begin{eqnarray}
V &=& \frac{\pi}{24} [ (S-B)^2 +(S-C)^2 +C(B-C) \nonumber \\
 & & \mbox{}+\frac{(2S^2-CB)(C+B)-2S^3}{\sqrt{S^2-CB}}]
\end{eqnarray} 
To write $V$ in a simpler form , we introduce the following variables,
$$
 x_1= S-B, \ \ \ x_2 = S-C
$$ 
and let 
$$
a\equiv x_1+x_2 , \ \ \ b\equiv x_1 x_2
$$
Using $a$ , $b$ and $S$ as variable $V$ takes the following form,
\beq
V= a^2-b +(s-a)[S+\sqrt{aS-b}]-\frac{bS}{\sqrt{aS-b}}
\eeq

Let us call this volume $V_m$.

Now the conditions $p_2\in J^+(p_1)$ can be written in terms of $x_1$
,$x_2$ and $S$
as
\beq
 x_1+x_2\le 0
\eeq
On the other hand we have
$$
t_1^2 - r_1^2 \le 0 , \ \ \   t_2^2-r_2^2 \ge 0 
$$
it follows that
\beq
x_2 \le x_1 
\eeq
We conclude that $ x_2\le 0$.

Now, as a consistency check for the above result, we will evaluate it for
three special cases.

a) If point $p_1$ coincides with $p_0$ ( in the origin) and $p_2$ is
arbitrary we have,
$$
b =0, \ \ \mbox{ and} \ \ a= -C
$$ 
substituting in eqt(A.21) we get
$$
V= \frac{\pi}{24} C^2 =V_A(0,2)
$$
  As it should be.

b)Another a less trivial check, is for the case when, $p_1$ and $p_2$ are
null related in this case the volume should turn out to  be also $
V(2,0)$.
Using the fact that in this case 
$$
a=0, \ \ \ b=-x_2^2=-x_1^2
$$
it is easy to show that $V=V(2,0)$.

c) An other nontrivial check is when $p_2$ and $ p_0$ are null related, in
this case the volume should turn out to be just the volume of Alexandroff
volume between $p_1$ and $p_2$, which can be easily checked from the
above
formula by substituting for $ x_2 =S$ . 

 \subsection{Max , Min, and Link conditions}

In this section we will use the previous results to deduce the volumes
that are needed to  ensure Max, Min , Link.[Here we will evaluate the
volume for two cases ]

Let $\Sigma$ be a space-like hypersurface with equation $ t=T/2$ and
let  $p_1$ and
$p_1$ be two points with coordinates $ (t_1,r_1, \theta_1,\phi_1)$ and $
(t_2,r_2,
\theta_2,\phi_2) $, such that,
\beq
p_1 \in J^-(p_2) 
\eeq
 \beq
p_1\in J^-(H)\cap J^-(\Sigma) 
\eeq
\beq
 p_2 \in J^+(H)\cap J^+(\Sigma) 
\eeq
\beq
u_1 =t_1-r_1 \le 0  \mbox{and} \ v_1=t_1+r_1\ge 0
\eeq

Where $H$ stands for the future light cone of the origin (Horizon).

The aim is to calculate the volume defined by
\beq
V = V(J^+(p_1)\cap J^-(p_2) \cup J^+(p_1)\cap J^-(\Sigma) \cup
J^-(p_2)\cap J^+(H)\cap J^+(\Sigma))
\eeq

Corresponding to the condition that,
\beq
p_1 \ \mbox{Max in} \ J^-{\Sigma},\ p_2\ \mbox{Min in}\ J^+(H)\cap
J^+(\Sigma), \ p_1\ \mbox{linked to}\ p_2
\eeq

To do this we first have to distinguish three cases  depending on the
intersections
of the past light cone of $p_2$ and the boundary of the
future light cone of $p_1$.

1) If the two intersections  happen to be in the past
of $\Sigma$ Fig(A.5),  the volume is simply given by 
\beq
V= V(J^+ (p_1)\cap J^-(\Sigma)) + V(J^-(p_2)\cap J^+(\Sigma))
\eeq

The necessary and sufficient  condition to ensure  that both
intersections are in the
past of $\Sigma$ can be deduced easily from
the light cones equations and can be written as,
\beq
\frac{t_1+t_2+\Delta r}{2} \le T/2
\eeq
\begin{figure}[t]
\epsfxsize=6cm
\centerline{\epsfbox{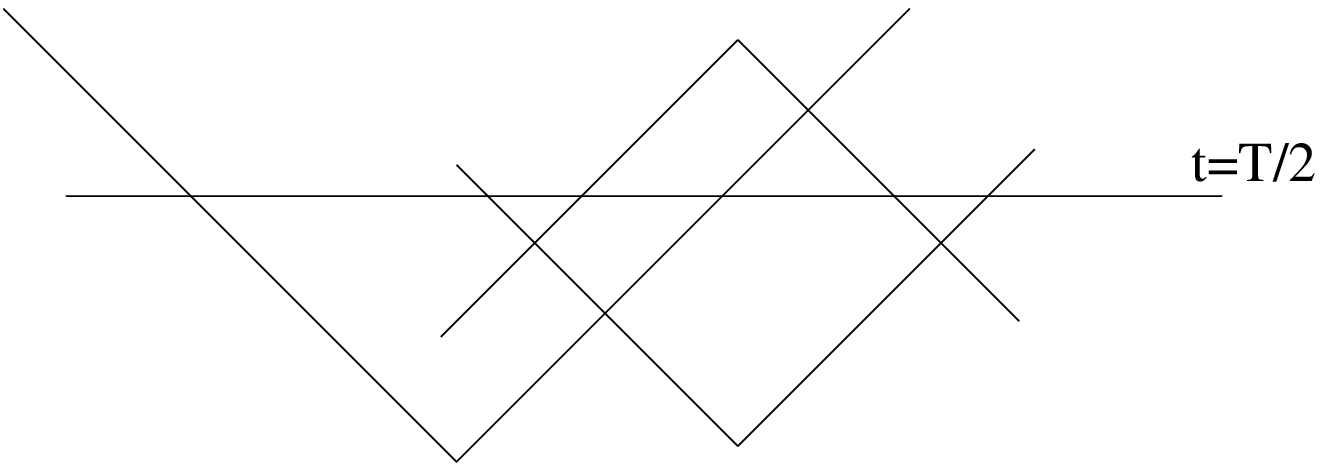}}
\caption{}
\end{figure}

 2) If the both intersections happen to be in the future of $\Sigma$,
Fig(A.6).
The necessary and sufficient condition is 
\beq
\frac{t_1+t_2-\Delta r}{2} \ge T/2
\eeq

Here also one has to distinguish three cases;

2-a)-$ v_2 \ge T$ and $ v_1 \le T$. Fig(A.6) case 2-a 

2-b)-$ v_2\ge T$ and $ v_1 \ge T$.Fig(A.6) case 2-b
2-c)-$v_2 \le T$ and $ v_1\le T$.Fig(A.6) case 2-c

\begin{figure}[h]
\epsfxsize=9cm
\centerline{\epsfbox{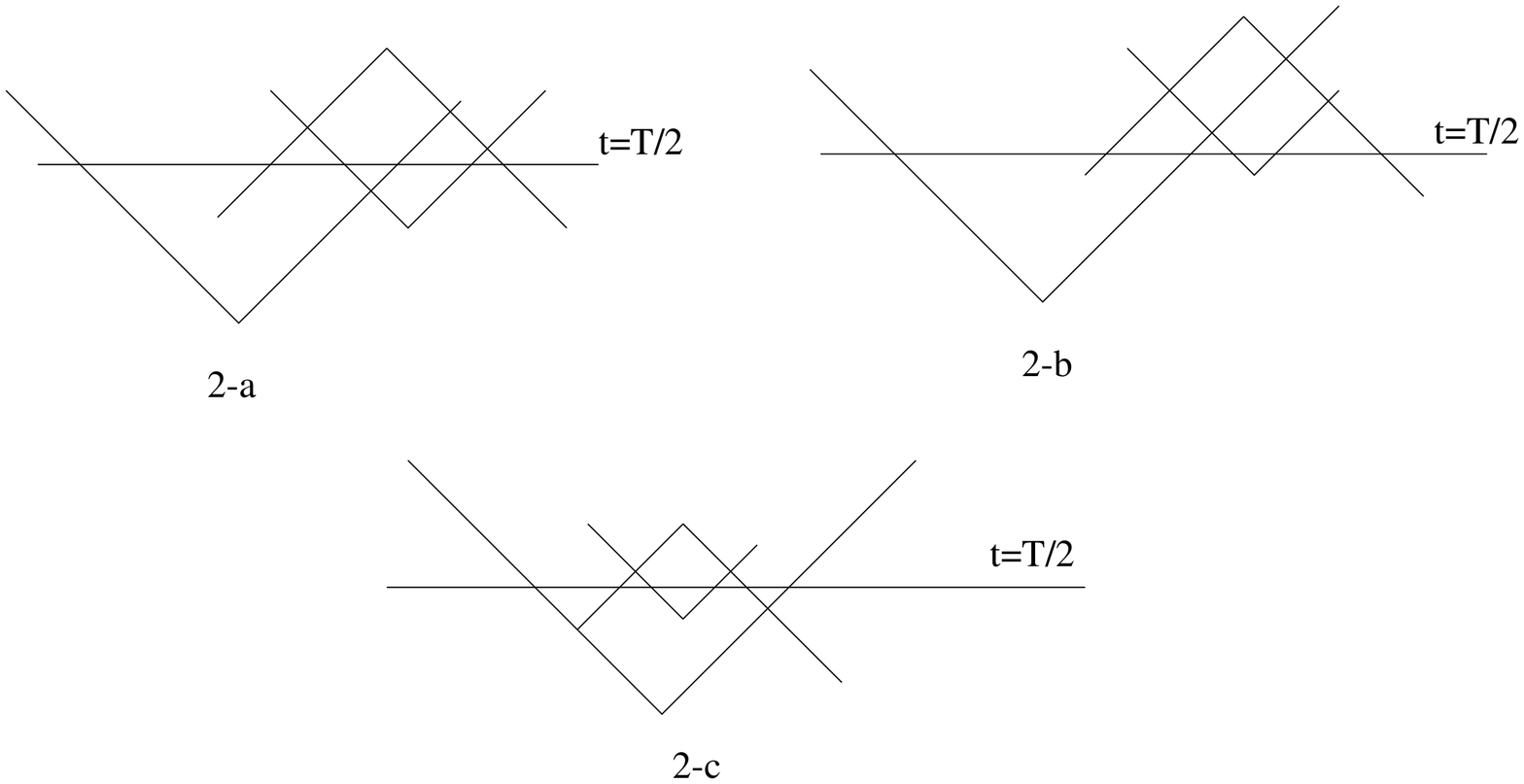}}
\caption{}
\end{figure}               

3)If one intersection is in the future of $\Sigma $ and the other in the
past, the necessary and sufficient condition for this is.
\beq
\frac{t_1+t_2-\Delta r}{2} \le T \le \frac{t_1+t_2+\Delta
r}{2} 
\eeq
One has to distinguish also three cases

3-a)-$ v_2 \ge T$ and $ v_1 \le T$. Fig(A.7) case 3-a

3-b)-$ v_2\ge T$ and $ v_1 \ge T$.Fig(A.7) case 3-b

3-c)-$v_2 \le T$ and $ v_1\le T$.Fig(A.7) case 3-c

\begin{figure}[h]
\epsfxsize=9cm
\centerline{\epsfbox{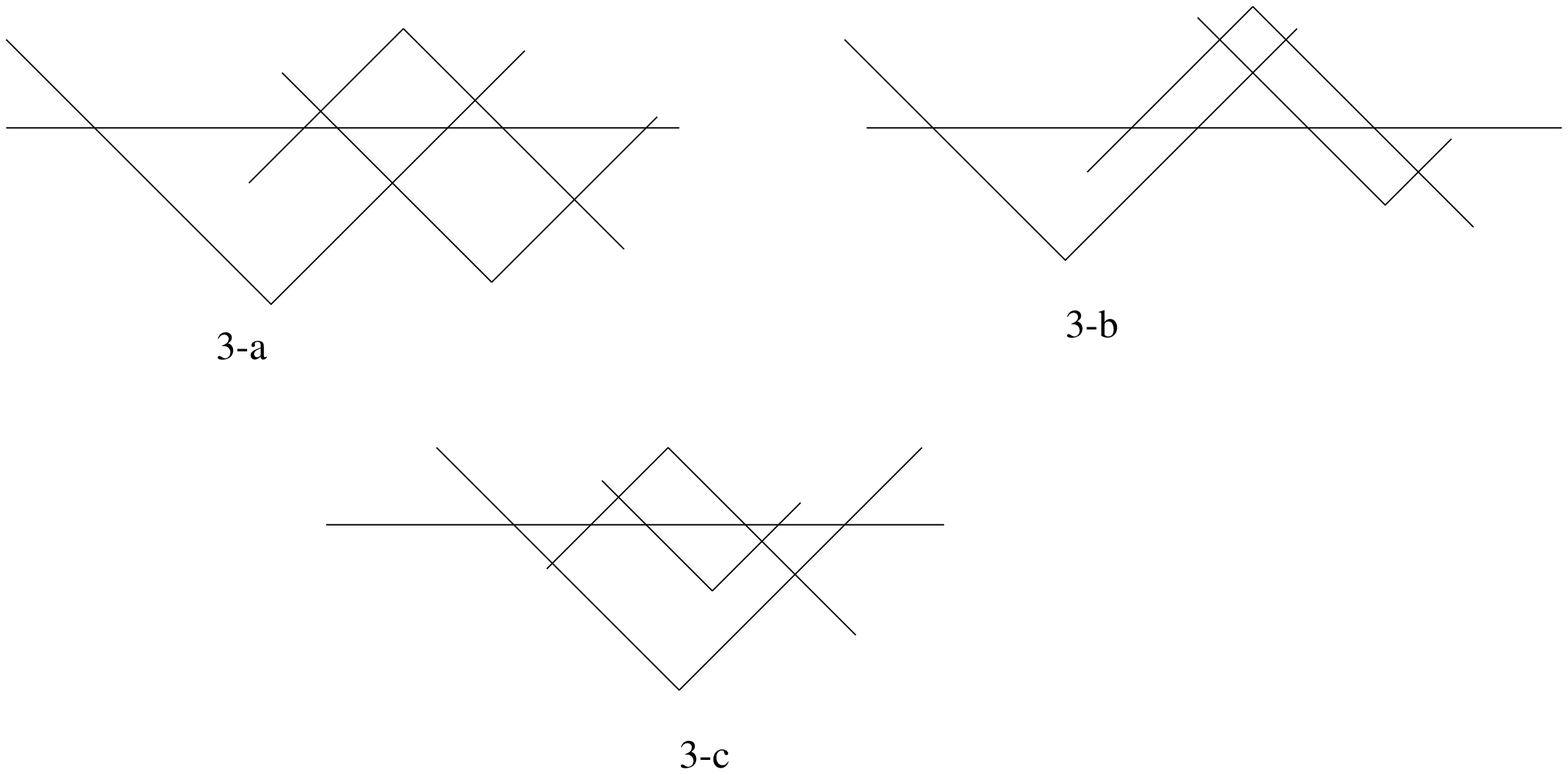}}
\caption{}
\end{figure}               
\subsection{Case 2-a}
In this section we will use the results of the previous section to deduce
the volume for the case 2-a. To the exception of  case 3-a, the other
cases may be deduced in similar fashion. For
the case 3-a, the procedure developed  here fails and one may need to
calculated some intersection of three spheres explicitly.

To calculate the volume in case 2-a we choose a coordinate system such
that $p_1$ and $p_2$ lay in the $z$-plane and having the same coordinate
$x$\footnote{Given three points in the 3-d space ,there is always a
plane containing all of them, if we choose one of them to be the origin,
than we can rotate the $x-y$-plane in such way $y$ axe is parallel to the
line joining the other tow points}

So we have following Cartesian coordinate for $p_1$ and $p_2$
$$
p_1 =(\tau_1, a,b,0) \\\ p_2=(\tau_2, a,c,0) 
$$

Since we are interested only on the cases which would give a non-suppressed
contribution, we will assume that $t_{02} \le T/2$.

Using the above coordinate the problem is practically reduced to 2+1 problem
, and the fig (A.8)\footnote{Had not we assumed that $t_{02}\le T/2$
fig(A.8)
would be inadequate in describing the evolution of the intersections.}
show
the evolution of the section
$z=0$ with time, from which the real 3+1 evolution can be easily deduced.
And important parameter of the evolution is the time in which the three
$z=0$ section of the three cones intersect at the same point fig(A.8).
This is happened in general in two different times, $t_-$ and $t_+$,
these two times
can be calculated given the equations of the three light cones,
their explicit form is a little complicated and involve a very long
calculation to deduce, even though I will give their explicit form
here,
we will not need them explicitly  for calculating the volume $V_{2.a}$.

\beq
t_{\mp} = \frac{B\mp \sqrt{\Delta}}{4A}  
\eeq
Where 
\beq
\Delta = 4[r_1^2r_2^2
-(\vec{r_1}\vec{r_2})^2][(l_1\vec{r_2}-l_2\vec{r_1})^2 -(\tau_1 l_2-\tau_2
l_1)^2] 
\eeq
\beq
B= 2(\tau_1 \vec{r_1}-\tau_2\vec{r_1})(l_1\vec{r_2}-l_2\vec{r_1}) 
\eeq
\beq
A = (\tau_1 \vec{r_1}-\tau_2\vec{r_1})^2 
+(\vec{r_1}\vec{r_2})^2 -r_1^2 r_2^2
\eeq

\begin{figure}[t]
\epsfxsize=11cm
\centerline{\epsfbox{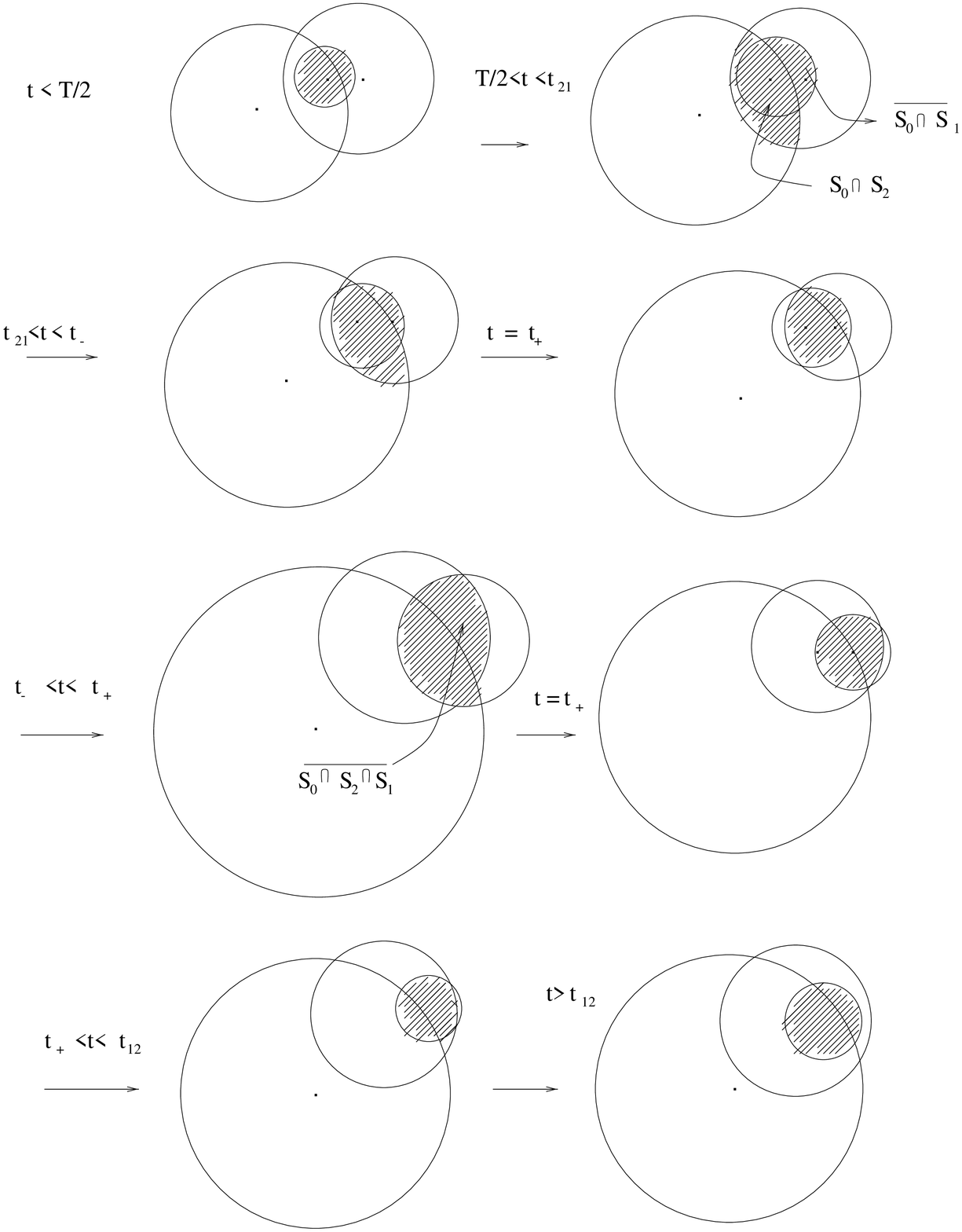}}
\caption{This diagramm illustrates the time evolution of section $z=0$
,for 
the case 2-a}
\end{figure}           

Using fig(A.8), the volume $V_2$ can be written as
\begin{eqnarray}
V_b &=& \int_{\tau_1}^{T/2} V(S_1) +\int_{T/2}^{t_-}V(S_2\cap S_0)
+\int_{T/2}^{t_-} V(\overline{S_1\cap S_0}) \nonumber \\
& & \mbox{} +
\int_{t_-}^{t_+} V(S_2\cap S_0) +\int_{t_-}^{t_+}V(\overline{S_1\cap
S_0\cap
S_2}) +\int_{t_+}^{\tau_2} V( S_1\cap S_2) dt  
\end{eqnarray}
On the other hand, if we use this coordinate system ,$V_m$ calculated in
the previous section,  can be written as
\begin{eqnarray}
V_m &=& \int_{0}^{t_{02}}V(S_0) +\int_{t_{02}}^{t_-} V(S_2\cap
S_0)+\int_{\tau_1}^{t_{01}} V(S_1) +\int_{t_{01}}^{t_-}
V(\overline{S_1\cap S_0}) \nonumber \\
& & \mbox{} +\int_{t_-}^{t_+} V(S_2\cap S_0)
+\int_{t_-}^{t_+}V(\overline{S_1\cap
S_0\cap S_2}) +\int_{t_+}^{\tau_2} V( S_1\cap S_2) dt  
\end{eqnarray}
Combining eqt(A.38) and eqt(A.39) we deduce that
\beq
V_b = V_m +\int_{T/2}^{t_{02}}V(S_2\cap S_0) -\int_{T/2}^{t_{01}}V(S_1\cap
S_0) -\int_{0}^{t_{02}} V(S_0)
\eeq

Where 

$$
t_{01}=\frac{t_1+r_1}{2} \ \ \ t_{02}=\frac{t_2-r_2}{2}
$$
Using eqt(A.5), the integrals appearing in eqt(A.40) can easily be
evaluated and we quote only the final answer,
\beq
\int_{T/2}^{t_{01}}V(S_1\cap S_0) dt=
\pi\frac{(t_1+r_1-T)^3(Tr_1+r_1^2-2t_1^2-t_1r_1)}{48r_1}
\eeq

\beq
\int_{T/2}^{t_{02}}V(S_2\cap S_0)dt
=\pi \frac{(t_2-r_2)^2(T-t_2+r_2)((T-t_2)^2 -r_2(t_2+T))}{24 r_2}
\eeq

\beq
\int_{0}^{t_{02}} V(S_0)  = \pi \frac{(t_2-r_2)^4}{48}
\eeq


\chapter{Evaluation of some integrals}

\section{Leading order of the Shawrzshild contribution}
We evaluate 
the  leading order of the following integral, which appeared 
chapter 3.
\beq
I(a)=\int_0^\infty dx \frac{x}{a-x} \int_a^x dy \frac{y}{y-a} \int_a^y
e^{z^2} dz
\eeq

let 
$$
I_1(y)= \int_a^y e^{z^2} dz 
$$
$$
I_1 = \sum_{n=0}^{\infty} \frac{y^{2n+1}-a^{2n+1}}{(2n+1)n!}
$$

myltiplying $I_1$ by $\frac{y}{y-a}$
we get:

$$
\frac{y}{y-a} I_1 \equiv I_2(y)=\frac{y}{0!} + \frac{y^3+a y^2
+a^2y}{3\times1!}
+\frac{y^5+ay^4 +a^2y^3+a^3y^2+a^4y}{5\times 2!} +\cdots
$$

Now  let 
$$
I_3(x)\equiv \frac{1}{x-a} \int_a^x I_2(y)
$$
It is easy to show that
\begin{eqnarray*}
I_3(x)  & = & \frac{a}{1\times
0! }(\frac{ (x/a)+1}{2}) \\
& & \mbox{} \frac{a^3}{3\times 1!}( \frac{
(x/a)^3+(x/a)^2+(x/a)+1}{4} \\
&  & \mbox{} +\frac{(x/a)^2+(x/a)+1}{3} +\frac{(x/a)+1}{2} ) \\
&  & \mbox{} +\frac{a^5}{5\times 2!}(\frac{(x/a)^5 +(x/a)^4+\cdots+1}{6}\\ 
&  & \mbox{}+ 
\frac{(x/a)^4 +(x/a)^3+\cdots +1}{5}+\cdots + \frac{(x/a)+1}{2}) \\
  &   & \mbox{}+\cdots
\end{eqnarray*}

Now the full integral $I(a)$ can be written as
\begin{eqnarray*}
I(a) & = & \frac{a}{1\times 0!} (\frac{J_2+J_1}{2}) \\ 
&   & \mbox{} +\frac{a^3}{3\times 1!}(\frac{J_4+J_3+J_2+J_1}{4} 
+\frac{J_3+J_2+J_1}{3}+\frac{J_2+J_1}{2}) \\
&  & \mbox{} +\frac{a^5}{5\times
2!}(\frac{J_6+J_5+\cdots+J_1}{6}+\frac{J_5+\cdots+J_1}{5}+\cdots+\frac{J_2+J_1}
{2})
\end{eqnarray*}

where 
$$
J_n\equiv \frac{1}{a^{n-1}}\int_a^\infty x^n e^{-x^2} dx.
$$
It is easy to show that,
$$
J_n = \frac{1}{2} e^{-a^2} +\frac{n-1}{2a^2} J_{n-2 }, \\\
J_1=\frac{1}{2}e^{-a^2}.
$$

The $J_n$'s with odd $n$ can be evaluted exactly, the ones with
$n$ even admit an expansion in $1/a^2$.
For generic $J_{2n}$ we can write
\begin{eqnarray*}
J_{2n}& = & J_1( 1+\frac{2n-1}{2a^2}
+\frac{(2n-1)(2n-3)}{(2a^2)^2}+ \cdots \\
&   &\mbox{} +\frac{(2n-1)!!}{2a^2)^n}(1-\frac{1}{(2a^2)^2} +\frac{1\times
3}{(2a^2)^2}+\cdots \frac{(-1)^k 1\times 3\cdots
(2k-3)}{(2a^2)^{k-1}}+\cdots))
\end{eqnarray*}

For genric $J_{2n+1}$
$$
J_{2n+1}= J_1( 1+ \frac{2n}{2a^2}+
\frac{(2n)(2n-2)}{(2a^2)^2}+\cdots+\frac{(2n)!!}{(2a^2)^n}) 
$$
Let us know write $I(a)$ as
$$
I(a)= J_1( I_0 +I_2 +I_4 +\cdots)
$$

Where $I_{2n}$ is obtained by summing terms of order
$\frac{1}{a^{2n}}$.

By doing that it can easily be shwon 	 that ,

$$
I_0 = a e^{a^2},\ \ \ I_1= a (\frac{1}{2^2}+ \frac{1}{(2a)^2 }) e^{a^2}
$$

$$
I_2= a(\frac{1}{3^2}+ \frac{7}{ 72a^2} -\frac{1}{8a4} )e^{a^2} 
$$

$$
I_3= a(\frac{1}{4^2} +\frac{5}{96a^2}
-\frac{1}{12a^4}+\frac{3}{16a^6})e^{a^2} 
$$
$$
I_4=a(\frac{1}{5^2} +\frac{13}{400a^2} -\frac{143}{2400a^4}+\frac{17}{96
a^6} -\frac{15}{32a^8})e^{a^2}
$$

Now it is easy to read the behavior of the leading term , and the
expanison of $I$ may be wriiten as

$$
I=\frac{a}{2} (\sum_{n=1}^{\infty} \frac{1}{n^2} +\frac{1}{a^2}\sum 
-\frac{1}{a^4} \sum +\frac{1}{a^6}\sum -\frac{1}{a^8}\sum +\cdots)
$$

The leading order term is simply given by 
$$
\frac{a}{2}\sum_{n=1}^{\infty} \frac{1}{n^2} = \frac{a\pi^2}{12} 
$$
The other higher order terms are given by other series and as can be seen
are raplidly convergent, and the overall series is alternating, howvere it
should be noted that this expansion  makes sense only for large
$a$.\footnote{R.Sorkin has provided an independent check of this result
which is reproduced here.} 

\section{ Evaluation of One 4-dimensional Contribution}
                                             
This appendix is devoted to the evaluation of the integral which appeared in
section chapter3 equation() .

Let
\beq
I = \int_{\cal{D}} e^{-\frac{\pi}{3}(x^4-y^4)} dV
\eeq

In spherical coordinates $dV$ is given by
\beq
dV= r_1^2 r_2^2 s_1 s_2 d\theta_1 d\theta_2 d\phi_1 d\phi_2
dr_1 dr_2 dx dy, \ \ \ s_i \equiv \sin\theta_i
\eeq
\beq
\cal{D} := \left\{ \begin{array}{c}
 0 \le\theta_i\le \pi, \ \ \  0\le\phi_i\le 2\pi \\
r_1\ge x+a/2, \ \ \ r_2\le y+a/2 \\
x\le 0,\ \ \ y\ge 0 \\

\Delta r+x+y\le 0, \ \ \ \Delta r^2-(x-y)^2 \le 0\\  
\end{array} \right.
\eeq

We make the following change of variables.
$$
(x,y,r_1,r_2, \theta_1, \theta_2,\phi_1, \phi_2)\ria
(x,y,r_1,r_2,\theta_1,\theta_2,S,\phi)
$$
Where,
$$
S =r_1 r_2 s_1 s_2 cos (\phi_2-\phi_1), \ \ \
\phi_1+\phi_2=\phi
$$
$$ 
\Ria dV \ria  \frac{1}{2}\frac{ r_1^2 r_2^2 s_1s_2}{\sqrt{r_1^2 r_2^2s_1^2 s_2^2
-S^2}}
$$

In term of the new variables the domain of integration $\cal{D}$ can be
written as

\beq
\cal{D} := \left\{ \begin{array}{c}  
0\le\phi\le 4\pi, \\
r_1\ge x+a/2, \ \ \ r_2 \le y+a/2 \\
-r_1 r_2 s_1 s_2 \le S\le r_1 r_2 s_1 s_2 \\
2S\ge r_1^2 +r_2^2 -(x+y)^2 \\
x+y\le 0 \\
2S \ge r_1^2 +r_2^2 -(x-y)^2\\
\end{array} \right.
\eeq            

Note that 
$$
xy\le 0\Ria
r_1^2 +r_2^2 -(x+y)^2 \ge  r_1^2 +r_2^2 -(x-y)^2
$$
So we need only to impose the stronger condition on $S$ namely
\beq
2S\ge r_1^2 +r_2^2 -(x+y)^2
\eeq

Let us proceed now.

The integration over $\phi$ gives just $4 \pi$ .

To integrate  $\theta_1$ we note that from the condition
$$
 - r_1 r_2 s_1 s_2 \le S \le r_1 r_2 s_1 s_2
$$
we deduce
\beq
  -\sqrt{1-\frac{S^2}{(r_1r_2)^2 s_2^2}}\le \cos\theta_1\le
\sqrt{1-\frac{S^2}{(r_1r_2)^2 s_2^2}} 
\eeq

\beq
\Ria  \theta_-\equiv \arccos(-\sqrt{1-\frac{S^2}{(r_1r_2)^2}}) \le \theta_2\le
\arccos(\sqrt{1-\frac{S^2}{(r_1r_2)^2}}) \equiv \theta_+
\eeq

It is easy to show that 
\beq
\int_{\theta_-}^{\theta_+} \frac{s_1}{\sqrt{(r_1r_2 s_1s_2)^2-S^2}} =
\frac{\pi}{r_1 r_2
s_2}
\eeq
The  integration over $\theta_2$ becomes trivial and  we end up with
\beq
I = 2\pi^2 r_1 r_2 [\pi -2 \arccos(\sqrt{1-\frac{S^2}{(r_1r_2)^2}})] \ e^{-V} 
\ dS dr_1 dr_2 dx dy 
\eeq

To integrate over $S$ we have to distinguish many case, depending on the
range of the other variables, this can be done by studying the domain of
integration $\cal{D}$, which is not hard, once that is done, the domain
$\cal{D}$ splits into  the following domains.

-A): $ S\ge 0$

\beq
{\cal{D}}(A.1):\left \{\begin{array}{c}
2 r_1 r_2\ge 2S\ge r_1^2+r_2^2-(x+y)^2\ge 0\\
r_2-x-y\ge r_1\ge r_2+x+y \\ 
 y+a/2 \ge r_2\ge a/2-y\\
-y\ge x\ge -a/2\\
a/2\ge y\ge 0\\
\end{array}\right.
\eeq

\beq
{\cal{D}}(A.2) :\left \{\begin{array}{c}
2 r_1 r_2\ge 2S\ge r_1^2+r_2^2-(x+y)^2\ge 0\\
r_2-x-y\ge r_1\ge r_2+x+y \\
 y+a/2 \ge r_2\ge -x-y\\
 x\le -a/2\\
\end{array}\right.
\eeq                                          

\beq
{\cal{D}} (A.3) :\left \{\begin{array}{c}
2 r_1 r_2\ge 2S\ge r_1^2+r_2^2-(x+y)^2\ge 0\\
r_2-x-y\ge r_1\ge x+ a/2\\
 2x+y+a/2 \le r_2\le a/2-y\\
 -x\ge y\ge  -3/2 x-a/4\\
-a/2\le x\le-a/6\\
\end{array}\right.
\eeq

\beq
{\cal{D}} (A.4): \left \{\begin{array}{c}
2 r_1 r_2\ge 2S\ge r_1^2+r_2^2-(x+y)^2\ge 0\\
r_2-x-y\ge r_1\ge x+ a/2\\
 2x+y+a/2 \le r_2\le a/2-y\\
 -x\ge y\ge  0\\
0\ge x\ge-a/6\\
\end{array}\right.
\eeq                         

\beq
{\cal{D}} (A.5) :\left \{\begin{array}{c}
2 r_1 r_2\ge 2S\ge r_1^2+r_2^2-(x+y)^2\ge 0\\
r_2-x-y\ge r_1\ge x+ a/2\\
-x-y \le r_2\le a/2-y\\
 -3/2x-a/4\ge y\ge  0\\
-a/6\ge x\ge-a/2\\              
\end{array}\right.
\eeq         

\beq
{\cal{D}} (A.6):\left \{\begin{array}{c}  
2r_1r_2 \ge 2S\ge 0\ge r_1^2+r_2^2 -(x+y)^2\\
x+a/2\le 0\le r_1\le \sqrt{(x+y)^2-r_2^2}\\
0\le r_2\le -x-y, y+a/2\\
x\le-a/2\\
\end{array}\right.
\eeq           

\beq
{\cal{D}} (A.7):\left \{\begin{array}{c}
2r_1r_2 \ge 2S\ge 0\ge r_1^2+r_2^2 -(x+y)^2\\ 
x+a/2\le r_1\le \sqrt{(x+y)^2-r_2^2}\\ 
y\le a/2, x\le -a/4\\
\end{array}\right.
\eeq                   

\beq
{\cal{D}} (A.7):\left \{\begin{array}{c} 
2r_1r_2 \ge 2S\ge 0\ge r_1^2+r_2^2 -(x+y)^2\\
x+a/2\le r_1\le \sqrt{(x+y)^2-r_2^2}\\ 
y\ge a/2
\end{array}\right.
\eeq              

-B): $S\le 0$.
\beq
{\cal{D}} (B.1):\left \{\begin{array}{c}  
0\ge 2S\ge -2r_1 r_2 \ge r_1^2 +r_2^2 -(x+y)^2\\
0\le x+a/2 \le r_1\le -r_2 -x-y\\
0\le r_2\le -2x-y-a/2\\
0\le y\le -2x-a/2 \\
 x\le -a/4
\end{array}\right. 
\eeq
\beq
{\cal{D}} (B.2):\left \{\begin{array}{c}
0\ge 2S\ge -2r_1 r_2 \ge r_1^2 +r_2^2 -(x+y)^2\\ 
 x+a/2\le 0\le r_1\le -r_2 -x-y\\ 
x\le-a/2
\end{array}\right.
\eeq
\beq
{\cal{D}} (B.3):\left \{\begin{array}{c} 
0\ge 2S  \ge r_1^2 +r_2^2 -(x+y)^2 \ge -2r_1 r_2\\  
\sqrt{(x+y)^2-r_2^2} \ge r_1\ge -r_2-x-y\\
0\le r_2 \le -x-y\\
 x\le -a/2\\
\end{array}\right.  
\eeq

\beq
{\cal{D}} (B.4):\left \{\begin{array}{c}
0\ge 2S  \ge r_1^2 +r_2^2 -(x+y)^2 \ge -2r_1 r_2\\
\sqrt{(x+y)^2-r_2^2} \ge r_1\ge -r_2-x-y\\ 
0\le r_2\le y+a/2\\
x\le -a/4
\end{array}\right.
\eeq          
\beq
{\cal{D}} (B.5):\left \{\begin{array}{c}    
\sqrt{(x+y)^2-r_2} \ge r_1 \ge x+a/2\\
-2x-y-a/2 \le r_2\le y+a/2\\
y\ge a/2
\end{array}\right.
\eeq

Now except the domain ${\cal{D}}(A.1)$ and ${\cal{D}}(A.4)$, the others give
a exponentially suppressed contributions.This can be seen as follow.
If $|x|$ or $y$ $\ge c a$ (for some $c$ of order 1), the volume in the
exponential is always
bigger than some thing like $a^4$ which will always factor out at the end of
the calculation and since the large $x$ and $ y$ are exponentially suppressed (no
null contribution in this case), the final result will be at most some thing
like $ poly(a)e^{-a^4}$, or more precisely $ a^2 e^{-a^4}$.

Let us now evaluate the contribution   ${\cal{D}}(A.1)$ and
${\cal{D}}(A.4)$

We will use the following notation
\beq
 z= r_1,\ \ \alpha =r_2^2 -(x+y)^2 \ge 0, \ \ \ t =r_2 \ge 0
\eeq
Let 
\beq
I(z)=\int_{\frac{z^2 +\alpha}{2}}^{r_1r_2}  r_1 r_2 (
\pi-2\arcsin(\frac{S}{r_1
r_2})) dS
\eeq
This integral is easy to evaluate  and the result is,
 \beq
 I(z)= z t
\left[\frac{\pi}{2}[z^2+\alpha]-[z^2+\alpha]\arcsin[\frac{z^2+\alpha}{
2zt}
] -\sqrt{4z^2t^2 -(z^2+\alpha)^2} \right]
\eeq

To integral over $z=r_1$, we let
\beq
I_1 =\int(\frac{z^2+\alpha}{2} )z dz
\eeq
\beq
I_2 =\int z(z^2+\alpha ) \arcsin(\frac{z^2+\alpha}{2z\beta}) dz
\eeq
\beq
I_3 =\int z\sqrt{4z^2\beta^2-(z^2+\alpha)^2} dz
\eeq

All the above integrals can be evaluated  and  we  only quote the result

\beq
I_1= \frac{z^2}{4}(\frac{z^2}{2}+\alpha)
\eeq
\begin{eqnarray}
I_2 &=& \frac{z^2}{2} (\frac{z^2}{2}+\alpha)
\arcsin(\frac{z^2+\alpha}{2z t}) \nonumber \\
& & -\frac{1}{8}[2(3t^4-2t^2 \alpha -\alpha^2)\arcsin[\frac{1}{2}
\frac{z^2+2t^2-\alpha}{\sqrt{t^2(t^2-\alpha)}}] \nonumber \\
& & +\frac{1}{2}
(\frac{\alpha}{2}-3t^2-\frac{z^2}{2})\sqrt{4z^2t^2-(z^2+\alpha)^2 }]
\end{eqnarray}
\begin{eqnarray}
I_3 &=& t^2(\beta^2-\alpha)
\arcsin[\frac{z^2+2t^2-\alpha}{t^2(t^2-\alpha)}]\nonumber \\
& &+\frac{1}{4} (z^2-2t^2 +\alpha)\sqrt{4z^2t^2-(z^2+\alpha)^2}
\end{eqnarray}

Now
\begin{eqnarray}
\int I(z)dz &=&  I_1-I_2-I_3 = \frac{1}{4}
z^2[\frac{z^2}{2}+\alpha][\pi-2\arcsin(\frac{z^2+\alpha}{2zt})]\nonumber\\
& & -\frac{1}{4} [t^2-\alpha]^2\arcsin \left[\frac{z^2-2t^2+\alpha}{2\sqrt{t^2(t^2-\alpha)
}}\right] \nonumber \\
& & +\frac{1}{16}[2t^2-3\alpha
-5z^2]\sqrt{4z^2t^2-(z^2+\alpha)^2}
\end{eqnarray}

We start with  the case ${\cal{D}}(A.4)$

We introduce the following notation
\beq
\alpha_+= a/2 -y, \ \ \ \alpha_-= 2x+y+a/2
\eeq
\begin{eqnarray}
\int_{\alpha_-}^{\alpha_+} I_z dz &=& 
\frac{1}{8} (\frac{\alpha_+-\alpha_-}{2})^4 t(\pi- \arcsin
\left[\frac{\alpha_-\alpha_+-t^2}{t(\alpha_+-\alpha_-)}\right])\nonumber\\
& & +\frac{1}{128} t(\alpha_-+\alpha_+)^2[8t^2
-\alpha_-^2-\alpha_+^2+6\alpha_-\alpha_+] (\pi \nonumber \\ 
&&-2\arcsin\left[\frac{\alpha_+\alpha_-+t^2}{t(\alpha_+ +\alpha_+)}\right]) \nonumber
\\
&& -\frac{1}{16}[\frac{1}{2} \alpha_-^2 +\frac{1}{2} \alpha_+^2 +4\alpha_+\alpha_-
+t^2] \sqrt{(\alpha_+^2-t^2)(t^2-\alpha_-^2)}
\end{eqnarray}

Let
\beq
f_0= \frac{1}{128} t(\alpha_-+\alpha_+)^2[8t^2
-\alpha_-^2-\alpha_+^2+6\alpha_-\alpha_+] 
\eeq
\beq
f_1 =-\frac{t}{64} (\alpha_-+\alpha_+)^2[
8t^2-\alpha_+^2-\alpha_-^2
+6\alpha_-\alpha_+]\arcsin\left[\frac{\alpha_-\alpha_+
+t^2}{t(\alpha_-+\alpha_+)}\right]
\eeq

\beq
f_2= -\frac{1}{4} (\frac{\alpha_+-\alpha_-}{2})^4 t \arcsin
\left[\frac{\alpha_-\alpha_+
-t^2}{t(\alpha_+ -\alpha_-)}\right] 
\eeq
\beq
f_3 =  -\frac{1}{16}[\frac{1}{2} \alpha_-^2 +\frac{1}{2} \alpha_+^2 +4\alpha_+\alpha_-
+t^2] \sqrt{(\alpha_+^2-t^2)(t^2-\alpha_-^2)}    
\eeq

\beq
f_4 =\frac{\pi}{8} t(\frac{\alpha_+-\alpha_-}{2})^4 
\eeq

We have
\beq
\int f_0 =\frac{t^2}{256}(\alpha_-+\alpha_+)^2
[4t^2-\alpha_-^2-\alpha_+^2+6\alpha_-\alpha_+]
\arcsin\left[\frac{\alpha_-\alpha_+ +t^2}{t(\alpha_- +\alpha_+)}\right]\
\eeq
\begin{eqnarray}
\int f_1 dt& =&  -\frac{t^2}{128}(\alpha_-+\alpha_+)^2
[4t^2-\alpha_-^2-\alpha_+^2+6\alpha_-\alpha_+]
\arcsin\left[\frac{\alpha_-\alpha_+ + t^2}{t(\alpha_-
+\alpha_+)}\right]\nonumber \\
&& 
+\frac{1}{256}(\alpha_+^2-\alpha_-^2)^2(\alpha_+^2+\alpha_-^2+4\alpha_-\alpha_+
)\arcsin\left[\frac{
2t^2 -\alpha_-^2-\alpha_+^2}{\alpha_+^2-\alpha_-^2}\right]\nonumber \\
&& -\frac{1}{128} (\alpha_-+\alpha_-)^2 [\alpha_-^2+\alpha_-^2+
\alpha_-\alpha_++t^2]\sqrt{(\alpha_-^2-t^2)(t^2-\alpha_2^2)}
\end{eqnarray}
\begin{eqnarray}
\int f_2 &=&-\frac{1}{8}(\frac{\alpha_--\alpha_+}{2})^2 t^2 \arcsin
\left[\frac{\alpha_-\alpha_+-t^2}{t(\alpha_+-\alpha_-)}\right]\nonumber \\
&& +\frac{1}{32}(\frac{\alpha_+-\alpha_-}{2})^2(\alpha_-+\alpha_-)^2\arcsin
\left[\frac{2t^2-\alpha_-^2 -\alpha_+^2}{\alpha_+^2-\alpha_-^2}\right]\nonumber \\
&& -\frac{1}{8} (\frac{\alpha_+-\alpha_-}{2})^4
t^2\arcsin\left[\frac{\alpha_-\alpha_+-t^2}{t(\alpha_+-\alpha_-)}\right] \nonumber \\
&& -poly(t) \sqrt{(t^2-\alpha_-^2)(\alpha_+-t^2)}
\end{eqnarray}
\begin{eqnarray}
\int f_3 &=& -\frac{1}{64}
(\frac{\alpha_+^2-\alpha_-^2}{2})^2[\alpha_-^2+\alpha_+^2+4\alpha_-\alpha_+]\arcsin\left[
\frac{2
 t^2 -\alpha_-^2 -\alpha_+^2}{\alpha_+^2-\alpha_-^2}\right]\nonumber \\
&& +poly(t) \sqrt{(t^2-\alpha_-^2)(\alpha_+^2-t^2)}
\end{eqnarray}

\beq
\int f_4= \frac{\pi}{16} (\frac{\alpha_+-\alpha_-}{2})^4 t^2
\eeq

Evaluating these integrals between $\alpha_-$ and $\alpha_+$, we get
\beq
\int_{\alpha_-}^{\alpha_+}(f_0 +f_1)=
\frac{\pi}{256}(\alpha_+^2-\alpha_-^2)^2(\alpha_-^2+\alpha_+^2+4\alpha_-\alpha_+)
\eeq
\beq
\int_{\alpha_-}^{\alpha_+} f_2 =\frac{\pi}{32}(\frac{\alpha_+-\alpha_-}{2})^4 \left(
(\alpha_-+\alpha_+)^2 +2(\alpha_-^2+\alpha_+^2)\right)
\eeq
\beq
\int_{\alpha_-}^{\alpha_+}
f_3=-\frac{\pi}{256}(\alpha_+^2-\alpha_-^2)^2(\alpha_-^2+\alpha_+^2+4\alpha_-\alpha_+)
\eeq

\beq
\int_{\alpha_-}^{\alpha_+} f_4 = \frac{\pi}{16} (\frac{\alpha_+-\alpha_-}{2})^4
(\alpha_+^2-\alpha_-^2)
\eeq
Summing up all the contribution we end up

\beq
I({\cal{D}}(A.4))\equiv I_1 = \frac{\pi}{32}
(\frac{\alpha_--\alpha_+}{2})^4\left[(\alpha_-+\alpha_+)^2+4\alpha_+^2\right]
\eeq
Thus the contribution of ${\cal{D}}(A.4)$ to the expected number of links
is

\begin{eqnarray}
<n>_4 &=&\frac{\pi^3}{8} \int_{-a/6}^{0} \int_{0}^{-x} (x+y)^4
[(x+a/2)^2+(a/2-y)^2]
 e^{-\frac{\pi}{3}(x^4+y^4)}dx dy\nonumber \\
& &= \frac{\pi^3}{8} \int_{0}^{a/6} \int_{0}^{x} (x-y)^4
[(x-a/2)^2+(a/2-y)^2]
 e^{-\frac{\pi}{3}(x^4+y^4)}dx dy
\end{eqnarray}
It is not hard to see that this integral will have the following
expansion
\beq
<n>_4 = \frac{\pi^3}{16}a^2 ( c +(1/a))
\eeq
Where 
\beq 
c= \int_{0}^{\infty} dx \int_{0}^{x} (x-y)^4 \
e^{-\frac{\pi}{3}(x^4+y^4)} \ dy
\eeq

For the case  ${\cal{D}} (A.1)$ we have

\beq
I_z(t+x+y)-I_z(t-x-y) =\frac{\pi}{4}(x+y)^4 t
\eeq

The contribution of ${\cal{D}} (A.1)$ to $<n>$ is given by

\begin{eqnarray}
<n>_1 &=& \frac{\pi}{2} \int_{-a/2}^{0} dx \int_{0}^{-x} (x+y)^4\ dy
\int_{a/2-y}^{a/2+y}  t \ e^{-V(x,y)} dt \nonumber \\
& & = \frac{\pi^3 a}{2} \int_{-a/2}^{0} dx \int_{0}^{-x} y(x+y)^4\
e^{-\frac{\pi}{3} x^4 -\frac{\pi}{3}y^4} \ dy\nonumber \\
& & =  \frac{\pi^3 a}{2} \int_{0}^{a/2} dx \int_{0}^{x}  y(x-y)^4
e^{-\frac{\pi}{3} (x^4 +y^4)} \ dy  
\end{eqnarray}
Now it is not difficult to see that the result must has the following form
\beq
<n>_1 = \frac{\pi^3a}{2} ( c_1 +(1/a)) 
\eeq

where 
\beq
c_1 = \int_{0}^{\infty} dx \int_{0}^{x}  y(x-y)^4
e^{-\frac{\pi}{3} (x^4 +y^4)} \ dy        
\eeq

Summing up,
\beq
<n> = \frac{\pi^3}{16} a^2 ( c + (1/a))
\eeq


\clearpage
\renewcommand{\baselinestretch}{1}

\end{document}